\documentclass[aps, prd, floatfix, nofootinbib, showpacs, twocolumn, superscriptaddress]{revtex4-1}

\usepackage{amsmath,amsfonts,amssymb,bm}
\usepackage{graphicx}
\usepackage{color}
\usepackage{enumitem}
\usepackage{multirow}
\usepackage{subfigure}
\usepackage{slashed}
\usepackage{textcomp}

\usepackage{CJKutf8}

\usepackage[mathscr,scaled=1.15]{urwchancal}
\DeclareFontFamily{OT1}{pzc}{}
\DeclareFontShape{OT1}{pzc}{m}{it}%
{<-> s * [1.15] pzcmi7t}{}
\DeclareMathAlphabet{\mathpzc}{OT1}{pzc}{m}{it}

\definecolor{purple}{rgb}{0.5,0,0.5}
\definecolor{blue}{rgb}{0.0,0,0.9}
\definecolor{prdblue}{rgb}{0.133,0.118,0.498}
\usepackage[colorlinks=true, pdfstartview=FitV, linkcolor=prdblue, citecolor= prdblue, urlcolor=prdblue]{hyperref}

\DeclareMathOperator\arctanh{atanh}

\begin{document}

\begin{CJK}{UTF8}{song}

\title{$\,$\\[-7ex]\hspace*{\fill}{\normalsize{\sf\emph{Preprint no}. NJU-INP 051/21}}\\[1ex]
Revealing pion and kaon structure via generalised parton distributions}

\author{K.~Raya}
\affiliation{School of Physics, Nankai University, Tianjin 300071, China}
\affiliation{Departamento de F\'isica Te\'orica y del Cosmos, Universidad de Granada, E-18071, Granada, Spain}
\affiliation{Instituto de Ciencias Nucleares, Universidad Nacional Aut\'onoma de M\'exico, Apartado Postal 70-543, CDMX 04510, M\'exico}
\author{Z.-F. Cui} 
\affiliation{School of Physics, Nanjing University, Nanjing, Jiangsu 210093, China}
\affiliation{Institute for Nonperturbative Physics, Nanjing University, Nanjing, Jiangsu 210093, China}
\author{L. Chang} 
\email[]{leichang@nankai.edu.cn}
\affiliation{School of Physics, Nankai University, Tianjin 300071, China}
\author{\\J. M. Morgado}
\affiliation{Department of Integrated Sciences and Center for Advanced Studies in Physics, Mathematics and Computation; University of Huelva, E-21071 Huelva; Spain.}
\author{C.~D.~Roberts}
\email[]{cdroberts@nju.edu.cn}
\affiliation{School of Physics, Nanjing University, Nanjing, Jiangsu 210093, China}
\affiliation{Institute for Nonperturbative Physics, Nanjing University, Nanjing, Jiangsu 210093, China}
\author{J.~Rodr\'{\i}guez-Quintero}
\email[]{jose.rodriguez@dfaie.uhu.es}
\affiliation{Department of Integrated Sciences and Center for Advanced Studies in Physics, Mathematics and Computation; University of Huelva, E-21071 Huelva; Spain.}

\date{2021 October 13}

\begin{abstract}
Clear windows onto emergent hadron mass (EHM) and modulations thereof by Higgs boson interactions are provided by observable measures of pion and kaon structure, many of which are accessible via generalised parton distributions (GPDs).  Beginning with algebraic GPD \emph{Ans\"atze}, constrained entirely by hadron-scale $\pi$ and $K$ valence-parton distribution functions (DFs), in whose forms both EHM and Higgs boson influences are manifest, numerous illustrations are provided.  They include the properties of electromagnetic form factors, impact parameter space GPDs, gravitational form factors and associated pressure profiles, and the character and consequences of all-orders evolution.  The analyses predict that mass-squared gravitational form factors are stiffer than electromagnetic form factors; reveal that $K$ pressure profiles are tighter than $\pi$ profiles, with both mesons sustaining near-core pressures at magnitudes similar to that expected at the core of neutron stars; deliver parameter-free predictions for $\pi$ and $K$ valence, glue, and sea GPDs at the resolving scale $\zeta=2\,$GeV; and predict that at this scale the fraction of meson mass-squared carried by glue and sea combined matches that lodged with the valence degrees-of-freedom, with a similar statement holding for mass-squared radii.
$\,$\\[1ex]
\end{abstract}




\maketitle

\end{CJK}


\section{Introduction}
Uncovering the source of the vast bulk of visible mass in the Universe is one of the highest priorities in modern physics.  It is the quest to explain the emergence of hadron mass (EHM) \cite{Denisov:2018unj, Aguilar:2019teb, Chen:2020ijn, Arrington:2021biu, Roberts:2020udq, Roberts:2020hiw, Roberts:2021nhw, Chang:2021utv}.  Two questions define the heart of this problem \cite{Roberts:2016vyn}: how does a proton, built from three valence quarks whose masses are comparable to that of an electron, acquire a mass $m_p \simeq 1$\,GeV; and how does the pion, built from similar valence degrees-of-freedom, remain nearly massless?  Within the Standard Model, the answers are expected to lie within quantum chromodynamics (QCD) and relate to the physical expressions of the scale anomaly in this theory.

It has been argued that low-energy elastic scattering of the heavy-mesons $V=J/\psi, \Upsilon$ from the proton can provide insights into connections between the proton mass and QCD's scale anomaly \cite{Krein:2020yor}.  Direct access to such reactions is difficult.  So, proposals for measurements of the scale anomaly have been based on the supposition that the photoproduction of such mesons from the proton, $e + p \to e^\prime + V + p$, provides a route to $V+p \to V+p$ \cite{Anderle:2021wcy, AbdulKhalek:2021gbh}.  However, it is now known that this assumption is ill-founded; and this realisation eliminates any model-independent link between vector-meson photoproduction and the in-proton scale anomaly \cite{Du:2020bqj, Xu:2021mju}.

In contrast, increasingly many robust connections are being drawn between EHM and the properties of pions and kaons, revealed in both phenomenological sketches and theoretical predictions \cite{Roberts:2020udq, Roberts:2020hiw, Roberts:2021nhw, Chang:2021utv, Fanelli:2016aqc, Lan:2019rba, Joo:2019bzr, deTeramond:2018ecg, Chang:2020kjj, Sufian:2020vzb, Kaur:2020vkq, Kock:2020frx, Han:2020vjp, Wan:2021bfx, Li:2016dzv, Gao:2021hvs}.  Importantly, the links can be verified empirically \cite{Denisov:2018unj, Aguilar:2019teb, Chen:2020ijn, Arrington:2021biu}.  Such is the context for the analysis of $\pi$- and $K$-meson generalised parton distributions (GPDs) \cite{Diehl:2003ny, Guidal:2013rya, Mezrag:2016hnp} described herein, which explores their capacity to reveal novel expressions of EHM in measures of pion and kaon structure and is also both a modernisation and extension of the studies in Refs.\,\cite{Zhang:2020ecj, Zhang:2021mtn}.

Our report is arranged as follows.
Section~\ref{sec:PDFsDAs} recapitulates known properties of meson light-front wave functions (LFWFs) and explains features that will subsequently be exploited.  Of particular importance are the connections drawn between LFWFs, distribution amplitudes (DAs), and distribution functions (DFs).
This is followed in Sec.\,\ref{LFWFalgebraic} by elucidation of our approach to the construction of realistic  \emph{Ans\"atze} for such LFWFs.

Section~\ref{SecOverlap} exploits the overlap representation \cite{Diehl:2000xz} to define GPDs, $H(x,\xi,t)$, on the DGLAP domain ($|x|\geq \xi$) \cite{Dokshitzer:1977sg, Gribov:1971zn, Lipatov:1974qm, Altarelli:1977zs}; uses it to develop algebraic forms for $\pi$ and $K$ GPDs; and explicates some of their consequences, with results that include predictions for meson electromagnetic form factors.

Impact parameter space (IPS) GPDs \cite{Burkardt:2000za} for the $\pi$ and $K$ are discussed in Sec.\,\ref{SecIPSGPD}, with insights developed by capitalising on the algebraic character of our GPDs.

Pion and kaon gravitational form factors and associated Breit-frame pressure profiles are reported in Sec.\,\ref{SecPressure}.  The latter require an extension of our overlap GPDs onto the ERBL domain ($|x|<\xi$) \cite{Lepage:1979zb, Efremov:1979qk, Lepage:1980fj}; so an explanation of the procedure we employ is also included.  It is based on the Radon transform scheme described in Refs.\,\cite{Chouika:2017dhe, Chouika:2017rzs}.

Section~\ref{SecAll} explains our approach to evolving the hadron scale ($\zeta_H$) distributions described in the preceding sections to any other scale $\zeta>\zeta_H$.  We benchmark our scheme by delivering predictions for all pion DFs (valence, glue and sea) and comparing the valence-quark DF with a modern analysis of relevant data \cite{Conway:1989fs, Aicher:2010cb}.
This method of evolution enables us to calculate and report in Sec.\,\ref{SecPartition} the meson mass-squared fractions carried by a given parton species at $\zeta=\zeta_2:=2\,$GeV, the value typical of contemporary data-based fits to parton distribution functions, \emph{viz}.\ to separate the meson mass-squared into valence, glue, and sea components at this scale.
The scheme is similarly exploited to calculate analogous results for the mass-squared radii.

Section~\ref{SecIPSEvolved} discusses the effect of $\zeta_H\to\zeta_2$ evolution on the IPS GPDs reported in Sec.\,\ref{SecIPSGPD}.

The body of the presentation concludes with a summary and perspective in Sec.\,\ref{epilogue}.
This is followed by an appendix that contains supplementary remarks on LFWFs.

\section{Light-front wave functions and parton distributions}
\label{sec:PDFsDAs}
At leading twist, the LFWF for the $u$ quark in a ${\mathsf P}=u\bar h$ pseudoscalar meson can be written:
\begin{align}
\label{EqLFWFG}
%
\psi_{\mathsf P}^u(x,k_\perp^2;\zeta) & =
{\rm tr}_{\rm CD} Z_2 \int \frac{dk_3 dk_4}{\pi}
\delta(n\cdot k - x n\cdot P_{\mathsf P}) \nonumber \\
& \qquad \times \gamma_5 \gamma\cdot n \chi_{\mathsf P}(k-P_{\mathsf P}/2,P_{\mathsf P};\zeta)\,.
\end{align}
The trace is over colour and spinor indices;
$Z_2$ is the dressed-quark wave function renormalisation constant;
%
$k=(k_1,k_2,k_3,k_4)$,
$x$ is the $u$-quark light-front fraction of the meson's total momentum, $P_{\mathsf P}$,  and $k_\perp=(k_1,k_2)$ is its momentum in the light-front transverse plane;
%
%
$n$ is a light-like four-vector, $n^2=0$, $n\cdot P_{\mathsf P} = -m_{\mathsf P}$ in the meson's rest frame, with $m_{\mathsf P}$ its mass;
$\chi_{\mathsf P}$ is the Poincar\'e-covariant Bethe-Salpeter wave function describing the meson's structure;
and $\zeta$ is the renormalisation scale.
The LFWF is invariant under light-front Lorentz boosts.  Thus, when solving bound-state scattering problems using a light-front formulation, compressed or contracted objects are not encountered \cite{Terrell:1959zz, Penrose:1959vz, Weisskopf}, \emph{e.g}., the cross-section for the meson+proton Drell-Yan process does not depend on whether the proton is at rest or moving.

Eq.\,\eqref{EqLFWFG} follows Ref.\,\cite{tHooft:1974pnl} in defining the LFWF via projection of the meson's Bethe-Salpeter wave function onto the light-front.  The approach is efficacious for strong-interaction bound-states \cite{Chang:2013pq}, yielding a LFWF expressed in a quasiparticle basis whose character is defined by the renormalisation scale.  At $\zeta=\zeta_{\mathpzc H}$, the hadron scale, the quark$+$antiquark quasiparticle pair embody all properties of ${\mathsf P}$ \cite{Ding:2019qlr, Ding:2019lwe, Cui:2019dwv, Cui:2020dlm, Cui:2020tdf, Roberts:2021nhw}; for instance, they are invested with all the meson's charge, mass, and light-front momentum.  Owning a LFWF, one has access to many key hadron structure measures, including form factors, DAs and DFs.
For some such measures, evolution -- DGLAP \cite{Dokshitzer:1977sg, Gribov:1971zn, Lipatov:1974qm, Altarelli:1977zs} or either ERBL \cite{Lepage:1979zb, Efremov:1979qk, Lepage:1980fj} -- is necessary before comparisons become possible with quantities inferred from experiment.  As described in Sec.\,\ref{SecAll}, that is achieved herein by using QCD's process-independent effective charge to integrate the relevant evolution equations \cite{Ding:2019qlr, Ding:2019lwe, Cui:2019dwv, Cui:2020dlm, Cui:2020tdf, Roberts:2021nhw}.

A meson's Bethe-Salpeter wave function can be obtained by solving an appropriate coupled set of gap and Bethe-Salpeter equations.  Realistic and practicable kernels for these equations are being developed \cite{Eichmann:2016yit, Fischer:2018sdj, Qin:2020rad, Qin:2020jig} and results are available for spectra, DAs, DFs, and form factors \cite{Shi:2015esa, Ding:2015rkn, Li:2016mah, Gao:2017mmp, Chen:2018rwz, Ding:2018xwy, Binosi:2018rht, Ding:2019qlr, Ding:2019lwe, Cui:2019dwv, Cui:2020dlm, Cui:2020tdf, Roberts:2021nhw}.  Accurate projection of $\chi_{\mathsf P}$ onto the light-front requires the use of perturbation theory integral representations (PTIRs) \cite{Nakanishi:1969ph} for all Schwinger functions involved.  Construction of PTIRs is time consuming and case specific.  So, following Refs.\,\cite{Xu:2018eii, Zhang:2021mtn}, we take a different and, for now, more insightful path.

Profiting from continuum and lattice analyses that have generated coherent predictions for meson DAs and DFs \cite{Roberts:2021nhw}, we work in reverse to form LFWFs that replicate such mutually consistent results for the pion and kaon.  The procedure begins by recalling a relationship between LFWFs and leading-twist meson DAs \cite{Brodsky:1989pv}:
\begin{equation}
\label{eq:DA}
f_{\mathsf P} \varphi^{u}_{\mathsf P}(x,\zeta_{\mathpzc H})
=  \int \frac{dk^2_\perp}{16 \pi^3} \psi_{{\mathsf P}}^{u}\left(x,k_\perp^2;\zeta_{\mathpzc H} \right)\,,
\end{equation}
where $f_{\mathsf P}$ is the meson's leptonic decay constant, which is an order parameter for dynamical chiral symmetry breaking (DCSB), itself a corollary of EHM.
As written, the DA is unit normalised.  That connected with the $\bar{h}$ quasiparticle is
\begin{equation}
\varphi_{\mathsf P}^{\bar h}(x;\zeta_{\mathpzc H}) = \varphi_{\mathsf P}^u(1-x;\zeta_{\mathpzc H})\,.
\end{equation}
Importantly, the valence quark DFs can also be defined through the LFWFs \cite{Brodsky:1989pv, Cui:2020dlm, Cui:2020tdf, Roberts:2021nhw}:
\begin{subequations}
\label{eq:uMzetaH}
\begin{align}
{\mathpzc u}^{\mathsf P}(x;\zeta_{\mathpzc H}) & = \int \frac{d^2k_\perp}{16\pi^3} \left| \psi_{{\mathsf P}}^{u}\left(x,k_\perp^2;\zeta_{\mathpzc H} \right) \right|^2  \,,
\label{eq:uMzetaHa}\\
\overline{\mathpzc h}^{\mathsf P}(x;\zeta_{\mathpzc H})& ={\mathpzc u}^{\mathsf P}(1-x;\zeta_{\mathpzc H})\,.
\end{align}
\end{subequations}
Eqs.\,\eqref{eq:DA}\,--\,\eqref{eq:uMzetaH} can reliably be used to constrain \emph{Ans\"atze} for $\pi$ and $K$ LFWFs; and two complementary approaches will subsequently be used to develop efficacious parametrisations for the LFWF in Eq.\,\eqref{EqLFWFG}.

\section{Algebraic models for LFWFs}
\label{LFWFalgebraic}
\subsection{Spectral representation}
PTIRs have successfully been used to explore the character of meson parton quasidistributions \cite{Xu:2018eii}. Following that approach, the LFWF can be expressed thus:\footnote{
This expression differs from its analogue \cite[Eq.\,(10)]{Xu:2018eii} by a factor $f_{\mathsf{P}}$.  The mismatch is explained by a different normalisation convention; therein, $\int dx \int d^2{\bf k}_\perp \psi_{{\mathsf P}}^{u}(x,k_\perp^2;\zeta_{\mathpzc H})=16 \pi^3$.
}
\begin{equation}
\label{psiLFWF}
\psi_{{\mathsf P}}^{u}(x,k_\perp^2;\zeta_{\mathpzc H}) =
12 \left[ M_u (1-x) + M_{\bar h} x \right] {\mathpzc X}_{\mathsf P}(x;\sigma_\perp^2) \,,
\end{equation}
where:
$M_{u,\bar h}$ are dressed-quark/-antiquark mass-scales, whose values should be commensurate with the infrared size of the relevant running masses \cite[Fig.\,2.5]{Roberts:2021nhw};
$\sigma_\perp=k_\perp^2+\Omega_{\mathsf P}^2$,
{\allowdisplaybreaks
\begin{subequations}
\label{Inputs}
\begin{align}
\nonumber
\Omega_{\mathsf P}^2  & = v M_u^2  + (1-v)\Lambda_{\mathsf P}^2 \nonumber \\
&\quad  + (M_{\bar h}^2-M_u^2)\left(x - \tfrac{1}{2}[1-w][1-v]\right) \nonumber \\
&\quad  + ( x [x-1] + \tfrac{1}{4} [1-v] [1-w^2]) \, m_{\mathsf P}^2\,;
\label{Omega}\\
\nonumber
{\mathpzc X}_{\mathsf P}(x;\sigma_\perp^2)  & =
\left[\int_{-1}^{1-2x} \! dw \int_{1+\frac{2x}{w-1}}^1 \!dv \right.\\
& \left. \quad + \int_{1-2x}^1 \! dw \int_{\frac{w-1+2x}{w+1}}^1 \!dv \right]
\frac{\rho_{\mathsf P}(w) }{{\mathpzc n}_{\mathsf P} } \frac{\Lambda_{\mathsf P}^2}{\sigma_\perp^2}\,; \label{X2c}
\end{align}
\end{subequations}
$\Lambda_{\mathsf P}$ is a mass parameter; and ${\mathpzc n}_{\mathsf P}$ is the Bethe-Salpeter wave function's canonical normalisation constant \cite{Nakanishi:1969ph}.
}

It is worth highlighting the $\Delta_{hu}^2=M_h^2-M_u^2$ term in Eq.\,\eqref{Omega}.  As indicated above, the mass scales in this difference are an expression of EHM through its corollary of DCSB.  $\Delta_{du}$ vanishes for the pion in the isospin-symmetry limit, which is a good approximation in Nature.  On the other hand, $\Delta_{su}$ is significant for the kaon; and this term plays an important role in recovering known flavour-symmetry breaking effects in kaon DAs and DFs \cite{Shi:2015esa, Li:2016mah, Cui:2020dlm, Cui:2020tdf, Roberts:2021nhw}.

\begin{table}[t!]
\caption{
\label{tab:params}
Used in Eqs.\,\eqref{psiLFWF} -- \eqref{eq:spectralw}, one reproduces the pion and kaon DAs described in Refs.\,\cite{Cui:2020dlm, Cui:2020tdf} and empirical values for the meson decay constants \cite{Zyla:2020zbs}: $f_\pi=0.092\,$GeV, $f_K = 0.11\,$GeV.  (Mass dimensioned quantities in GeV.)
}
\begin{tabular*}
{\hsize}
{
l@{\extracolsep{0ptplus1fil}}
|l@{\extracolsep{0ptplus1fil}}
l@{\extracolsep{0ptplus1fil}}
l@{\extracolsep{0ptplus1fil}}
l@{\extracolsep{0ptplus1fil}}
l@{\extracolsep{0ptplus1fil}}
l@{\extracolsep{0ptplus1fil}}
l@{\extracolsep{0ptplus1fil}}}\hline
${\mathsf P}$ & $ m_{\mathsf P}$ & $M_u$ & $M_h$ & $\Lambda_{\mathsf P}$ & $b_0^{\mathsf P}$ & $\omega_0^{\mathsf P}$ & $v_{\mathsf P}$  \\
\hline
$\pi$ & $0.14$ & 0.31 & $\phantom{1.2} M_u$ & $\phantom{3} M_u$ & $0.316$ & $1.23\phantom{5}$ & 0\phantom{.41} \\
\hline
$K$ &  $0.49$ & 0.31 & $1.2 M_u$  &$3 M_s$ &  $0.1\phantom{75}$ & $0.625$ & $0.41$ \\\hline
\end{tabular*}
\end{table}

The bridge from Eq.\,\eqref{psiLFWF} to a realistic LFWF is $\rho_H(w)$, the spectral weight.  When judiciously chosen \cite{Xu:2018eii}, results can be obtained for many hadron structure measures that are practically equivalent to the most sophisticated predictions currently available.  For $\pi$- and $K$-mesons, an optimal parametrisation is \cite{Xu:2018eii}:
\begin{align}
\rho_{\mathsf P}(\omega) 
& = \frac{1+\omega\; v_{\mathsf P}}{2a_{\mathsf P} b_0^{\mathsf P}} \nonumber  \\
 & \quad \times  \left[\mbox{sech}^2 \! \left(\frac{\omega-\omega_0^{\mathsf P}}{2b_0^{\mathsf P}}\right)
 +\mbox{sech}^2\! \left(\frac{\omega+\omega_0^{\mathsf P}}{2b_0^{\mathsf P}}\right)\right]
,
\label{eq:spectralw}
\end{align}
where the density's profile is controlled by $b_0^{\mathsf P}$, $\omega_0^{\mathsf P}$;
skewing, driven by differences in the current-masses of a meson's valence-quark/-antiquark constituents, is introduced by $v_{\mathsf P} \neq 0$;
and unit normalisation is ensured via $a_{\mathsf P}$.
This flexible, yet compact, form reproduces the $\pi$ and $K$ DAs described in Refs.\,\cite{Cui:2020dlm, Cui:2020tdf} when evaluated using the parameters in Table~\ref{tab:params} \cite{Zhang:2021mtn}.

\subsection{Factorised representation}
\label{SecFactorised}
As sketched in Appendix\,\ref{app:PTIRfac}, the LFWF specified by Eq.\,\eqref{psiLFWF} factorises up to corrections that vanish as $m_{\mathsf P}^2$, $M_{\bar h}^2 - M_{u}^2$.  This suggests that for the $\pi$ and $K$, especially so far as integrated quantities are involved, it is a good approximation to write:
\begin{subequations}
\label{eq:facLFWF}
\begin{align}
\psi_{{\mathsf P}}^{u}\left(x,k_\perp^2;\zeta_{\mathpzc H} \right)   & =  \tilde\varphi_{\mathsf P}^u(x;\zeta_{\mathpzc H})  \tilde\psi_{{\mathsf P}}^{u}\left(k_\perp^2;\zeta_{\mathpzc H} \right) \,, \\
\tilde\varphi_{\mathsf P}^u(x;\zeta_{\mathpzc H}) &
= \varphi_{\mathsf P}^u(x;\zeta_{\mathpzc H})/{\mathpzc n}_{\tilde\varphi_{\mathsf P}^u}\,,\\
{\mathpzc n}_{\tilde\varphi_{\mathsf P}^u}^2 & = \int_0^1 dx\, [\varphi_{\mathsf P}^u(x;\zeta_{\mathpzc H})]^2,
\end{align}
\end{subequations}
with the optimal choice for $\tilde\psi_{{\mathsf P}}^{u}\left(k_\perp^2;\zeta_{\mathpzc H} \right)$ decided by the application. This assumption is supported by analyses in Ref.\,\cite{Roberts:2021nhw}, which show that when Eqs.\,\eqref{eq:facLFWF} are used in calculating meson structure measures, the accuracy typically exceeds the precision of foreseeable experiments.

\begin{table}[t!]
\caption{
\label{tab:DFs}
When these powers and coefficients are used in Eq.\,\eqref{eq:param}, one obtains representations of $\pi$ and $K$ DFs that are in accord with all available data \cite{Cui:2020dlm, Cui:2020tdf}.
}
\begin{tabular*}
{\hsize}
{
l@{\extracolsep{0ptplus1fil}}
|l@{\extracolsep{0ptplus1fil}}
l@{\extracolsep{0ptplus1fil}}
l@{\extracolsep{0ptplus1fil}}
l@{\extracolsep{0ptplus1fil}}
l@{\extracolsep{0ptplus1fil}}}\hline
%
${\mathsf P}\ $ & $ n_{\mathsf P}$ & $\rho_{\mathsf P}$ & $\gamma_{\mathsf P}$ & $\alpha_{\mathsf P}$ & $\beta_{\mathsf P}$ \\ \hline
$\pi\ $ & $375.3$ & $-2.51$ & $\phantom{-}2.03$ & $1.0\phantom{00}$ & $1.0\phantom{00}$ \\\hline
$K\ $ &  $299.2$ & $\phantom{-}5.00$ & $-5.97$ & $0.064$ & $0.048$ \\\hline
\end{tabular*}
\end{table}

Referring to Eq.\,\eqref{eq:uMzetaHa}, Eq.\,\eqref{eq:facLFWF} yields
\begin{equation}
\label{PDFeqPDA2}
{\mathpzc u}^{\mathsf P}(x;\zeta_{\mathpzc H})
= [\tilde\varphi_{{\mathsf P}}^u(x;\zeta_{\mathpzc H})]^2 ;
\end{equation}
so, for factorised LFWFs, the $x$-dependence is completely prescribed by the forms of the $\pi$ and $K$ DFs at the hadron-scale.  Such were determined in Refs.\,\cite{Cui:2020dlm, Cui:2020tdf}:
\begin{align}\label{eq:param}
{\mathpzc u}^{\mathsf P}&(x;\zeta_{\mathpzc H}) = n_{\mathsf P} x^2 (1-x)^2 \nonumber \\
& \times \left[ 1 + \rho_{\mathsf P} x^\frac{\alpha_{\mathsf P}}{2} (1-x)^\frac{\beta_{\mathsf P}}{2} + \gamma_{\mathsf P} x^{\alpha_{\mathsf P}} (1-x)^{\beta_{\mathsf P}} \right]^2 \,,
\end{align}
with the coefficients and powers listed in Table~\ref{tab:DFs}.
The $k_\perp^2$-dependence remains to be determined; and as shown below, that can be achieved by using one or two additional pieces of empirical information.

\section{Generalised Parton Distribution: Overlap Representation}
\label{SecOverlap}
Employing the LFWFs just described, the analyses of one-dimensional $\pi$ and $K$ structure measures in Refs.\,\cite{Cui:2020dlm, Cui:2020tdf} can be extended to obtain three-dimensional images by using the GPD overlap representation \cite{Burkardt:2002hr, Diehl:2003ny}:
\begin{align}
%
& H^u_{\mathsf P}(x,\xi,t;\zeta_{\mathpzc H})  \nonumber \\
&= \int \frac{d^2{k_\perp}}{16 \pi^3}
\psi_{{\mathsf P}}^{u\ast}\left(x_-,{k}_{\perp -}^2;\zeta_{\mathpzc H} \right)
\psi_{{\mathsf P}}^{u}\left( x_+,{k}_{\perp +}^2;\zeta_{\mathpzc H} \right) \,,
\label{eq:overlap}
\end{align}
%
with: $P=(p^\prime+p)/2$, where $p^\prime$, $p$ are the final, initial meson momenta in the defining scattering process;
$\Delta = p^\prime - p$, $P\cdot \Delta = 0$, $t=-\Delta^2$;
the ``skewness''
\begin{equation}
\xi = [-n\cdot \Delta]/[2 n\cdot P]\,,\;\mbox{$|\xi|\leq 1$}\,;
\end{equation}
and
\begin{subequations}
\begin{align}
x_\mp & = (x \mp \xi)/(1\mp \xi)\,, \\
{k}_{\perp \mp} & = {k}_\perp \pm ({\Delta}_\perp/2)(1-x)/(1 \mp \xi) \,.
\end{align}
\end{subequations}
For subsequent use, we record a useful kinematic identity \cite[Eq.\,(4.33)]{Chouika:2018mbk}:
\begin{equation}
\label{EqNabil}
\Delta_\perp^2 = \Delta^2 (1-\xi^2) - 4 \xi^2 m_{\mathsf P}^2;
\end{equation}
hence, $\Delta^2=\Delta_\perp^2$ when $\xi=0$.
%
%
%
By construction, the overlap representation ensures positivity \cite{Pire:1998nw}; and since, owing to time-reversal invariance, $H^u_{\mathsf P}(x,\xi,t)  =H^u_{\mathsf P}(x,-\xi,t)$, we only consider $\xi \geq 0$ in the following.

The expression in Eq.\,\eqref{eq:overlap} defines the GPD on $|x|\geq \xi$, which is referred to as the DGLAP domain because the evolution equations in Refs.\,\cite{Dokshitzer:1977sg, Gribov:1971zn, Lipatov:1974qm, Altarelli:1977zs} express scale evolution thereupon.  Further, Eq.\,\eqref{eq:uMzetaH} is seen to follow as the GPD's forward ($\Delta^2=0$) limit once it is realised that $x \geq -\xi $ is the domain of support for the quark GPD, whereas the antiquark GPD is only nonzero on $x\leq\xi$.

A given quark's contribution to a target meson's elastic electromagnetic form factor is expressed through a zeroth GPD moment:
\begin{equation}\label{eq:FF}
F_{\mathsf P}^u(\Delta^2) = \int_{-1}^{1} dx \, H_{\mathsf P}^u(x,0,-\Delta^2;\zeta_{\mathpzc H})  \,.
\end{equation}
The answer is independent of $\xi$; hence, computation with $\xi=0$ is sufficient.  The complete meson form factor is obtained by summing over quark contributions:
\begin{equation}\label{eq:FFM}
F_{\mathsf P}(\Delta^2) = e_u F_{\mathsf P}^u(\Delta^2) + e_{\bar h} F_{\mathsf P}^h(\Delta^2) \,,
\end{equation}
where $e_u$, $e_{\bar h}$ are the valence-constituent charges in units of the positron charge.  As usual, the meson's charge radius is defined via
\begin{equation}
\label{EqRadius}
r^2_{\mathsf P} = -\left.[6/F_{\mathsf P}(0)]\, dF_{\mathsf P}(\Delta^2)/d\Delta^2 \right|_{\Delta^2=0}\,.
\end{equation}

\begin{figure}[t!]
\vspace*{3.5ex}

\leftline{\hspace*{0.5em}{\large{\textsf{A}}}}
\vspace*{-5ex}
\centerline{\includegraphics[clip, width=0.36\textwidth]{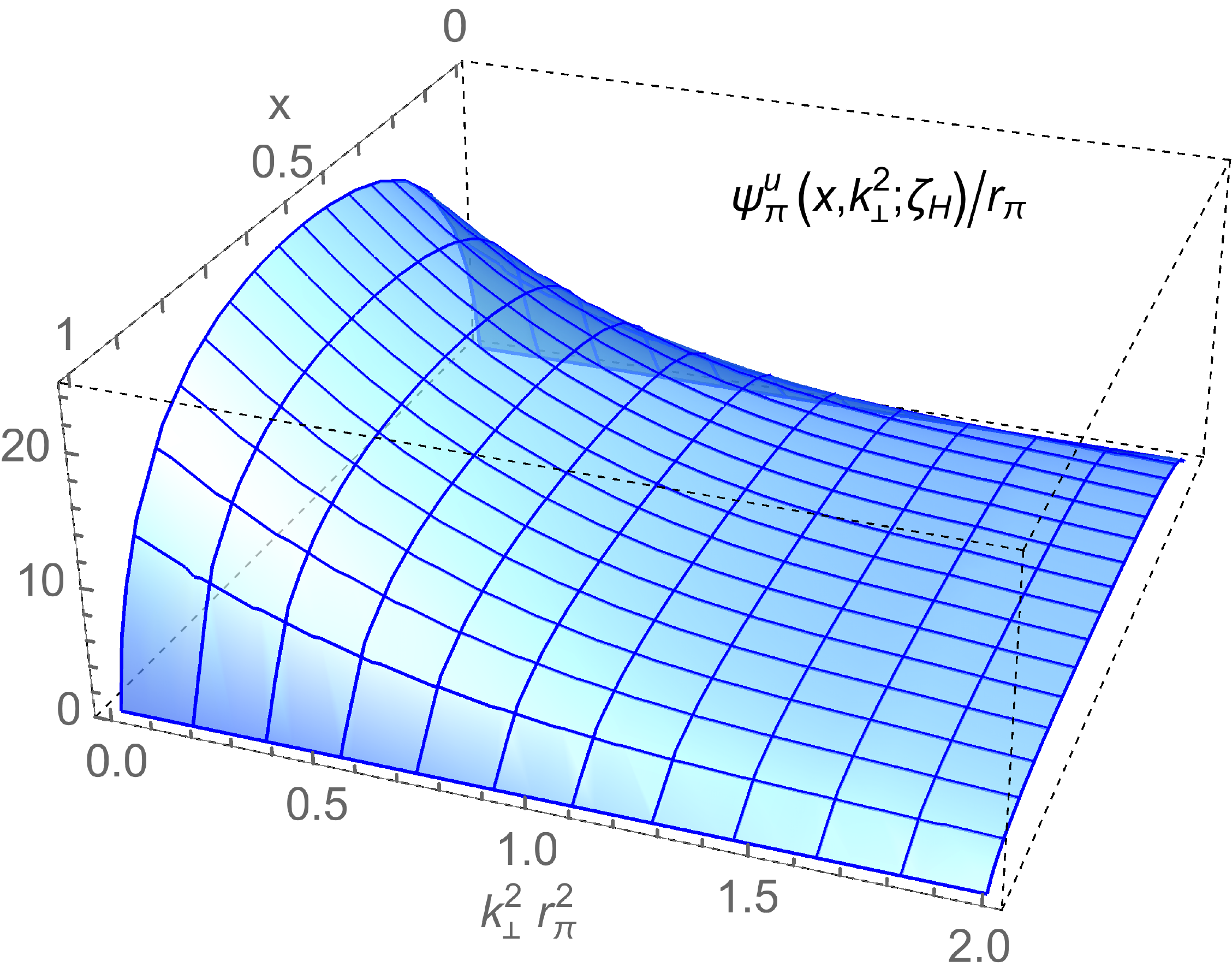}}
\vspace*{6ex}

\leftline{\hspace*{0.5em}{\large{\textsf{B}}}}
\vspace*{-5ex}
\centerline{\includegraphics[clip, width=0.37\textwidth]{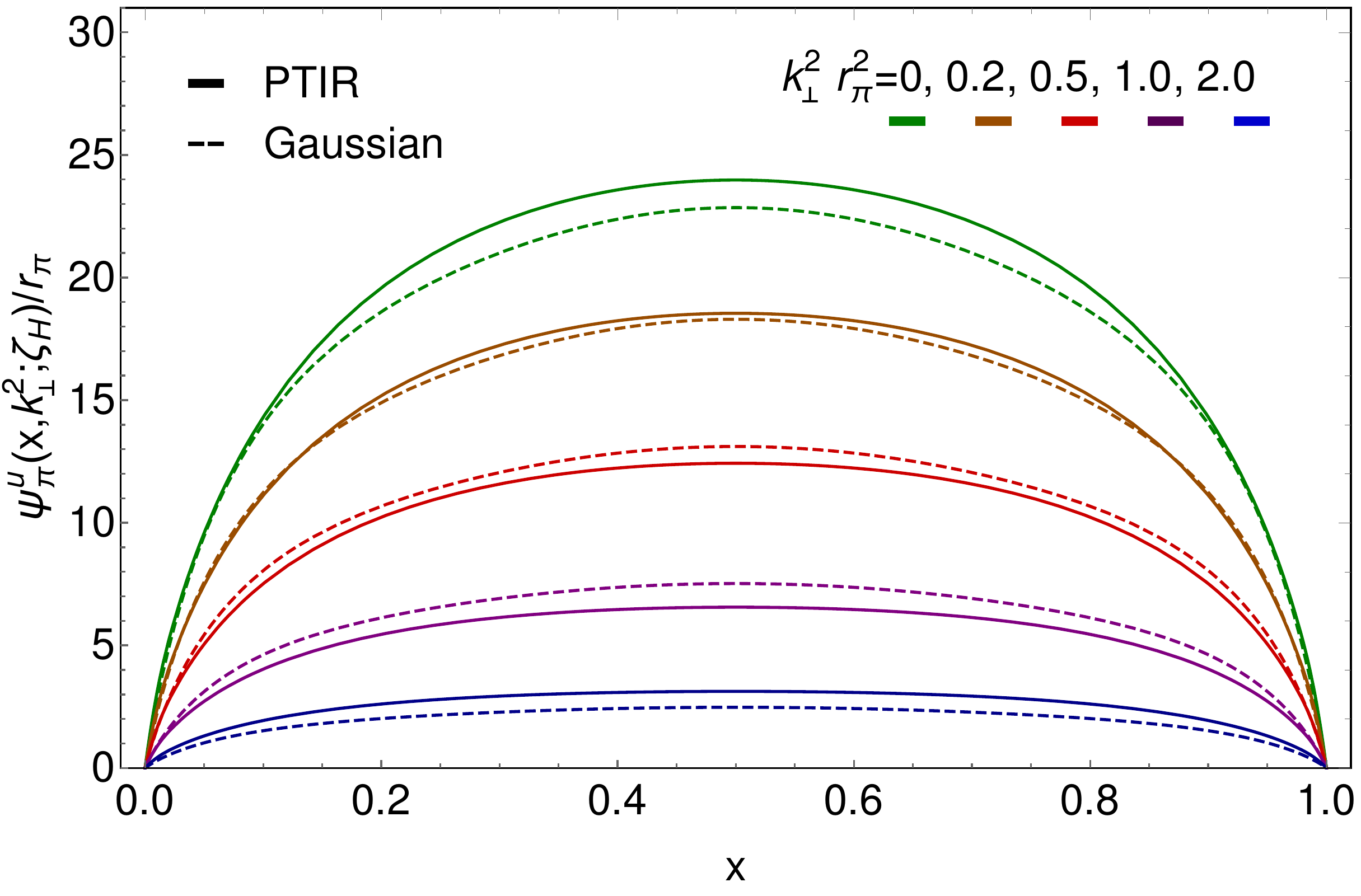}}
\caption{
\emph{Upper panel}\,--\,{\sf A}.
PTIR pion LFWF obtained using Eqs.\,\eqref{psiLFWF} -- \eqref{eq:spectralw} and Table~\ref{tab:params}.
%
\emph{Lower panel}\,--\,{\sf B}.
LFWF in Panel\,A (solid curves) contrasted with the factorised \emph{Ansatz}, Eq.\,\eqref{eq:LFWFgauss} (dashed curves), plotted as a function of $x$ on contours of constant $k_\perp^2 r_\pi^2$.
\label{fig:piLFWF}}
\end{figure}

An extension to the domain of ERBL evolution \cite{Lepage:1979zb, Efremov:1979qk, Lepage:1980fj}, $|x|\leq \xi$, is necessary to complete the definition of the GPD.  On this domain, the parton content of the initial and final state LFWFs differs by two.  Such an extension is challenging, but there is progress \cite{Chouika:2017dhe, Chouika:2017rzs, Zhang:2020ecj}.  We will therefore focus first on the GPD in the DGLAP domain before canvassing aspects of its ERBL extension.

\subsection{GPDs using the spectral representation}
\label{SecPTIR}
\begin{figure}[t!]
\vspace*{3.5ex}

\leftline{\hspace*{0.5em}{\large{\textsf{A}}}}
\vspace*{-5ex}
\centerline{\includegraphics[clip, width=0.36\textwidth]{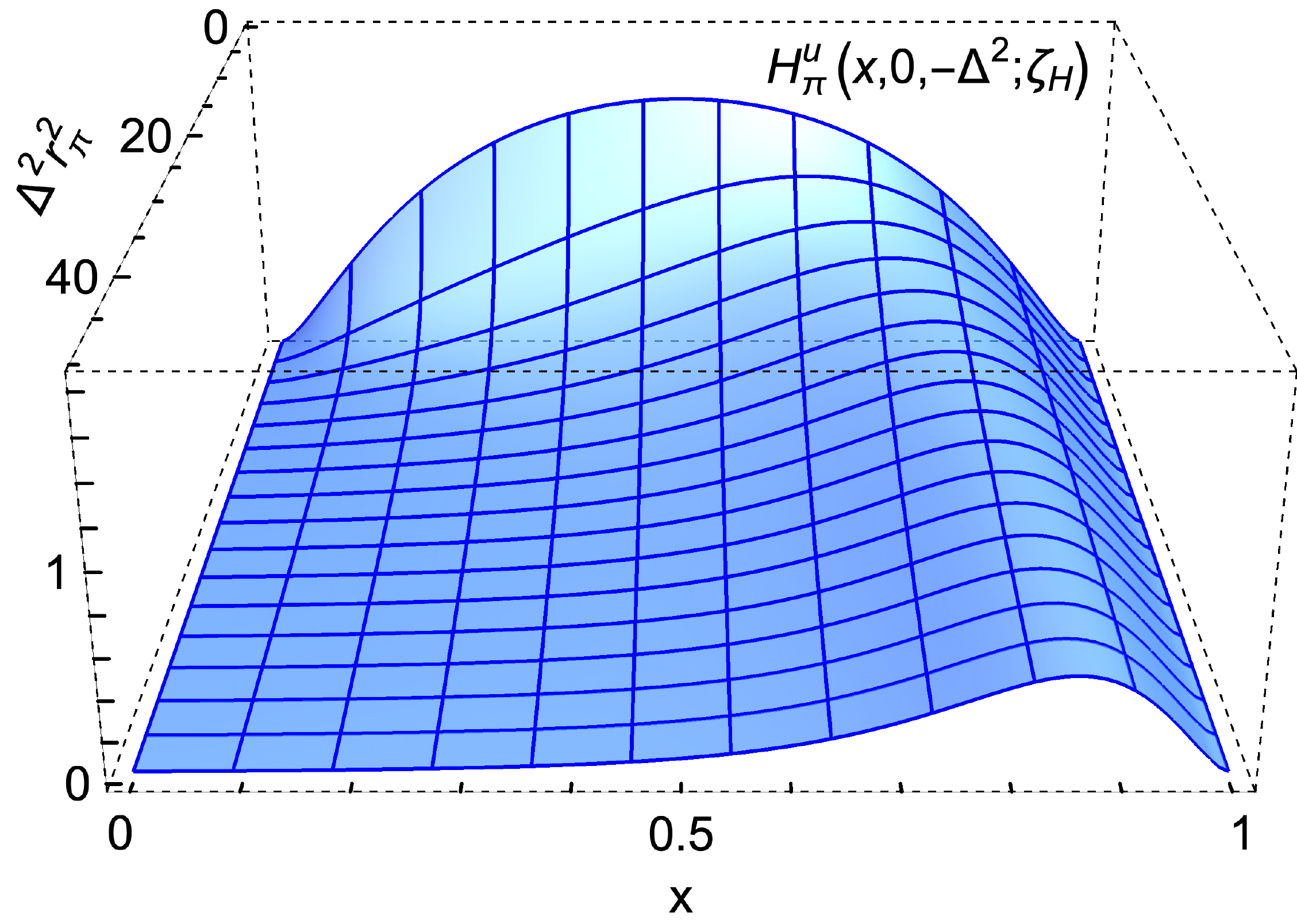}}
\vspace*{6ex}

\leftline{\hspace*{0.5em}{\large{\textsf{B}}}}
\vspace*{-5ex}
\centerline{\includegraphics[clip, width=0.36\textwidth]{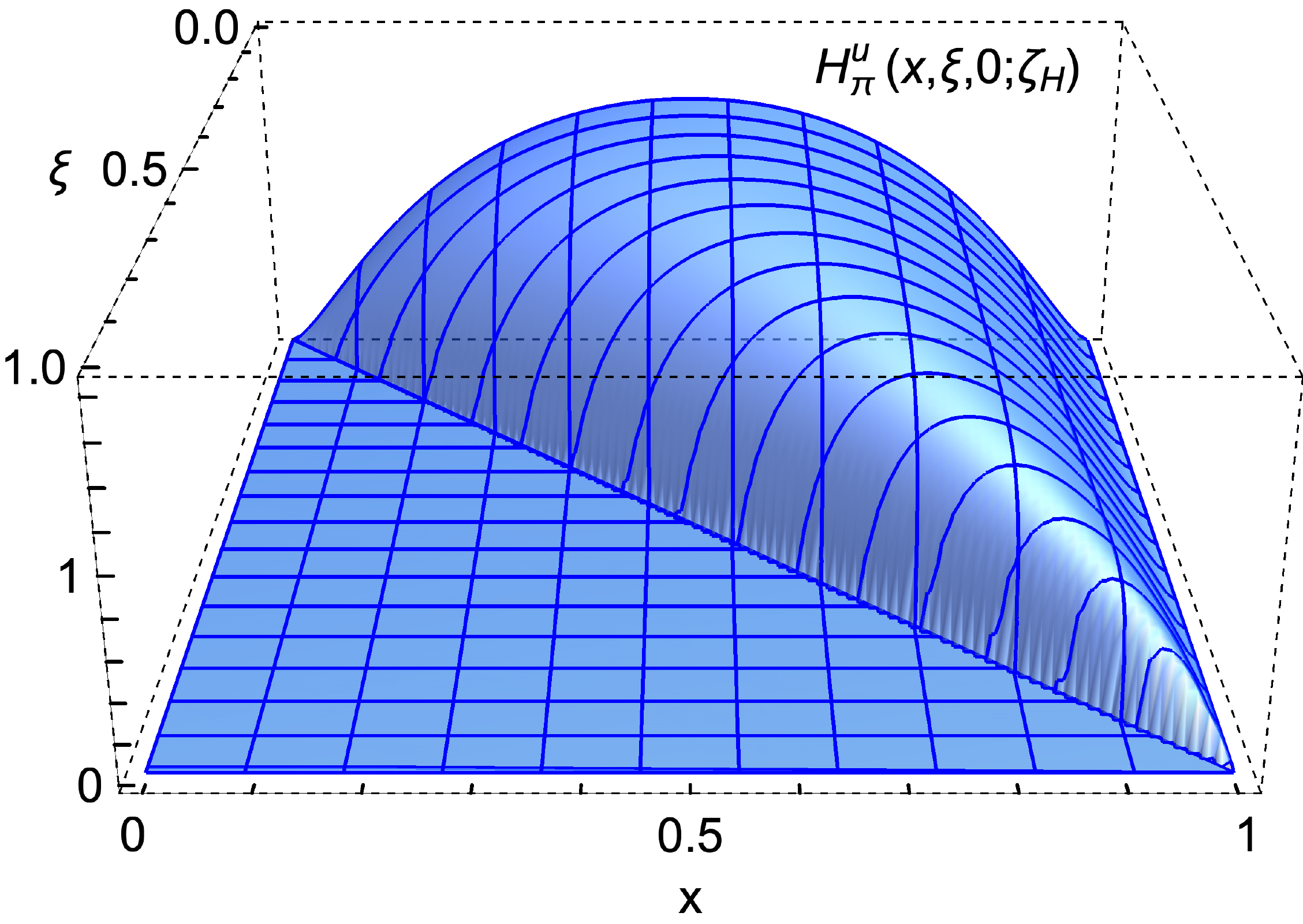}}
\caption{
Pion.  PTIR GPD obtained using Eq.\eqref{eq:overlap} and the LFWF in Fig.\,\ref{fig:piLFWF}A.
\emph{Upper panel}\,--\,{\sf A}.  $H^u_{\pi}(x,\xi=0,-\Delta^2;\zeta_{\mathpzc H}) $.
\emph{Lower panel}\,--\,{\sf B}.  $H^u_{\pi}(x,\xi,0;\zeta_{\mathpzc H}) $.
Here, $r_\pi \approx 0.69\,$fm, computed using Eq.\,\eqref{EqRadius}.  (Experiment \cite{Zyla:2020zbs, Cui:2021aee}: $0.659(4)\,$fm, $0.640(7)\,$fm, respectively.)
\label{fig:piGPD}}
\end{figure}

\begin{figure}[t!]
\vspace*{3.5ex}

\leftline{\hspace*{0.5em}{\large{\textsf{A}}}}
\vspace*{-5ex}
\centerline{\includegraphics[clip, width=0.36\textwidth]{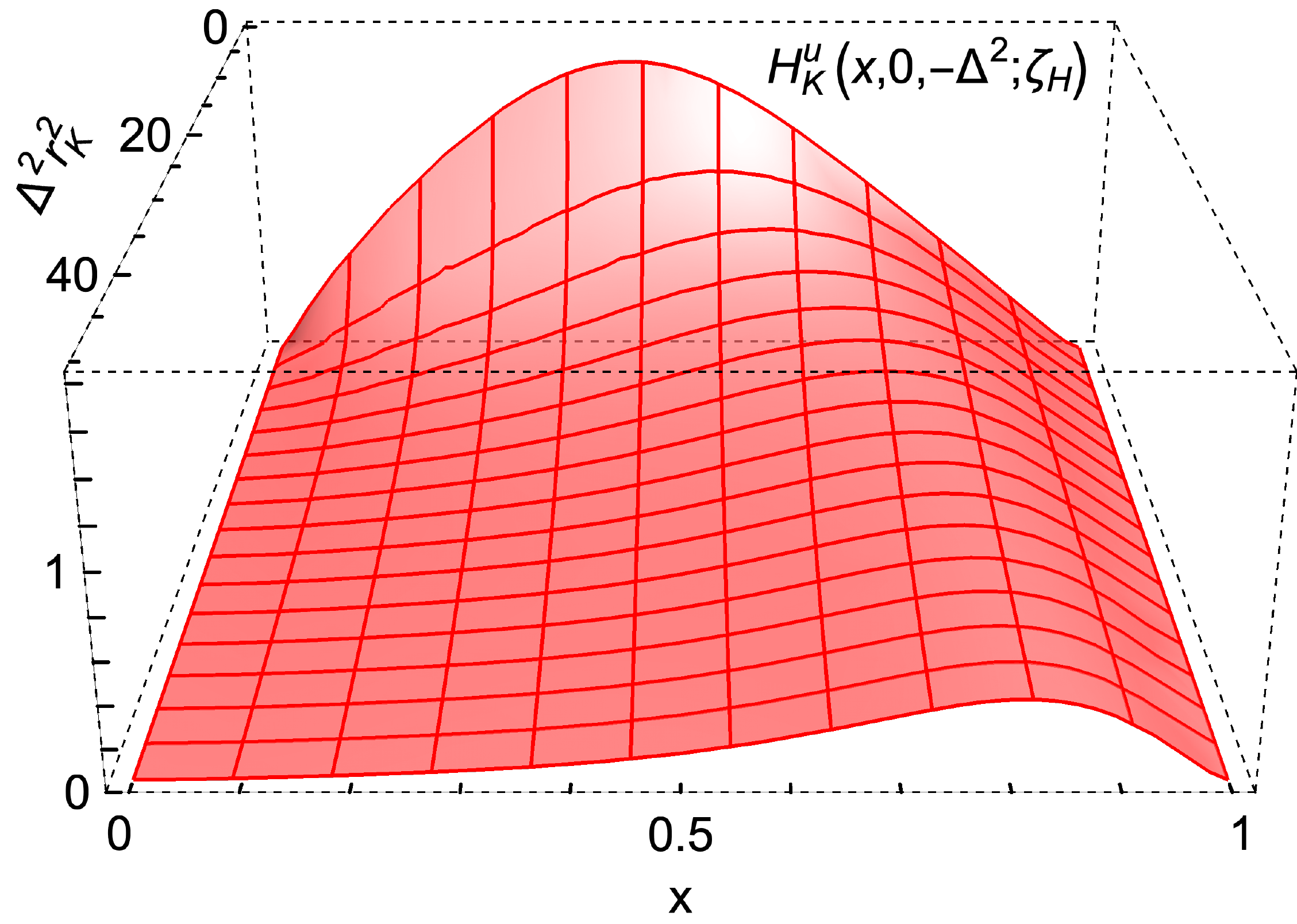}}
\vspace*{6ex}

\leftline{\hspace*{0.5em}{\large{\textsf{B}}}}
\vspace*{-5ex}
\centerline{\includegraphics[clip, width=0.36\textwidth]{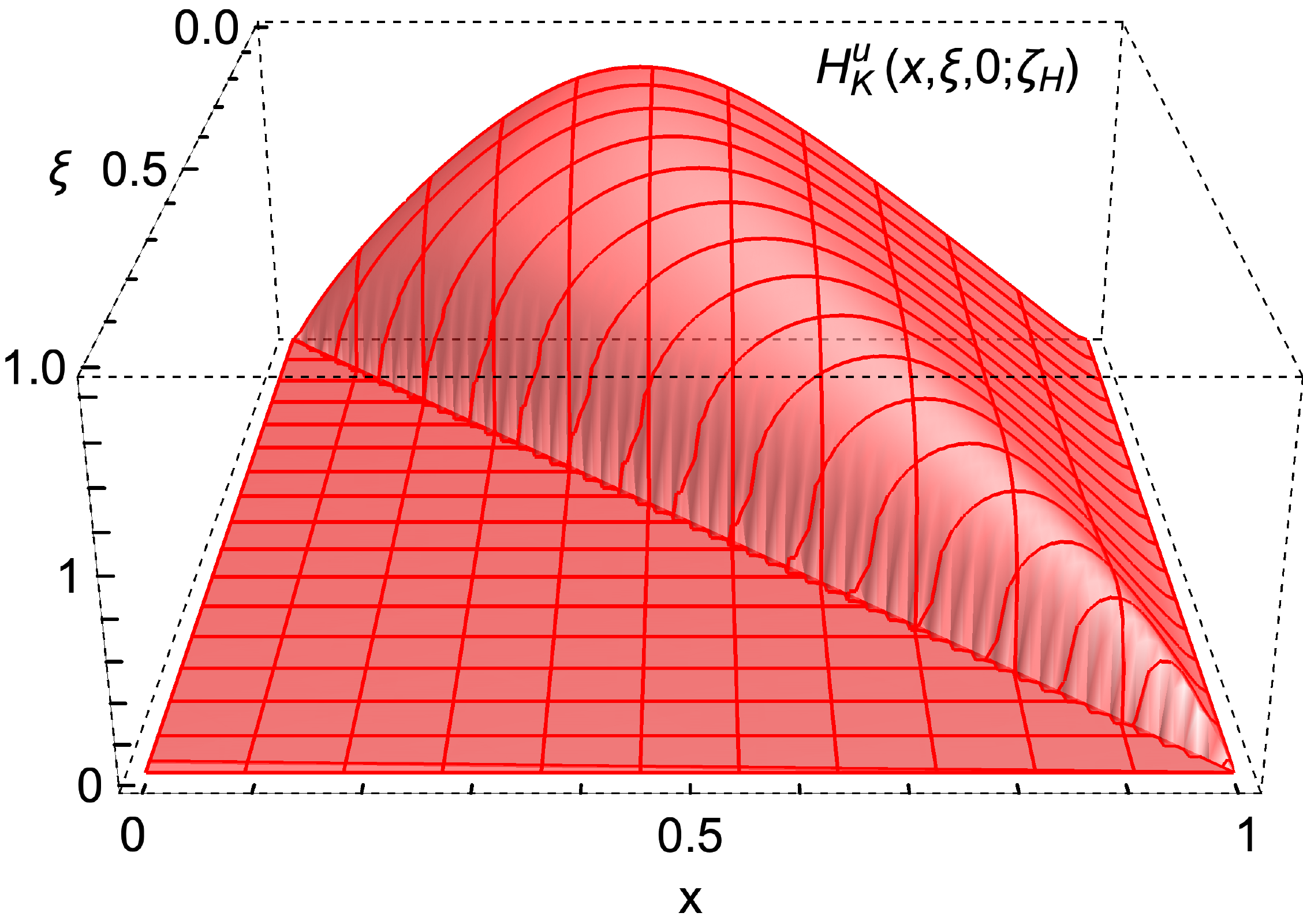}}
\caption{
Kaon.
PTIR GPD obtained using Eq.\eqref{eq:overlap} and the associated LFWF defined by Eqs.\,\eqref{psiLFWF}\,--\,\eqref{eq:spectralw} and Table~\ref{tab:params}.
\emph{Upper panel}\,--\,{\sf A}.  $H^u_{K}(x,\xi=0,-\Delta^2;\zeta_{\mathpzc H}) $.
\emph{Lower panel}\,--\,{\sf B}.  $H^u_{K}(x,\xi,0;\zeta_{\mathpzc H}) $.
Here, $r_K = 0.56\,$fm, computed using Eq.\,\eqref{EqRadius}.
(Experiment \cite{Zyla:2020zbs, Cui:2021aee}: $0.560(31)\,$fm, $0.53\,$fm, respectively.)
\label{fig:KGPD}}
\end{figure}

The pion LFWF obtained using Eqs.\,\eqref{psiLFWF}\,--\,\eqref{eq:spectralw} and the coefficients/powers in Table~\ref{tab:params} is depicted in Fig.\,\ref{fig:piLFWF}A.  Inserting this result and its kaon analogue into Eq.\,\eqref{eq:overlap}, one obtains the GPDs depicted in Figs.\,\ref{fig:piGPD} and \ref{fig:KGPD} by straightforward numerical integration.
The $\bar s$-quark in $K^+$ GPD is found by switching $M_u$ and $M_{\bar s}$ in Eqs.\,\eqref{psiLFWF}, \eqref{Omega} and changing $v_K \to -v_K$ in Eq.\,\eqref{eq:spectralw}.

Comparing Figs.\,\ref{fig:piGPD}A and \ref{fig:KGPD}A, one sees that $H^u_{\pi}(x,\xi=0,t=0;\zeta_{\mathpzc H}) $ is symmetric around $x=1/2$, whereas $H^u_{K}(x,\xi=0,t=0;\zeta_{\mathpzc H}) $ is skewed, peaking at $x<1/2$; and the peak magnitude of this kaon GPD is larger than that of the pion.
Similar patterns are seen when contrasting Figs.\,\ref{fig:piGPD}B, \ref{fig:KGPD}B, which also show that both GPDs vanish on $x<\xi$.

\subsection{GPDs using the factorised representation}
\label{SecFac}
Owing to the reliability of well-chosen factorised \emph{Ans\"atze} for LFWFs, discussed in Sec.\,\ref{SecFactorised}, we introduce such a form here so as to reveal additional insights \cite{Zhang:2021mtn}:
{\allowdisplaybreaks\begin{subequations}
\label{eq:HfacT}
\begin{align}
H^u_{\mathsf P}&(x,\xi,-\Delta^2;\zeta_{\mathpzc H}) =  \nonumber\\
&  \Theta(x_-) \sqrt{{\mathpzc u}^{\mathsf P}\left(x_-;\zeta_{\mathpzc H} \right) {\mathpzc u}^{\mathsf P}\left(x_+;\zeta_{\mathpzc H} \right)} \; \Phi_{\mathsf P}^u\left( z; \zeta_{\mathpzc H} \right) \,,
\label{eq:Hfac} \\ 
%
\Phi_{\mathsf P}^u& (z;\zeta_{\mathpzc H}) = \nonumber \\
&   \int \frac{d^2{k_\perp}}{16 \pi^3}
 \tilde \psi_{{\mathsf P}}^{u}\left({k}_\perp^2;\zeta_{\mathpzc H} \right)
 \tilde \psi_{{\mathsf P}}^{u}\left(\left({k}_\perp - {s}_\perp \right)^2;\zeta_{\mathpzc H} \right) \,,
 \label{eq:Phi}
\end{align}
\end{subequations}
where $\Theta$ is the Heaviside function,
$z={s}_\perp^2={\Delta}_\perp^2 (1-x)^2 /(1-\xi^2)^2$,
Eq.\,\eqref{EqNabil} is implicit,
and canonical normalisation 
guarantees $\Phi_{\mathsf P}^u(0;\zeta_{\mathpzc H})=1$ so that the GPD is properly normalised in the forward limit.
}

Recalling Eq.\,\eqref{eq:FF}, then Eqs.\,\eqref{eq:HfacT} can be used to obtain
{\allowdisplaybreaks
\begin{subequations}
\label{eq:PhifromFA}
\begin{align}
\label{eq:PhifromF}
\frac{\partial^n}{\partial z^n} \left. \Phi_{\mathsf P}^u(z;\zeta_{\mathpzc H}) \right|_{z=0} & =
\frac{1}{\langle x^{2n}\rangle_{\bar{h}}^{\zeta_{\mathpzc H}} }
\left. \frac{d^n F_{\mathsf P}^u(\Delta^2)}{d(\Delta^2)^n}  \right|_{\Delta^2=0} \,, \\
\langle x^{2n} \rangle_{\bar{h}}^{\zeta_{\mathpzc H}}  = \langle (1-x)^{2n} \rangle_{u}^{\zeta_{\mathpzc H}}
& = \int_0^1 dx (1-x)^{2n} {\mathpzc u}^{\mathsf P}(x;\zeta_{\mathpzc H}) \,.
\end{align}
\end{subequations}
Evidently, the $k_\perp^2$-overlap portion of a factorised \emph{Ansatz} for the $u$-in-${\mathsf P}$ GPD is fully determined by the $u$-quark's contribution to the elastic form factor and its DF.  Considering the simplest case, $n=1$:
\begin{subequations}
\label{eq:dphiMdz}
\begin{align}
& \left. \frac{\partial}{\partial z} \Phi_{\mathsf P}^u(z;\zeta_{\mathpzc H}) \right|_{z=0} = -\frac{r^2_{\mathsf P}}
{4{\mathpzc x}_{\mathsf P}^2(\zeta_{\mathpzc H})} \,,
\label{eq:dphiMKdzu} \\
& \left.\frac{\partial}{\partial z} \Phi_{\mathsf P}^{\bar h}(z;\zeta_{\mathpzc H}) \right|_{z=0} =  \left. (1 - {\mathpzc d}_{\mathsf P} ) \frac{\partial}{\partial z} \Phi_{\mathsf P}^u(z;\zeta_{\mathpzc H}) \right|_{z=0}   \,, \label{eq:dphiMdzh}
\end{align}
\end{subequations}
where
\begin{equation}
\label{definex}
{\mathpzc x}_{\mathsf P}^2(\zeta_{\mathpzc H}) = \langle x^2 \rangle_{\bar{h}}^{\zeta_{\mathpzc H}} + \tfrac{1}{2} (1 - {\mathpzc d}_{\mathsf P}) \langle x^2 \rangle_u^{\zeta_{\mathpzc H}}\,,
\end{equation}
and ${\mathpzc d}_{\mathsf P} \propto (M_{\bar h}-M_u )$ expresses the impact on the meson's charge distribution of any mass-difference between the valence constituents.  For the pion, in the isospin-symmetry limit, ${\mathpzc d}_\pi=0$; whereas for the kaon, ${\mathpzc d}_{K}>0$ because the $\bar s$-quark contribution to $F_K$ is stiffer (falls less rapidly with increasing $\Delta^2$) than that of the $u$-quark \cite{Chen:2012txa, Gao:2017mmp}.
}

\begin{figure}[t!]
\vspace*{3.5ex}

\leftline{\hspace*{0.5em}{\large{\textsf{A}}}}
\vspace*{-5ex}
\centerline{\includegraphics[clip, width=0.36\textwidth]{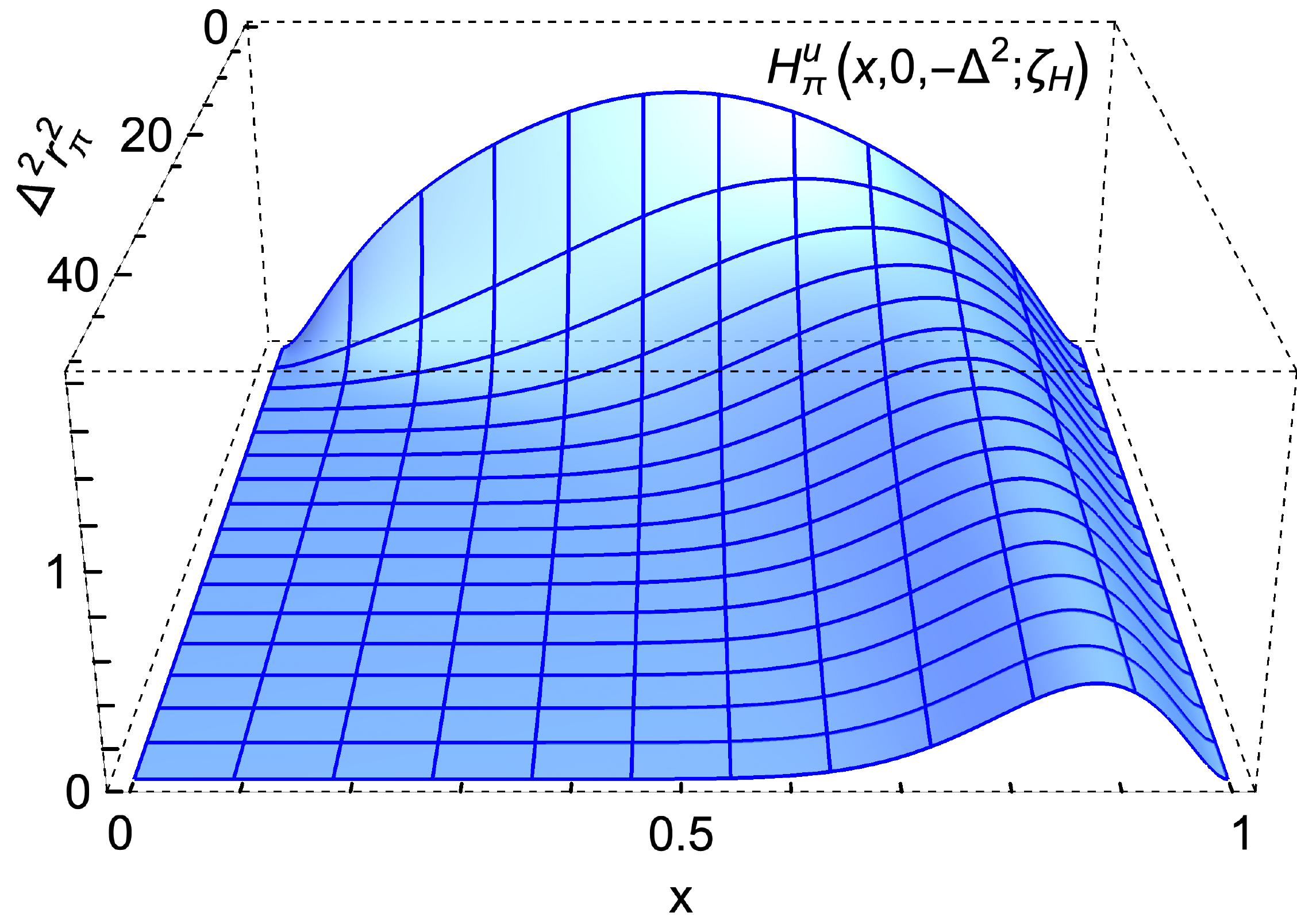}}
\vspace*{6ex}

\leftline{\hspace*{0.5em}{\large{\textsf{B}}}}
\vspace*{-5ex}
\centerline{\includegraphics[clip, width=0.36\textwidth]{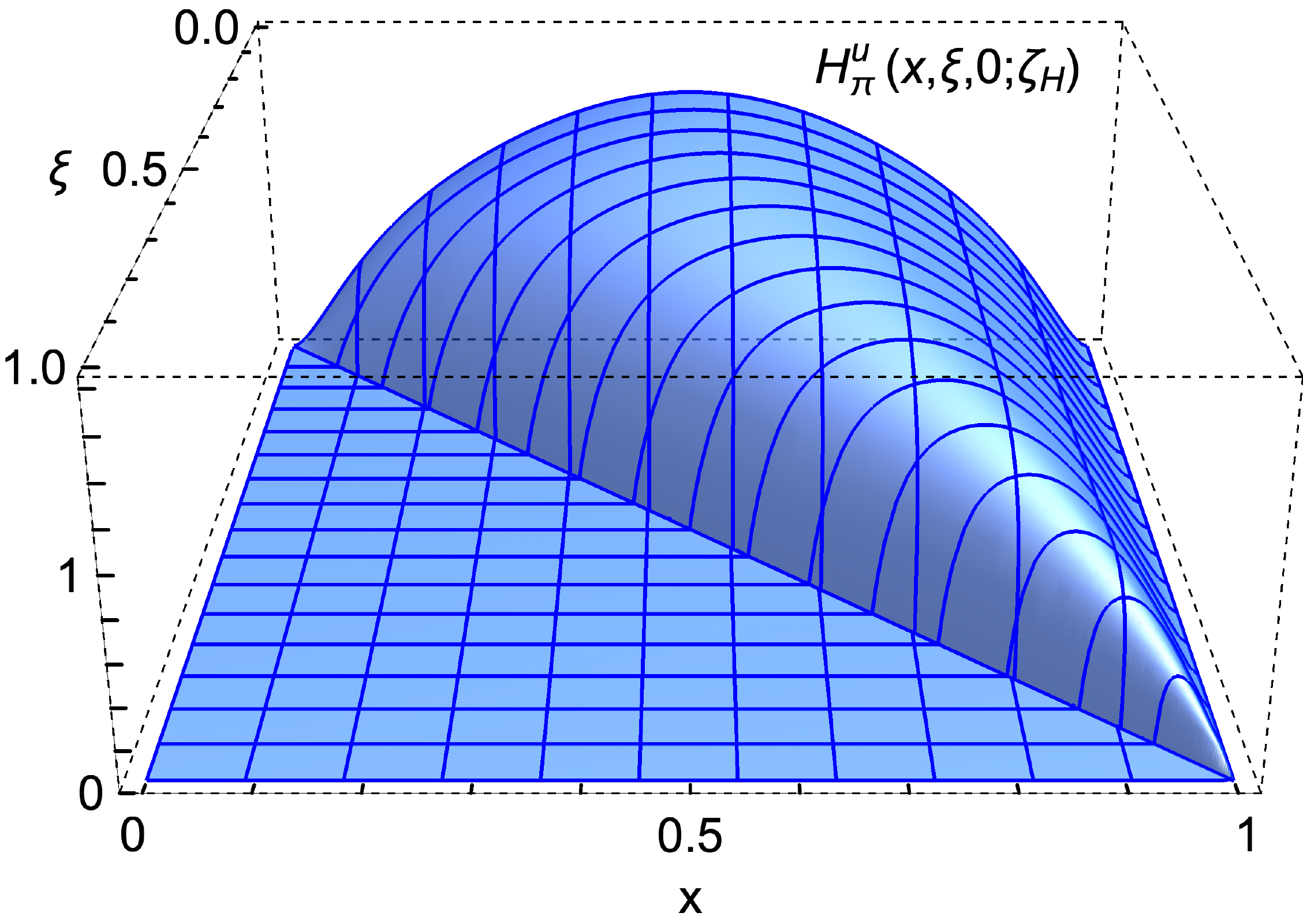}}
\caption{
Pion GPD obtained using Eq.\eqref{eq:overlap} and the factorised LFWF discussed in Sec.\,\ref{SecFac}.
\emph{Upper panel}\,--\,{\sf A}.  $H^u_{K}(x,\xi=0,-\Delta^2;\zeta_{\mathpzc H}) $.
\emph{Lower panel}\,--\,{\sf B}.  $H^u_{K}(x,\xi,0;\zeta_{\mathpzc H}) $.
\label{fig:piGPDFac}}
\end{figure}

It is especially useful to represent the $k_\perp^2$-dependence of the LFWF using a Gaussian because then Eqs.\,\eqref{eq:dphiMdz} completely constrain the pointwise behaviour:
\begin{align}
\psi_{{\mathsf P}}^{u}&\left(x,k_\perp^2;\zeta_{\mathpzc H} \right)  \nonumber \\
& = \left(\frac{16 \pi^2 r_{\mathsf P}^2}{{\mathpzc x}_{\mathsf P}^2(\zeta_{\mathpzc H})} \, u^{\mathsf P}(x;\zeta_{\mathpzc H})\right)^{1/2}
\exp{\left(-\frac{ r_{\mathsf P}^2 k_\perp^2}{2 {\mathpzc x}_{\mathsf P}^2(\zeta_{\mathpzc H})}\right)}  \,.
\label{eq:LFWFgauss}
\end{align}
This \emph{Ansatz} is compared with the more sophisticated spectral representation in Fig.\,\ref{fig:piLFWF}B: for practical purposes, the factorised form provides a satisfactory pointwise approximation.  Thus, as claimed above, it can be used to develop sound insights.

\begin{figure}[t!]
\vspace*{3.5ex}

\leftline{\hspace*{0.5em}{\large{\textsf{A}}}}
\vspace*{-5ex}
\centerline{\includegraphics[clip, width=0.36\textwidth]{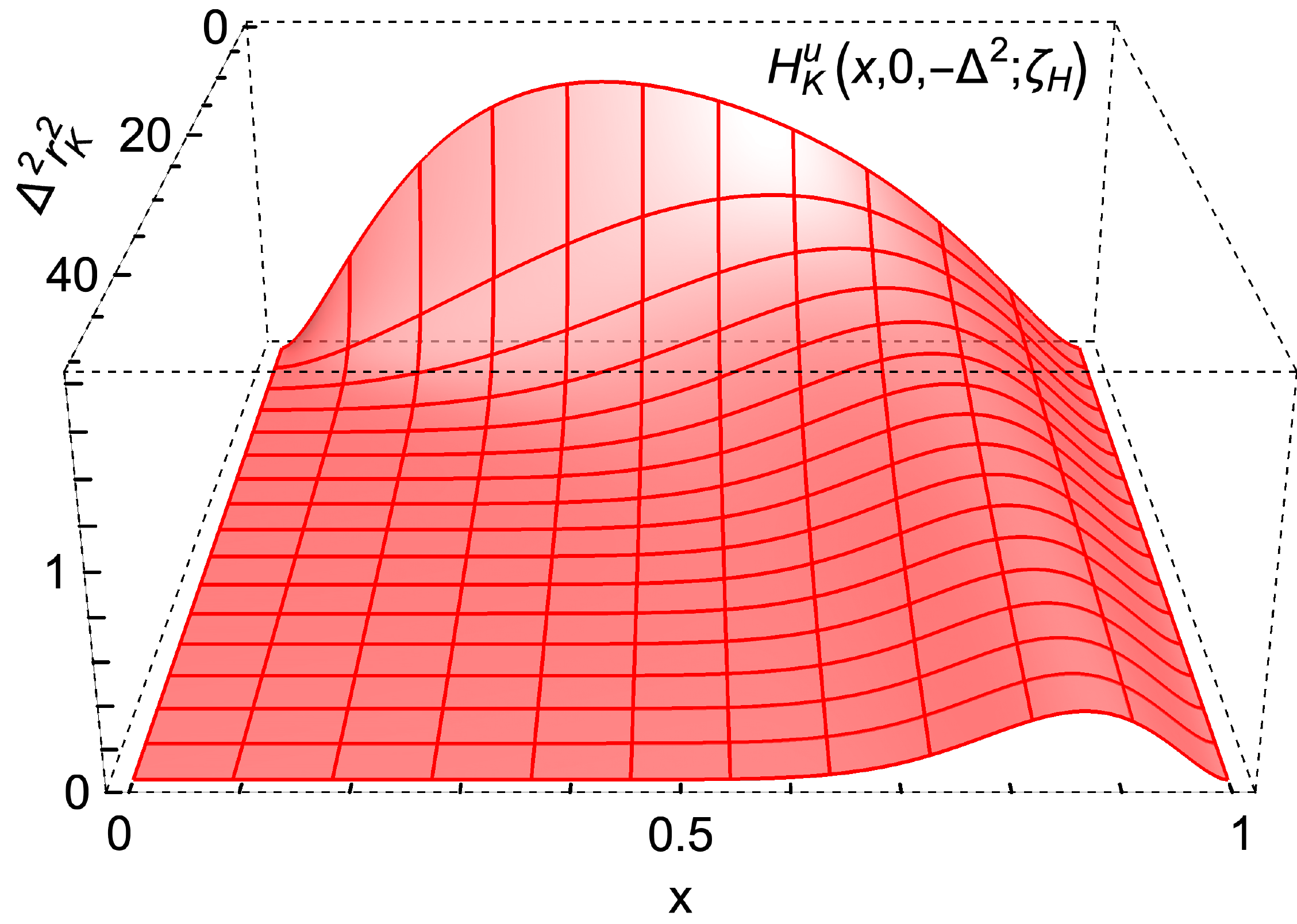}}
\vspace*{6ex}

\leftline{\hspace*{0.5em}{\large{\textsf{B}}}}
\vspace*{-5ex}
\centerline{\includegraphics[clip, width=0.36\textwidth]{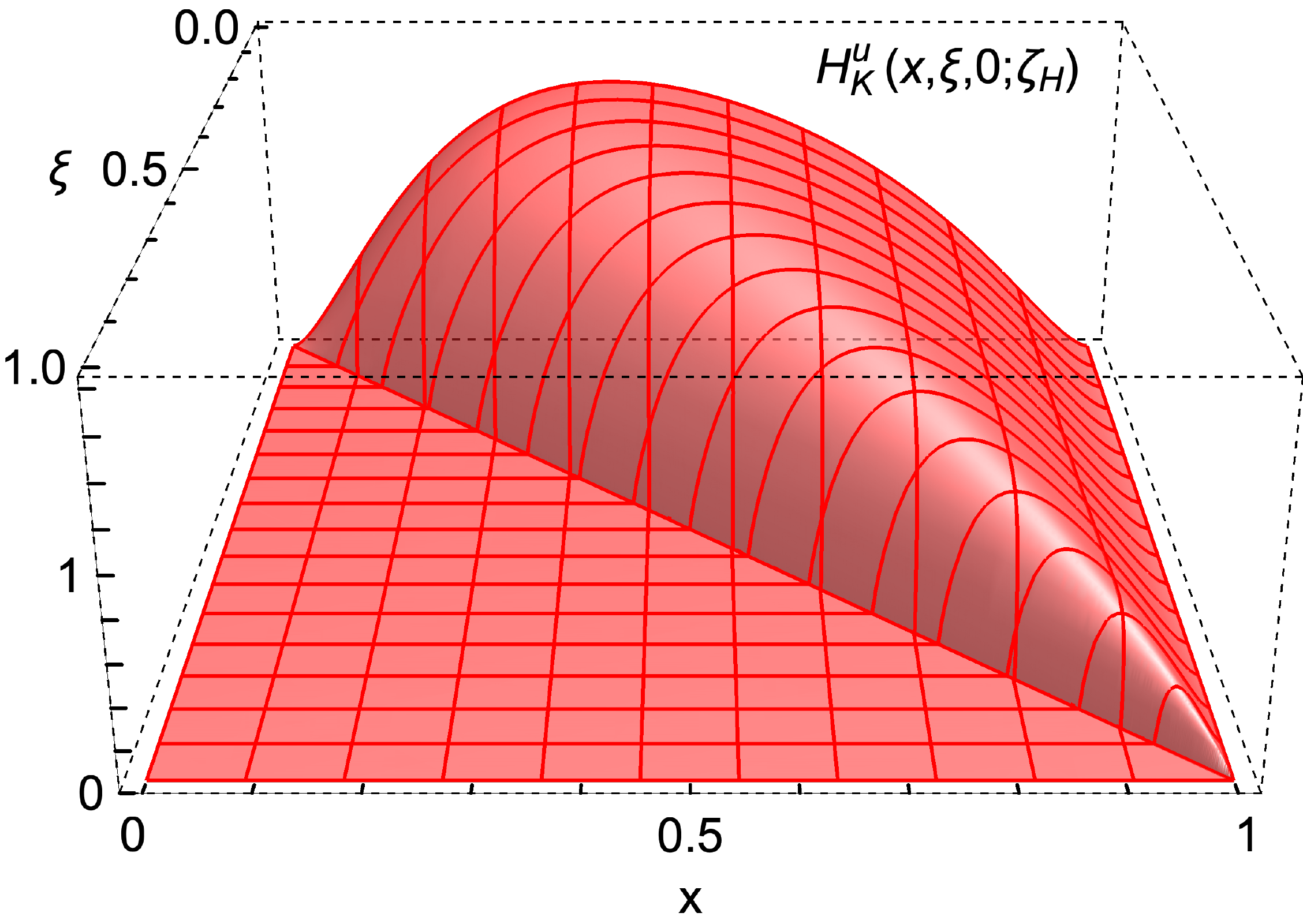}}
\caption{
Kaon GPD obtained using Eq.\eqref{eq:overlap} and the factorised LFWF discussed in Sec.\,\ref{SecFac}. %
\emph{Upper panel}\,--\,{\sf A}.  $H^u_{K}(x,\xi=0,-\Delta^2;\zeta_{\mathpzc H}) $.
\emph{Lower panel}\,--\,{\sf B}.  $H^u_{K}(x,\xi,0;\zeta_{\mathpzc H}) $.
\label{fig:KGPDFac}}
\end{figure}

Working with Eq.\,\eqref{eq:LFWFgauss}, the DGLAP-domain GPD is
\begin{align}
H_{\mathsf P}^u(x,\xi,-\Delta^2;&\zeta_{\mathpzc H}) =  \Theta(x-\xi) \sqrt{{\mathpzc u}^{\mathsf P}\left(x_-;\zeta_{\mathpzc H} \right) {\mathpzc u}^{\mathsf P}\left(x_+;\zeta_{\mathpzc H} \right)}
\nonumber \\
& \quad \times
\exp{\left( - \frac{\Delta_\perp^2 \, r_{\mathsf P}^2 (1-x)^2}{4 {\mathpzc x}_{\mathsf P}^2(\zeta_{\mathpzc H})(1-\xi^2)^2 }\right)} \,.
\label{eq:gaussH}
\end{align}
where Eq.\,\eqref{EqNabil} is understood.
Exploiting the behaviour of meson bound-states under charge conjugation, the $\bar{h}$-in-${\mathsf P}$ GPD is obtained by replacing $\Theta(x-\xi) \to -\Theta(-x-\xi)$, ${\mathpzc u}^{\mathsf P} \to \bar{{\mathpzc s}}^{\mathsf P}$, $r_{\mathsf P}^2 \to r_{\mathsf P}^2 (1-{\mathpzc d}_H)$ and $x \to |x|$ in Eq.\,\eqref{eq:gaussH}.

\begin{figure}[t]
\vspace*{3.5ex}

\leftline{\hspace*{0.5em}{\large{\textsf{A}}}}
\vspace*{-5ex}
\centerline{\includegraphics[clip, width=0.37\textwidth]{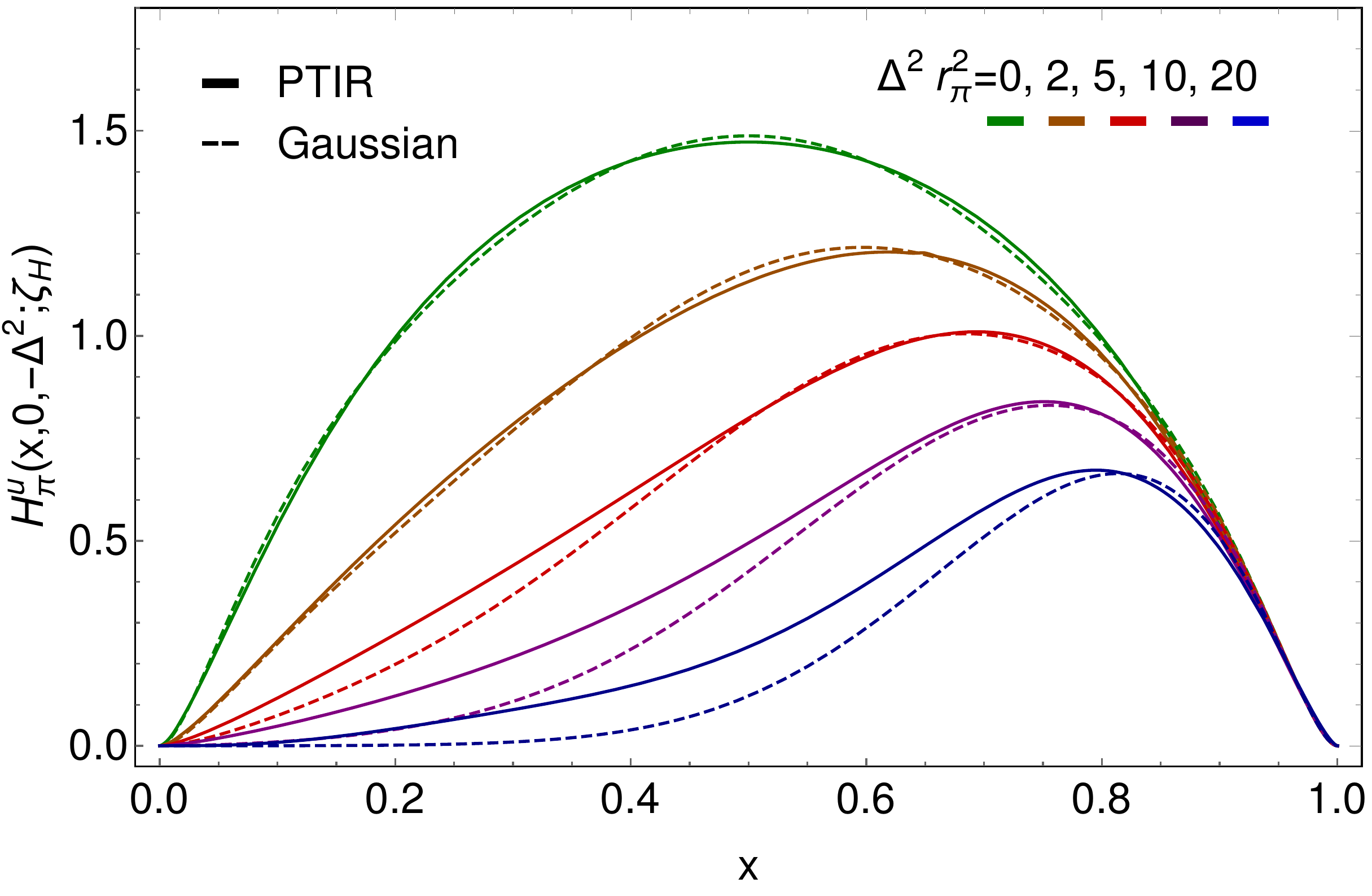}}
\vspace*{6ex}

\leftline{\hspace*{0.5em}{\large{\textsf{B}}}}
\vspace*{-5ex}
\centerline{\includegraphics[clip, width=0.37\textwidth]{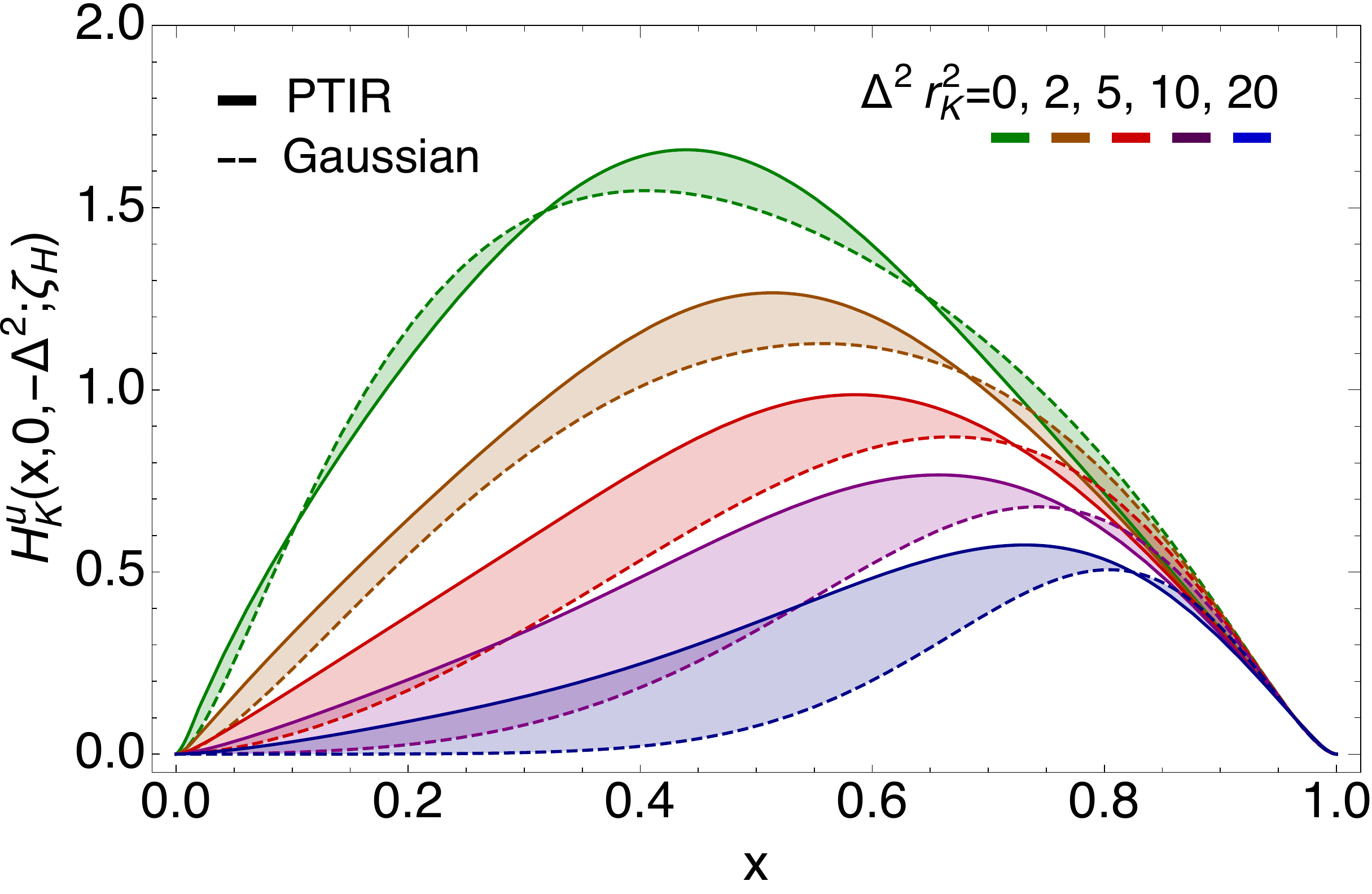}}
\caption{
\label{fig:GPDs}
\emph{Upper panel}\,--\,{\sf A}.  $u$-in-$\pi$ GPD, plotted as a function of $x$ on contours of constant $\Delta^2 r_\pi^2$: solid curves -- produced using PTIR LFWF, Eqs.\,\eqref{psiLFWF} -- \eqref{eq:spectralw} and Table~\ref{tab:params}; and dashed curves -- results from factorised LFWF, Eq.\,\eqref{eq:gaussH}.
\emph{Lower panel}\,--\,{\sf B}.  Kindred curves for $u$-in-$K$ GPD.  The factorised kaon LFWF is obtained with ${\mathpzc d}_K = 0.07$ in Eqs.\,\eqref{eq:dphiMdz}, \eqref{eq:LFWFgauss}.  (Shading highlights curves that should be compared.)
}
\end{figure}

The DGLAP-domain $\pi$ and $K$ GPDs produced by Eq.\,\eqref{eq:gaussH} are depicted in Figs.\,\ref{fig:piGPDFac}\,--\,\ref{fig:GPDs}.
Modest differences can be seen in the three-dimensional images, \emph{i.e}., in the following comparisons: Fig.\,\ref{fig:piGPD} with \ref{fig:piGPDFac} and Fig.\,\ref{fig:KGPD} with \ref{fig:KGPDFac};
and they are revealed with more definition in Fig.\,\ref{fig:GPDs}.

Consider Fig.\,\ref{fig:GPDs}A.  Plainly, the factorised \emph{Ansatz} provides a sound representation of the $\pi$ GPD on the complete domain of $\Delta^2$ depicted, although there is some degradation of pointwise accuracy as $\Delta^2$ increases.
Regarding Fig.\,\ref{fig:GPDs}B, Eqs.\,\eqref{psiLFWF}, \eqref{X2c} reveal, as already remarked, that pointwise differences between a LFWF and a well-constructed factorised approximation can grow with $m_{\mathsf P}^2$, $M_{\bar h}^2-M_u^2$.  Consequently, the accord between the PTIR GPD for the kaon and that produced by the factorised LFWF, Eq.\,\eqref{eq:LFWFgauss}, is poorer than for the pion.  Nonetheless, the evident semiquantitative agreement shows that the simple factorised \emph{Ansatz} can still yield a fair picture.


It is worth highlighting two qualitative features of the hadron-scale $\pi$ and $K$ GPDs in Fig.\,\ref{fig:piGPD} -- \ref{fig:GPDs}.
(\emph{i})
Given that $H_{\mathsf P}^u(x,0,0;\zeta_{\mathpzc H}) = {\mathpzc u}^{\mathsf P}(x;\zeta_{\mathpzc H})$, one recalls that the maximum of $H_\pi^u(x,0,0;\zeta_{\mathpzc H})$ is located at $x=1/2$ and that of $H_K^u(x,0,0;\zeta_{\mathpzc H})$ at $x=0.4$ \cite{Cui:2020dlm}.  This 20\% shift is standard for Higgs-boson modulation of EHM in the $s$-quark sector.
(\emph{ii})
The position of the peak shifts as the momentum transfer to the target increases: as $\Delta^2$ grows, the maximum of both the $\pi$ and $K$ GPDs shifts toward $x=1$ and its profile becomes narrower, \emph{i.e}., more tightly focused within the valence domain.  This aspect emphasises that hard probes reveal valence partons.

\subsection{Elastic electromagnetic form factors}
\begin{figure}[t]
\centerline{\includegraphics[clip, width=0.37\textwidth]{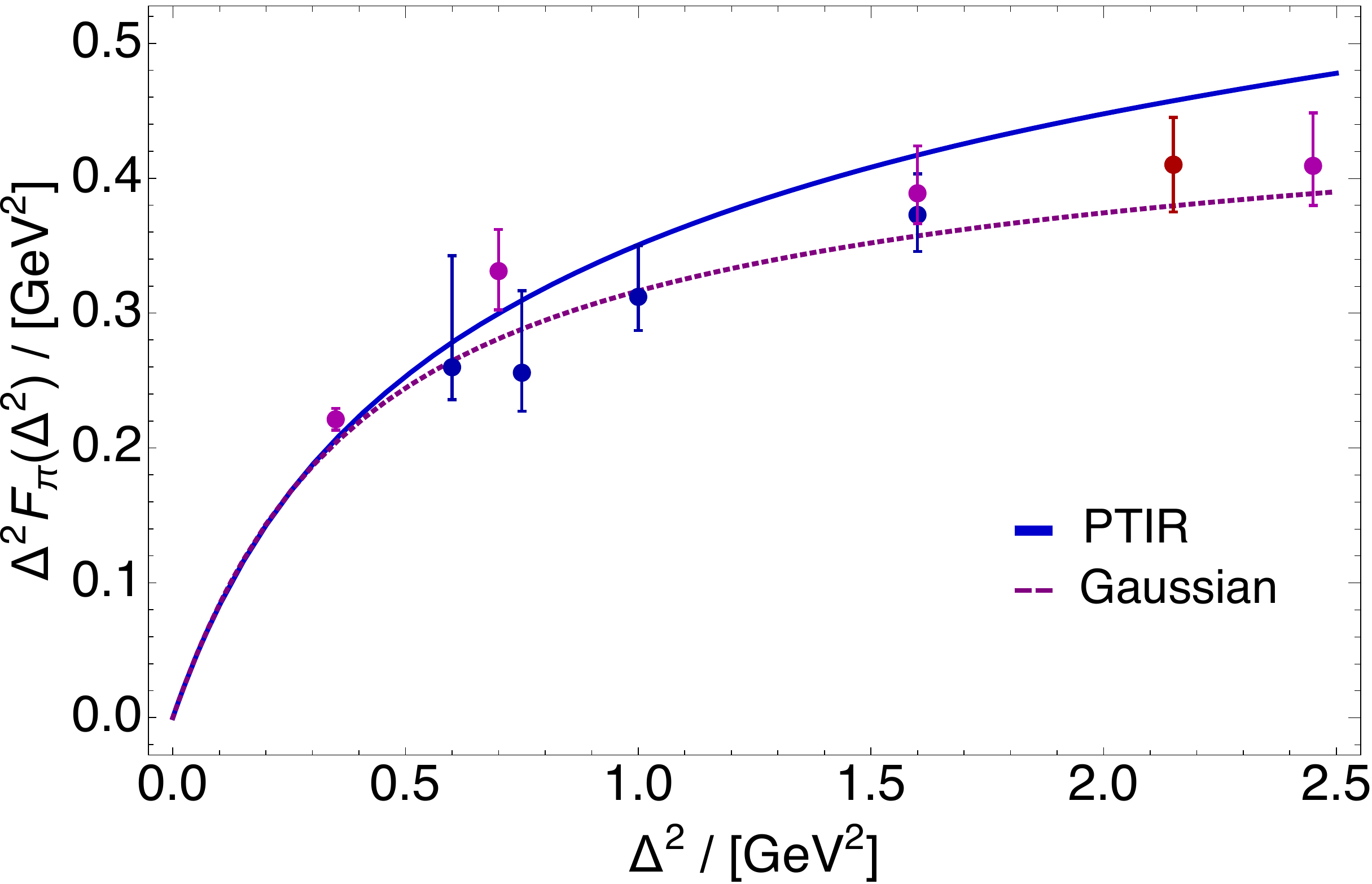}}
\caption{\label{fig:FFpi}
$\Delta^2 F_\pi(\Delta^2)$ obtained using the GPDs drawn in Fig.\,\ref{fig:piGPD}: PTIR (solid blue curve), Sec.\,\ref{SecPTIR}; and factorised LFWF (short-dashed magenta), Sec.\,\ref{SecFac}.  Data from Refs.\,\cite{Horn:2007ug, Huber:2008id}.
}
\end{figure}

\begin{figure}[t]
\vspace*{3.5ex}

\leftline{\hspace*{0.5em}{\large{\textsf{A}}}}
\vspace*{-5ex}
\centerline{\includegraphics[clip, width=0.37\textwidth]{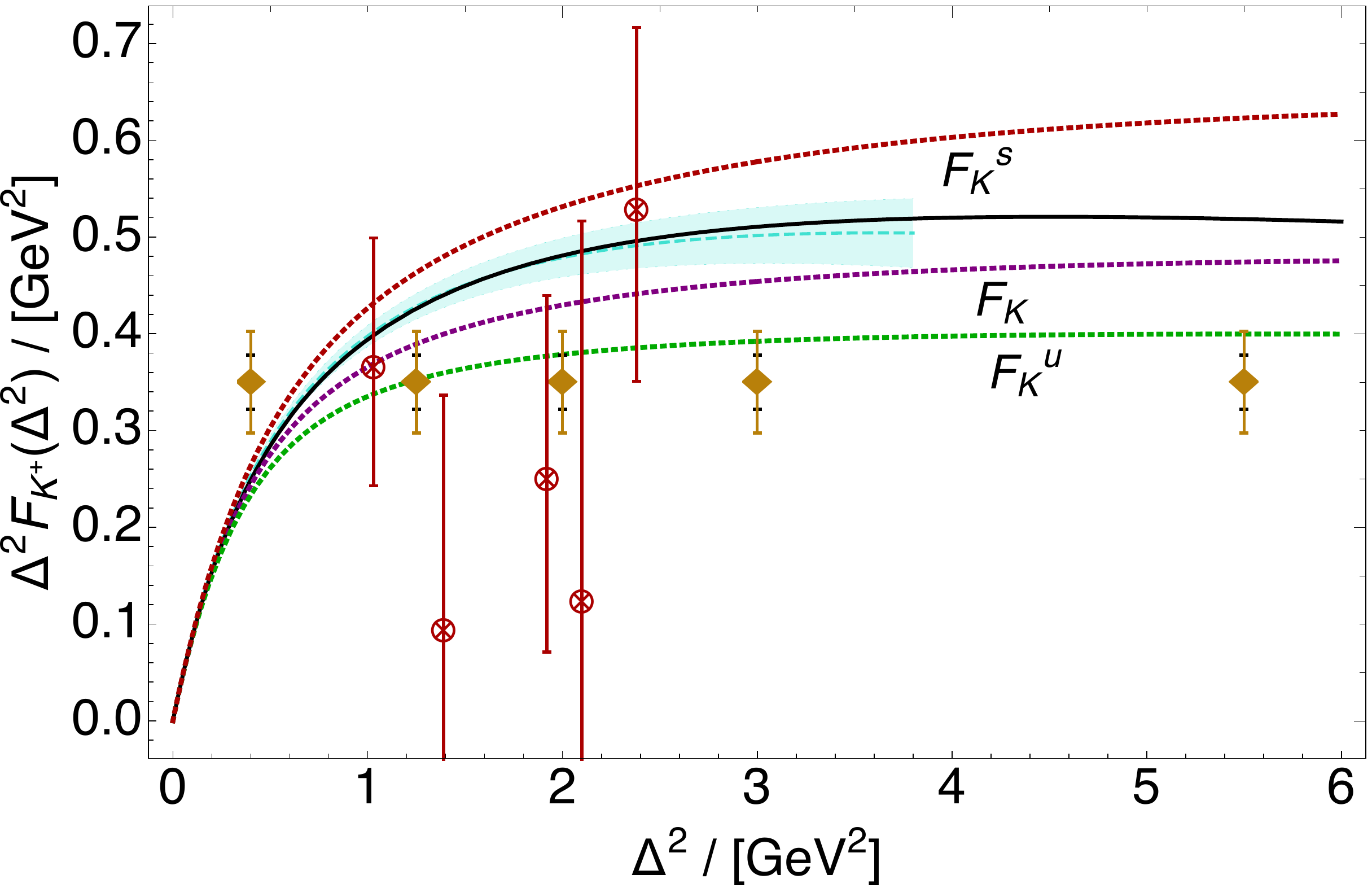}}
\vspace*{6ex}

\leftline{\hspace*{0.5em}{\large{\textsf{B}}}}
\vspace*{-5ex}
\centerline{\includegraphics[clip, width=0.37\textwidth]{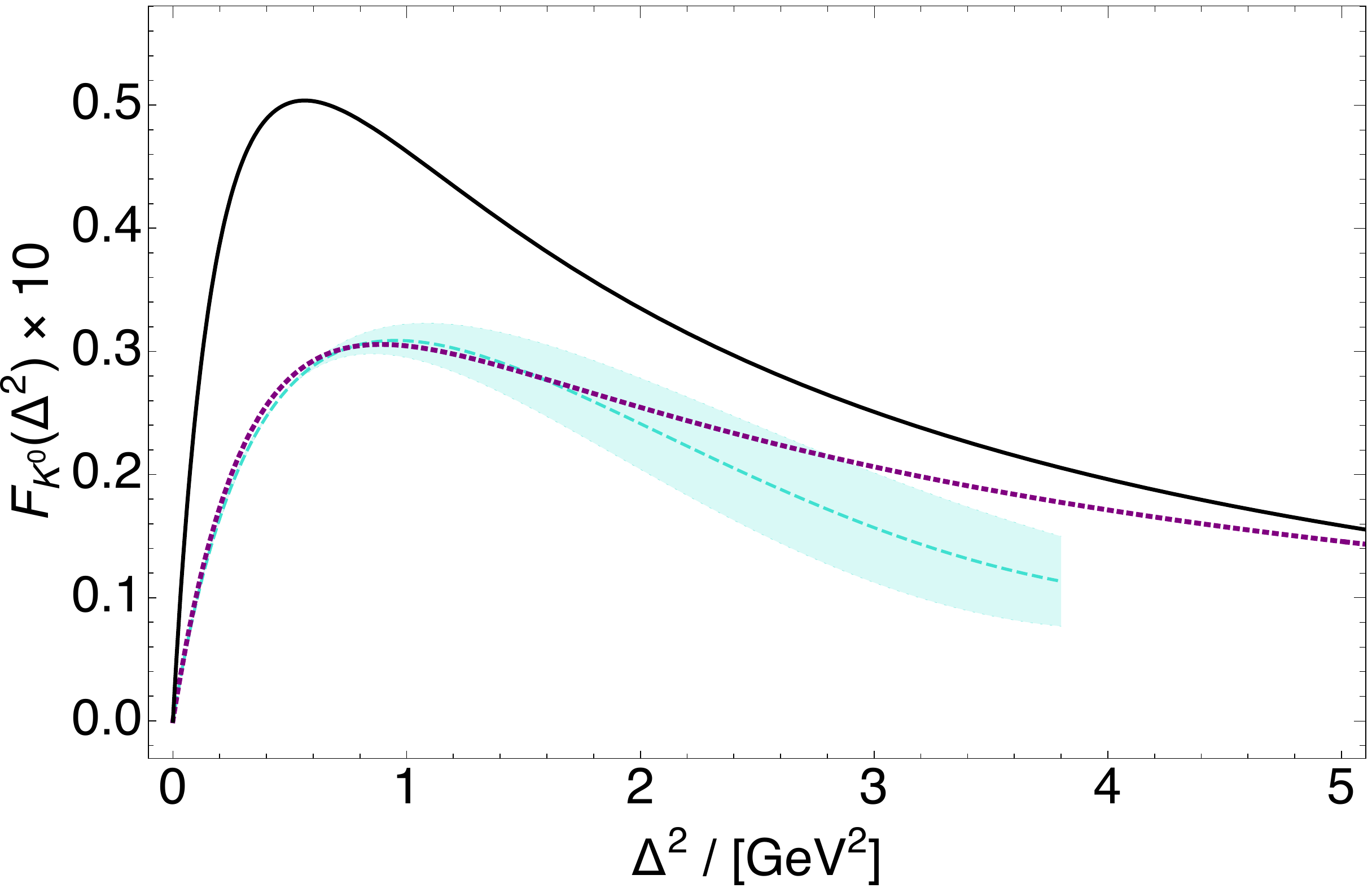}}
\caption{
\emph{Upper panel}\,--\,{\sf A}.  $\Delta^2 F_{K^+}(\Delta^2)$ calculated using the kaon GPD described in Sec.\,\ref{SecFac} -- short-dashed magenta curve.  Contributions from the individual valence-quark flavours are also drawn, each normalised to unity at $\Delta^2=0$.
The additional theory comparisons are as follows: prediction in Ref.\,\cite{Gao:2017mmp} -- solid black curve; and lQCD result from Ref.\,\cite{Davies:2019nut} -- dashed turquoise curve within like coloured band.
Data (crossed circles) from Ref.\,\cite{Carmignotto:2018uqj}, representing analyses of the $^1\!$H$(e,e^\prime K^+)\Lambda$ reaction.  The filled diamonds sketch anticipated data \cite{E12-09-011} (arbitrary normalisation): two error estimates are shown, which differ in their assumptions about the model- and $t$-dependence of the form factor extractions.
%
%
\emph{Lower panel}\,--\,{\sf B}.
$\Delta^2 F_{K^0}(\Delta^2)$, $F_{K^0}(\Delta^2) = (1/3) F_{K^+}^{\bar s}(\Delta^2)-(1/3)F_{K^+}^{u }(\Delta^2)$, computed using flavour-separated form factors in Panel~A.
Additional theory comparisons are as in Panel~A.
\emph{N.B}. Results multiplied by a factor of 10 to facilitate comparison with charged-kaon curves.
\label{fig:FFK}}
\end{figure}

Elastic electromagnetic form factors of the pion and kaon, computed using Eq.\,\eqref{eq:FF} and the GPDs described above, are drawn in Figs.\,\ref{fig:FFpi}, \ref{fig:FFK}.  For these calculations, one only need know the GPD on the DGLAP domain.
Regarding $F_\pi(\Delta^2)$ in Fig.\,\ref{fig:FFpi}, it is plain that PTIR and factorised-\emph{Ansatz} GPDs deliver practically equivalent predictions.  Importantly, the data in Fig.\,\ref{fig:FFpi} were not used to constrain either \emph{Ansatz} for the pion LFWF; instead, as previously remarked, all LFWFs used herein are entirely determined by the meson DAs described in Refs.\,\cite{Cui:2020dlm, Cui:2020tdf}.

Consider now the kaon elastic form factors in Fig.\,\ref{fig:FFK}.  Panel~A depicts $\Delta^2 F_{K^+}(\Delta^2)$ and the associated flavour separation as computed using the kaon GPD discussed in Sec.\,\ref{SecFac}, drawn in Fig.\,\ref{fig:KGPDFac}, and its partner for $H^{\bar s}_{K}(x,\xi,-\Delta^2;\zeta_{\mathpzc H})$.
The result obtained with this simple GPD \emph{Ansatz} is in fair agreement with the prediction in Ref.\,\cite{Gao:2017mmp}, which used a far more elaborate and computationally intensive approach, and also with a lattice-QCD (lQCD) result \cite{Davies:2019nut}, which is still preliminary and does not extend beyond $Q^2 \approx 4\,$GeV$^2$.
This GPD model, based on a factorised LFWF, also delivers agreement on the independent contributions from the $u$- and $\bar s$-quarks with those predicted in Ref.\,\cite{Gao:2017mmp}, \emph{e.g}., $F_{K^+}^{\bar s}/F_{K^+}^{u}  = 1.5$ at $\Delta^2=4\,{\rm GeV}^2$.  As $\Delta^2\to \infty$, this ratio approaches unity.

Analogous results for the neutral kaon are drawn in Fig.\,\ref{fig:FFK}B.  Again, there is semiquantitative agreement with both the Ref.\,\cite{Gao:2017mmp} prediction and the lattice-QCD result.
These outcomes are interesting because the neutral-kaon charge form factor is the difference between two curves that are identical at $Q^2=0$ and of similar magnitude thereafter; so, any loss of precision is magnified in the difference.

\section{Impact Parameter Space GPD -- Hadron Scale}
\label{SecIPSGPD}
The impact parameter space (IPS) GPD is obtained by considering the following specification of kinematics: $\Delta^2>0$, $\Delta\cdot P_{\mathsf P} = 0$, $\xi=0$, and then evaluating a two-dimensional Fourier (Hankel) transform with respect to the remaining two degrees-of-freedom, \emph{viz}.
\begin{equation}
{\mathpzc u}^{\mathsf P}(x,b_\perp^2;\zeta_{\mathpzc H}) = \int_0^\infty \frac{d\Delta}{2\pi} \Delta J_0(|b_\perp| \Delta) \, H_{\mathsf P}^u(x,0,-\Delta^2;\zeta_{\mathpzc H})\,,
\label{eq:IPDHgen}
\end{equation}
where $J_0$ is a cylindrical Bessel function.  This density reveals the probability of finding a parton within the light-front at a transverse distance $|b_\perp|$ from the meson's centre of transverse momentum.  The IPS GPD is completely determined by the GPD's properties on the DGLAP domain.

Once more, valuable insights can be obtained by evaluating Eq.\,\eqref{eq:IPDHgen} using the GPD obtained from a factorised LFWF.  Substituting Eq.\,\eqref{eq:HfacT} into Eq.\,\eqref{eq:IPDHgen}, yields:
\begin{align}
{\mathpzc u}^{\mathsf P}& (x, b_\perp^2;\zeta_{\mathpzc H}) = \nonumber \\
& \frac{{\mathpzc u}^{\mathsf P}(x;\zeta_{\mathpzc H}) }{(1-x)^2}  \int_0^\infty \frac{s ds}{2\pi} \, \Phi_{\mathsf P}(s^2;\zeta_{\mathpzc H}) J_0\left( \frac{b_\perp s}{1-x} \right) \,.
\label{eq:IPDH}
\end{align}
Now recall Eq.\,\eqref{PDFeqPDA2}, from which it is clear that using any hadron-scale valence-quark DF, the first factor increases as $x^2$ when $x\to 1$.
Considering next the second factor in Eq.\,\eqref{eq:IPDH} at fixed $x$, this Hankel transform takes its maximum value at $|b_\perp|=0$.
Consequently, one should expect the IPS GPD to peak at $|b_\perp|=0$ and the height of this peak to increase steadily as $x \to 1$ whilst simultaneously becoming narrower owing to the increasingly oscillatory behaviour of the integrand's Bessel function.
Thus, the global maximum of the IPS GPD is given by ${\mathpzc u}^{\mathsf P} (x=1, b_\perp^2=0;\zeta_{\mathpzc H})$.
Plainly, Eq.\,\eqref{eq:IPDH} defines a density that is rotationally invariant, \emph{i.e}., a function of $|b_\perp|$, not $\vec{b}_\perp$; so, it is usual to plot $2\pi |b_\perp| {\mathpzc u}^{\mathsf P}( x, b_\perp^2;\zeta_{\mathpzc H})$.  The peak in this function is shifted to $|b_\perp|>0$ by an amount that expresses aspects of bound-state dynamics.

Using the GPD developed from a factorised LFWF, one also obtains a simple expression for the longitudinal light-front distribution of the mean-square transverse light-front extent of $u$-in-${\mathsf P}$:
\begin{equation}
\label{eq:b2x}
\langle b_\perp^2(x;\zeta_{\mathpzc H}) \rangle_{u}^{\mathsf P}
=r_{\mathsf P}^2  \frac{(1-x)^2 {\mathpzc u}^{\mathsf P}(x;\zeta_{\mathpzc H})}{{\mathpzc x}_{\mathsf P}^2(\zeta_{\mathpzc H})}  \,,
\end{equation}
where ${\mathpzc x}_{\mathsf P}$ is given in Eq.\,\eqref{definex}.
%
The replacement ${\mathpzc u}^{\mathsf P}(x;\zeta_{\mathpzc H}) \to (1-{\mathpzc d}_{\mathsf P}) \bar{\mathpzc h}(x;\zeta_{\mathpzc H})$ in the numerator yields the ${\bar h}$ result.

%
%

\begin{figure}[t!]
\centerline{\includegraphics[clip, width=0.38\textwidth]{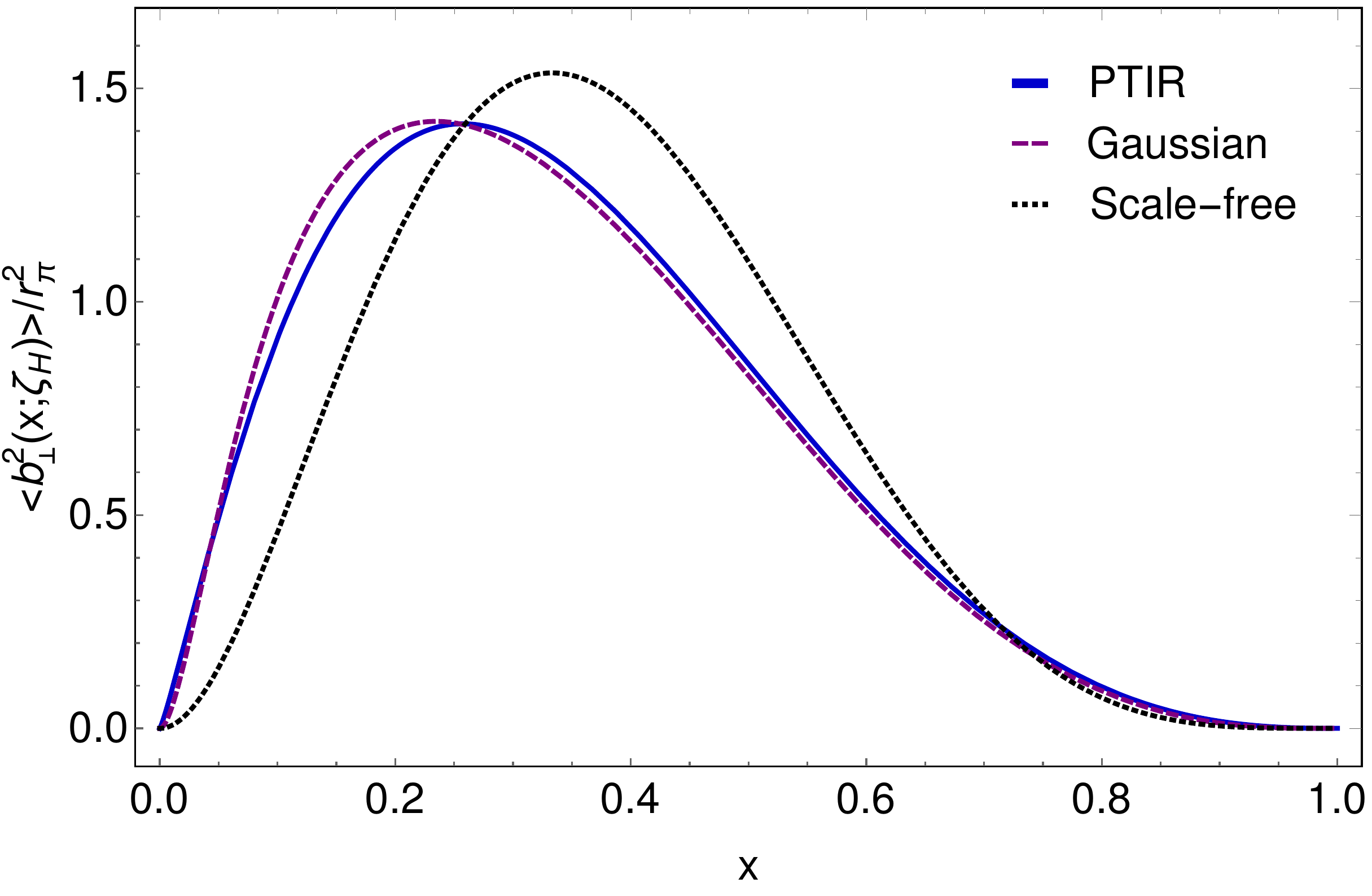}}
\caption{\label{fig:b2x}
$\langle b_\perp^2(x;\zeta_{\mathpzc H}) \rangle_{u}^{\pi}/r_\pi^2$, Eq.\,\eqref{eq:b2x}:
longitudinal light-front distribution of the mean-square transverse light-front extent of $u$-in-$\pi$.
Results obtained with PTIR GPD, Sec.\,\ref{SecPTIR}, and factorised GPD, Sec.\,\ref{SecFac}, are compared with that characterising a scale-free system, whose valence DF is given by Eq.\,\eqref{eq:scale-free}.
}
\end{figure}

Eq.\,\eqref{eq:b2x} indicates that the behaviour of $\langle b_\perp^2(x;\zeta_{\mathpzc H}) \rangle_{u}^{\mathsf P}$ can be read from that of the associated valence-quark DF.  In physical systems, owing to EHM as expressed in its DCSB corollary, such DFs are dilated and flattened with respect to the scale-free profile \cite{Chang:2014lva}:
\begin{equation}\label{eq:scale-free}
{\mathpzc q}_{\rm sf}(x;\zeta_H)=30 x^2 (1-x)^2 \,.
\end{equation}
It follows that $\langle b_\perp^2(x;\zeta_{\mathpzc H}) \rangle_{u}^{\mathsf P}$ for physical systems should peak at a lower value of $x$ than the result obtained with ${\mathpzc q}_{\rm sf}(x;\zeta_H)$, have a lower peak magnitude, and possess greater support at small- and large-$x$.  These expectations are borne out by the $\pi$-meson results plotted in Fig.\,\ref{fig:b2x}.
In detail, using Eq.\,\eqref{eq:param} and Table~\ref{tab:DFs}, one finds that $\langle b_\perp^2(x;\zeta_{\mathpzc H})\rangle_u^\pi $ is broadest on $x\simeq 0.23$ and becomes progressively narrower as $x\to 1$.

Integrating Eq.\,\eqref{eq:b2x} over $x$, one obtains the mean-square transverse light-front extent:
\begin{eqnarray}
\langle b_\perp^2 \rangle = \frac{ r^2_{\mathsf{P}} }{\mathpzc{x}^2_{\mathsf{P}}(\zeta_H)} 
\left\{
\begin{array}{lr}
\langle x^2 \rangle_{\bar{h}}^{\zeta_H} & u\\
\langle x^2 \rangle_u^{\zeta_H} (1-\mathpzc{d}_{\mathsf{P}}) & \bar{h}
\end{array}
\right. ;
\label{eq:b2int}
\end{eqnarray}
For the pion, so long as the $\zeta=\zeta_{\cal H}$ DF is symmetric around $x=1/2$, as it is in any sound treatment of the bound-state problem \cite{Ding:2019lwe}:
\begin{equation}
\langle b_\perp^2(\zeta_{\mathpzc H}) \rangle_{u}^\pi  = \frac{2}{3} r_\pi^2 = \langle b_\perp^2(\zeta_{\mathpzc H}) \rangle_{\bar d}^\pi \,.
\label{bperppion}
\end{equation}
Regarding the kaon, using the realistic DF specified by Eq.\,\eqref{PDFeqPDA2} and Table~\ref{tab:DFs}:
\begin{equation}
\label{SepBaryon}
\langle b_\perp^2(\zeta_{\mathpzc H}) \rangle_{u}^K = 0.71 r_K^2\,,
\langle b_\perp^2(\zeta_{\mathpzc H}) \rangle_{\bar s}^K = 0.58 r_K^2\,.
\end{equation}
Evidently, there is a separation of baryon number in a $u\bar h$ meson, with the lighter $u$-quark lying, on average, further from the system's centre of transverse momentum than the heavier $\bar h$-quark.


\begin{figure}[t!]
\vspace*{3.5ex}

\leftline{\hspace*{0.5em}{\large{\textsf{A}}}}
\vspace*{-5ex}
\centerline{\includegraphics[clip, width=0.42\textwidth]{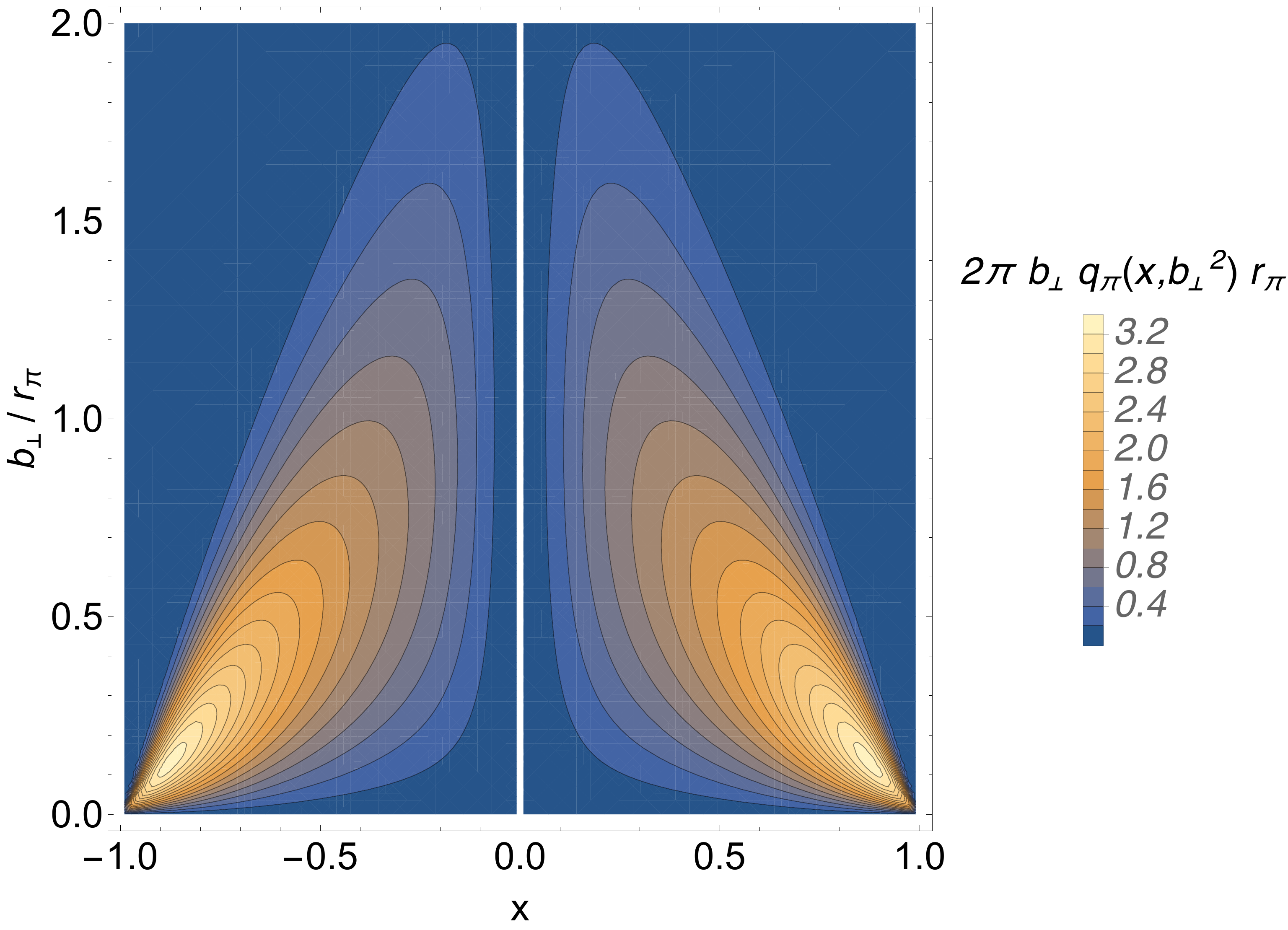}}
\vspace*{6ex}

\leftline{\hspace*{0.5em}{\large{\textsf{B}}}}
\vspace*{-5ex}
\centerline{\includegraphics[clip, width=0.42\textwidth]{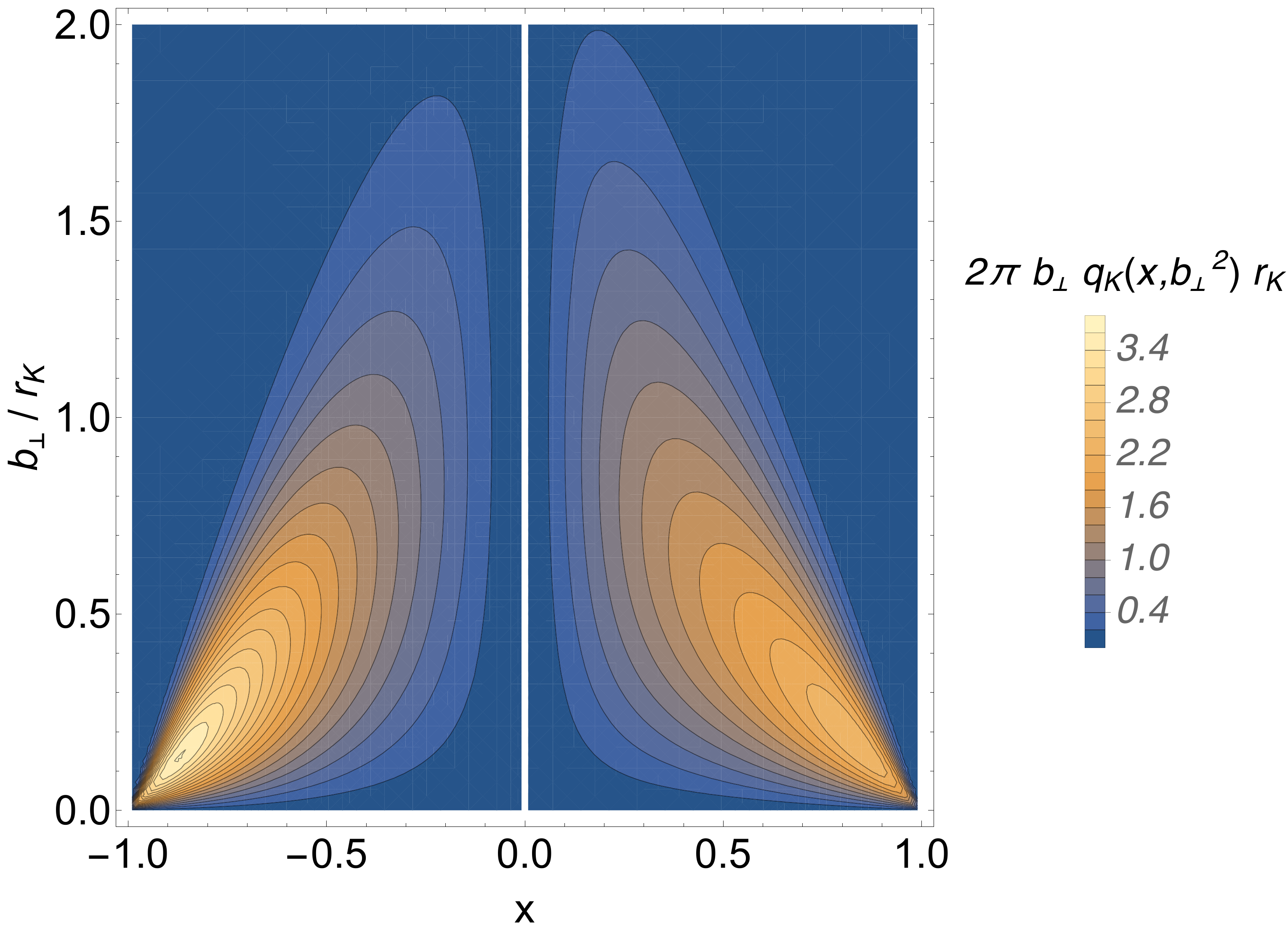}}
%
\caption{\label{fig:IPDGPDs} 
\emph{Upper panel}\,--\,{\sf A}.
Pion. IPS distributions for $\bar d$- (left) and $u$-quarks (right), charting the $x$-dependent probability density for locating these objects at distance $|b_\perp|$ from the pion's centre of transverse momentum
\emph{Lower panel}\,--\,{\sf B}.  Kaon.  Analogous profiles for $\bar s$ and $u$-quarks in the $K^+$.
Curves calculated using Eq.\,\eqref{eq:IPDHq} and its analogues (based on factorised LFWF \emph{Ans\"atze}, which yield the GPDs in Figs.\,\ref{fig:piGPDFac}, \ref{fig:KGPDFac}).
In both panels, $\zeta=\zeta_{\cal H}$.
}
\end{figure}

Using the Gaussian \emph{Ansatz} in Eq.\,\eqref{eq:gaussH}, one can also obtain an algebraic expression for the IPS GPD:
\begin{equation}
{\mathpzc u}^{\mathsf P}(x,b_\perp^2;\zeta_{\mathpzc H}) =
\frac{{\mathpzc x}_{\mathsf P}^2(\zeta_{\mathpzc H})}{\pi r_{\mathsf P}^2} \, \frac{{\mathpzc u}^{\mathsf P}(x;\zeta_{\mathpzc H})}{(1-x)^2}
 \exp{\left(- \frac{{\mathpzc x}_{\mathsf P}^2(\zeta_{\mathpzc H})}{(1-x)^2} \frac{b_\perp^2}{r_{\mathsf P}^2} \right)} \,.
\label{eq:IPDHq}
\end{equation}
The ${\bar h}$-in-${\mathsf P}$ GPD has nonzero support on $x\in (-1,0)$ and is obtained via the following replacements in Eq.\,\eqref{eq:IPDHq}: ${\mathpzc u}^{\mathsf P} \to \bar{{\mathpzc h}}^{\mathsf P}$, $r_{\mathsf P}^2 \to r_{\mathsf P}^2 (1-{\mathpzc d}_{\mathsf P})$, and $x \to |x|$.  The resulting pion and kaon IPS GPDs are drawn in Fig.\,\ref{fig:IPDGPDs}.

The images in Fig.\,\ref{fig:IPDGPDs} disclose some notable features of the three-dimensional distributions of valence degrees-of-freedom within pseudoscalar mesons.
\begin{enumerate}[label=(\emph{\roman*})]
%
\item All distributions are $|b_\perp|$-broad at small $|x|$, showing there is little probability of finding valence constituents on this domain at $\zeta=\zeta_{\mathpzc H}$.
%
\item As $|x|$ increases, each distribution acquires a clear maximum at some value of $|b_\perp|$, whose height increases with $|x|$ whilst its width diminishes.
%
\item Each probability density has a global maximum; and at $\zeta=\zeta_{\mathpzc H}$ their positions and magnitudes (${\mathpzc i}_M$ quoted from Fig.\,\ref{fig:IPDGPDs}) are:
%
\begin{subequations}
\label{IPSheights}
\begin{align}
\mbox{$\pi$:} & \quad (|x|,b_\perp/r_\pi)=(0.88,0.13)\,, \; {\mathpzc i}_\pi=3.29\,, \\
\mbox{$K_u$:} & \quad
(x,b_\perp/r_K)_u=(0.84,0.17)\,,\; {\mathpzc i}_K^u=2.38\,, \\
\mbox{$K_{\bar s}$:} & \quad
(x,b_\perp/r_K)_{\bar s}=(-0.87,0.13)\,, \; {\mathpzc i}_K^{\bar s}=3.61\,.
\end{align}
\end{subequations}
Clearly, in a bound-state generated by valence constituents with different masses, the heavier object plays a greater part in defining the system's centre of transverse momentum; hence, lies closer to this point.  For the kaon, the relative shift is small --  just 3\%, but the difference in magnitudes is 20\%, matching the size of $M_{\bar s}/M_u$, which is the typical scale for Higgs-modulation of EHM in this system.
It is worth noting, too, that the pion peak is 10\% larger than the mean of the $u$-in-$K$ and $\bar s$-in-$K$ heights.
\end{enumerate}

\begin{figure}[t!]
\vspace*{3.5ex}

\leftline{\hspace*{0.5em}{\large{\textsf{A}}}}
\vspace*{-5ex}
\centerline{\includegraphics[clip, width=0.39\textwidth]{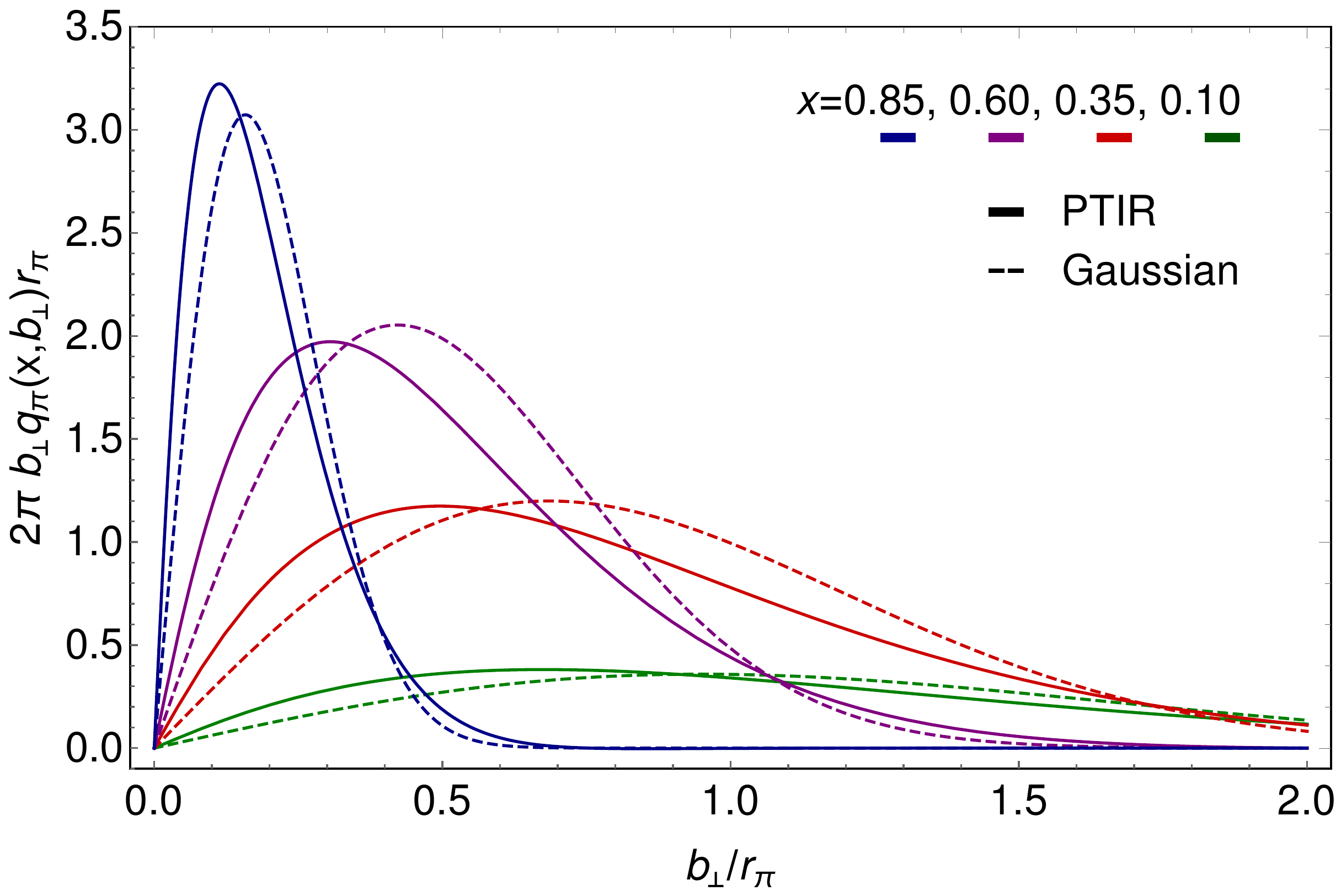}}
\vspace*{6ex}

\leftline{\hspace*{0.5em}{\large{\textsf{B}}}}
\vspace*{-5ex}
\centerline{\includegraphics[clip, width=0.39\textwidth]{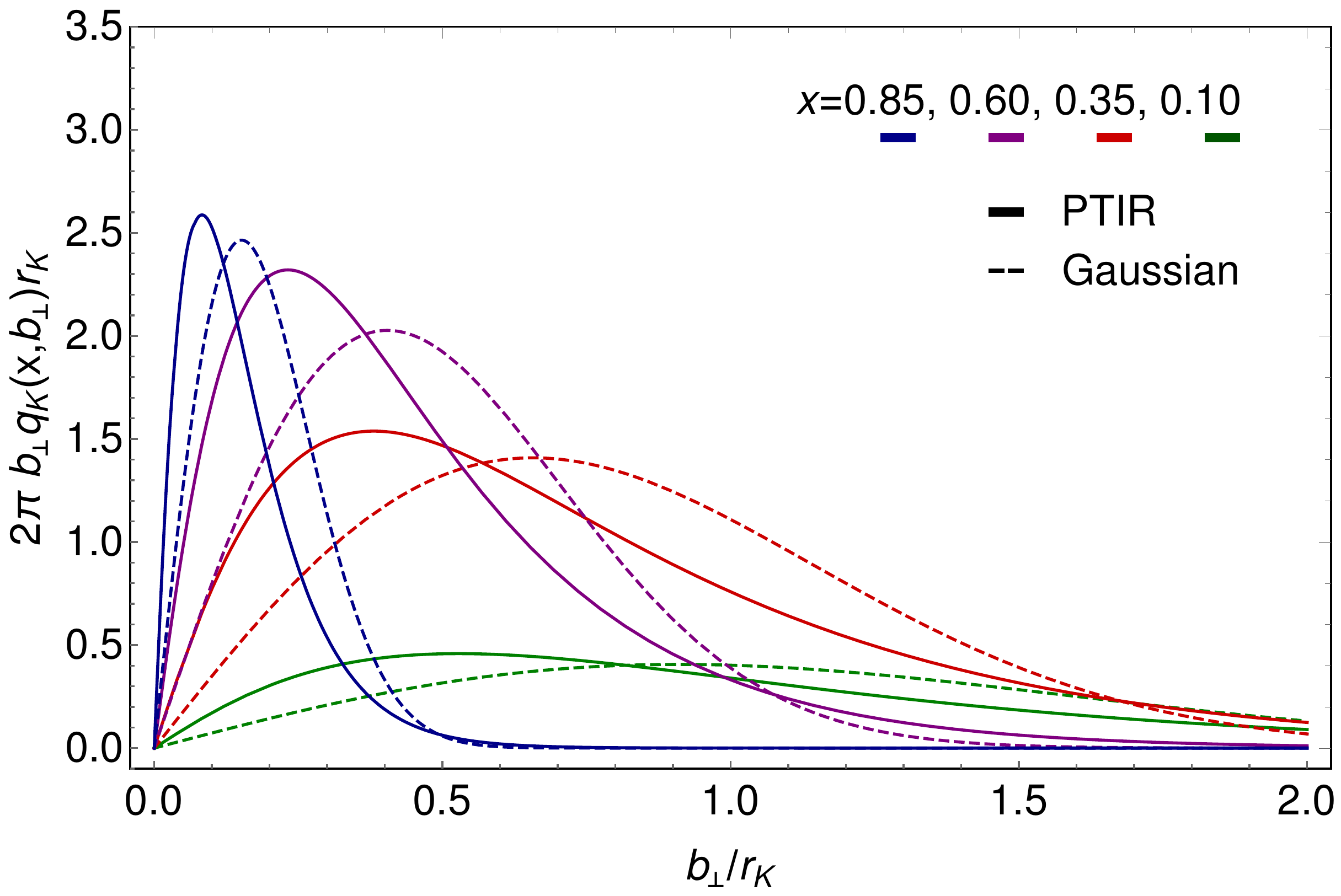}}
%
\caption{\label{fig:IPDGPDcomp} 
\emph{Upper panel}\,--\,{\sf A}.
Pion. $u$-quark IPS distribution slices, plotted as a function of $|b_\perp|$ at four $x$ values, as computed with: PTIR LFWF (solid curves), Sec.\,\ref{SecPTIR}; and factorised LFWF (short-dashed curves), Sec.\,\ref{SecFac}.  The peak height decreases with increasing $x$.
\emph{Lower panel}\,--\,{\sf B}.  Kaon.  Analogous profiles for $u$-quark in the $K^+$.
}
\end{figure}

Fig.\,\ref{fig:IPDGPDcomp} reveals that all these features persist in the results obtained using the more sophisticated PTIR LFWF \emph{Ans\"atze} discussed in Sec.\,\ref{SecPTIR}, although there are minor quantitative differences.
In this case, the probability for the $u$-in-$\pi$ to be found a distance $|b_\perp|$ from the center of transverse momentum peaks at $(x,b_\perp/r_\pi)$ = $(0.91,0.065)$, with $i_\pi=3.46$; whereas for the kaon -- $(x,b_\perp/r_K)_u=(0.83,0.094)$, ${\mathpzc i}_K^u=2.59$.
and
$(x,b_\perp/r_K)_{\bar s}=(-0.94,0.041)$, ${\mathpzc i}_K^{\bar s}=4.30$.

\section{Pressure Profiles}
\label{SecPressure}
\subsection{Gravitational form factors}
\label{SecGFF}
Meson gravitational form factors can be accessed via the first Mellin moment of their GPDs:
\begin{equation}
\label{Mom1GPD}
%
\int_{-1}^1 \! dx\, x\, H_{\mathsf P}^{\mathpzc q}(x,\xi,-\Delta^2;\zeta_{\cal H})  = \theta_2^{{\mathsf P}}(\Delta^2) - \xi^2 \theta_1^{{\mathsf P}}(\Delta^2)\,,
\end{equation}
where
$\theta_{1,2}$ are related, respectively, to the distributions of pressure and mass-squared within the meson.  QCD evolution will be canvassed in Secs.\,\ref{SecAll}, \ref{SecPartition}.  Here we only note that the individual form factors on the right-hand-side of Eq.\,\eqref{Mom1GPD} are scale invariant.  It is the left-hand-side that changes.

\begin{figure}[t!]
\centerline{\includegraphics[clip, width=0.42\textwidth]{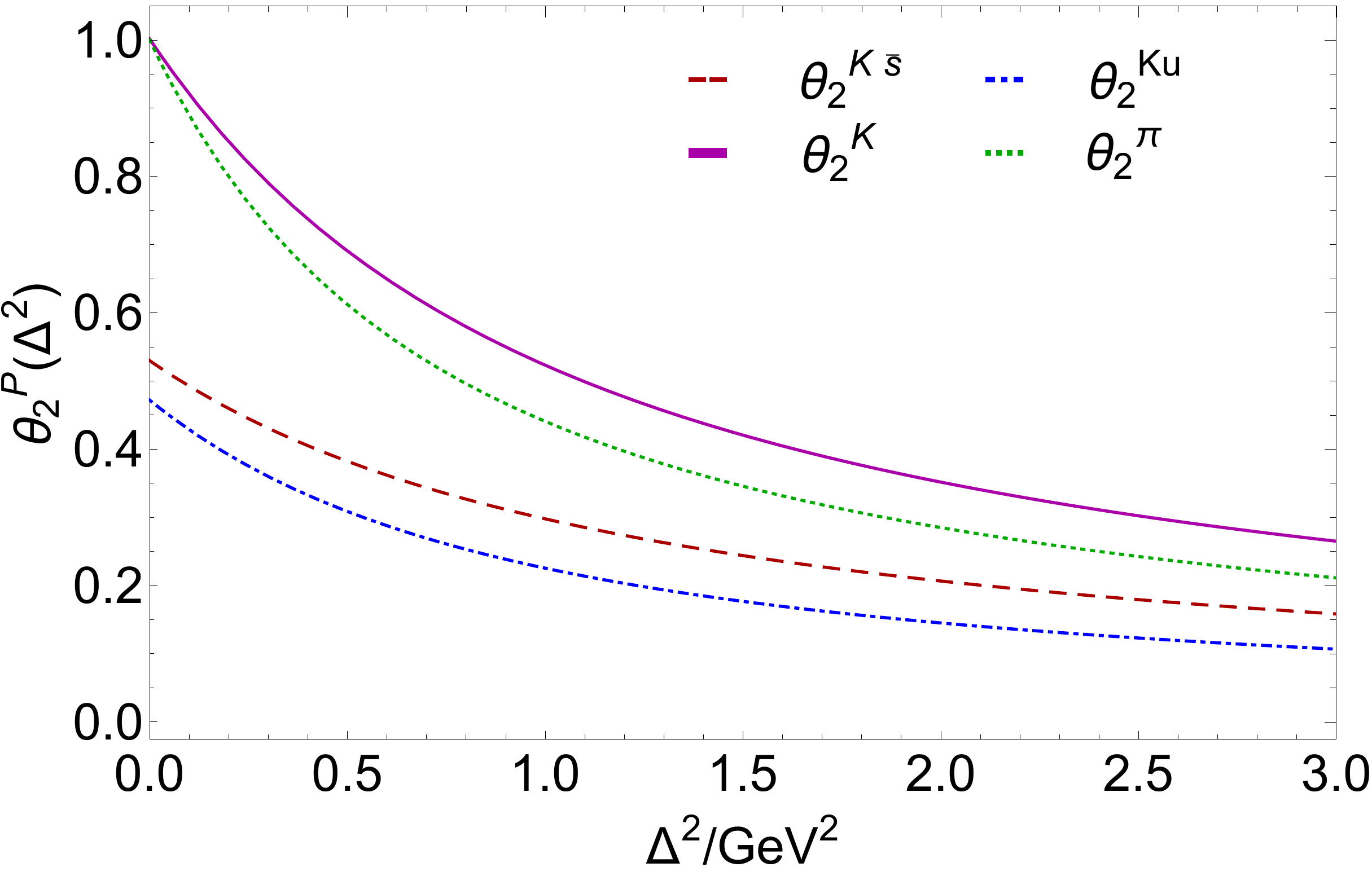}}
\caption{\label{fig:theta2P}
Mass-squared distribution form factors, $\theta_2$, for the pion and kaon, along with a $\zeta=\zeta_{\cal H}$ flavour separation of the latter, all computed using Eq.\,\eqref{eq:gaussH} (based on factorised LFWF) in Eq.\,\eqref{Mom1GPD} and $\xi=0$.
}
\end{figure}

Like an electromagnetic form factor, $\theta_2$ can readily be computed using the information above, \emph{viz}.\ setting $\xi=0$ and using the GPDs defined on the DGLAP domain.  Fig.\,\ref{fig:theta2P} depicts the results obtained with the GPDs in Sec.\,\ref{SecFac} including, for the kaon, its $\zeta=\zeta_{\cal H}$ flavour separation: at this scale, \begin{equation}
\theta_2^{K_u}(\Delta^2;\zeta_{\cal H}) + \theta_2^{K_{\bar s}}(\Delta^2;\zeta_{\cal H}) = \theta_2^K(\Delta^2)\,,
\end{equation}
so that the entirety of the meson's mass-squared is lodged with the dressed valence degrees-of-freedom.  (See Sec.\,\ref{SecPartition}.)  Evidently, following the pattern set by the electromagnetic form factors, $\theta_2^K$ is stiffer than $\theta_2^\pi$.  We return to this point in connection with Eq.\,\eqref{kaontheta2radii}.

The calculation of $\theta_1$ is not so straightforward because it is sensitive to properties of the GPD on the ERBL domain; thus, our \emph{Ans\"atze} must be extended.  
In principle, a covariant extension of the GPD can be accomplished by exploiting a connection between the Radon transform \cite{Radon:1986} and the GPD double distribution representation \cite{Radyushkin:1997ki, Mueller:1998fv, Polyakov:1999gs}.  For the pion, a practicable approach was explained in Refs.\,\cite{Chouika:2017dhe, Chouika:2017rzs}.  An issue is the ``D-term'' ambiguity \cite{Polyakov:1999gs, Teryaev:2001qm, Tiburzi:2004qr}; but building on Refs.\,\cite{Chouika:2017dhe, Chouika:2017rzs, Zhang:2020ecj}, Ref.\,\cite{Zhang:2021mtn} overcame this problem by writing the $u$-in-$\pi$ GPD as follows:
\begin{align}
\label{eq:GPDfull}
\widehat{H}_\pi^u&(x,\xi,-\Delta^2) = H_\pi^u(x,\xi,-\Delta^2) +   P_\pi(\Delta^2)  \nonumber \\
 &\times  \mbox{\rm sign}(\xi) \left[  D^{-}\left(\frac{x}{\xi}\right) \, + \,  \frac{1}{\xi} D^{+}\left(\frac{x}{\xi}\right) \right] \, ,
\end{align}
where, in general, $D^-(z)$ and $D^+(z)$ are odd and even functions, respectively, with support on $z\in [-1,1]$.  By construction, $\widehat{H}_\pi^u= H_\pi^u$ on the DGLAP domain; so, the former respects polynomiality if the latter does, irrespective of the choice for $P_\pi(\Delta^2)$.  This last function is a propagator for the scalar resonance that contributes on the ERBL domain owing to $\pi \pi$ rescatterings \cite{Theussl:2002xp, Zhang:2020ecj}.

With the following choices:
\begin{subequations}
\label{eq:Ds}
\begin{align}
\label{eq:Dm}
D^{-}(z) = \tfrac{1}{2} \, & \left[H_\pi^u(-z,1,0)  -   H_\pi^u(z,1,0) \right]  \, , \\
D^{+}(z) = \tfrac{1}{2} \, & \left[ \varphi_\pi^u\left( \frac{1+x}{2} \right) -  H_\pi^u(z,1,0)  \right.  \nonumber \\
& \qquad - \left. \widehat{H}_\pi^u(-z,1,0) \right]  \; ,
\label{eq:Dp}
\end{align}
\end{subequations}
%
then $\widehat{H}_\pi^u$ complies with the soft pion theorems \cite{Polyakov:1999gs}.
%
To ensure that all physical constraints are preserved, \emph{e.g}., GPD polynomiality, we complete an extension of $H_\pi^u$ to the ERBL domain, $|x| < \xi$, by adapting the algebraic PTIRs used elsewhere \cite{Mezrag:2014jka} to sketch the pion's GPD and employing the Radon transform approach \cite{Chouika:2017dhe, Chouika:2017rzs}.  In this case, one obtains an algebraic result, which, regarding Eq.\,\eqref{eq:Ds}, is only needed on $\xi\simeq 1$.

\begin{figure}[t!]
\centerline{\includegraphics[clip, width=0.42\textwidth]{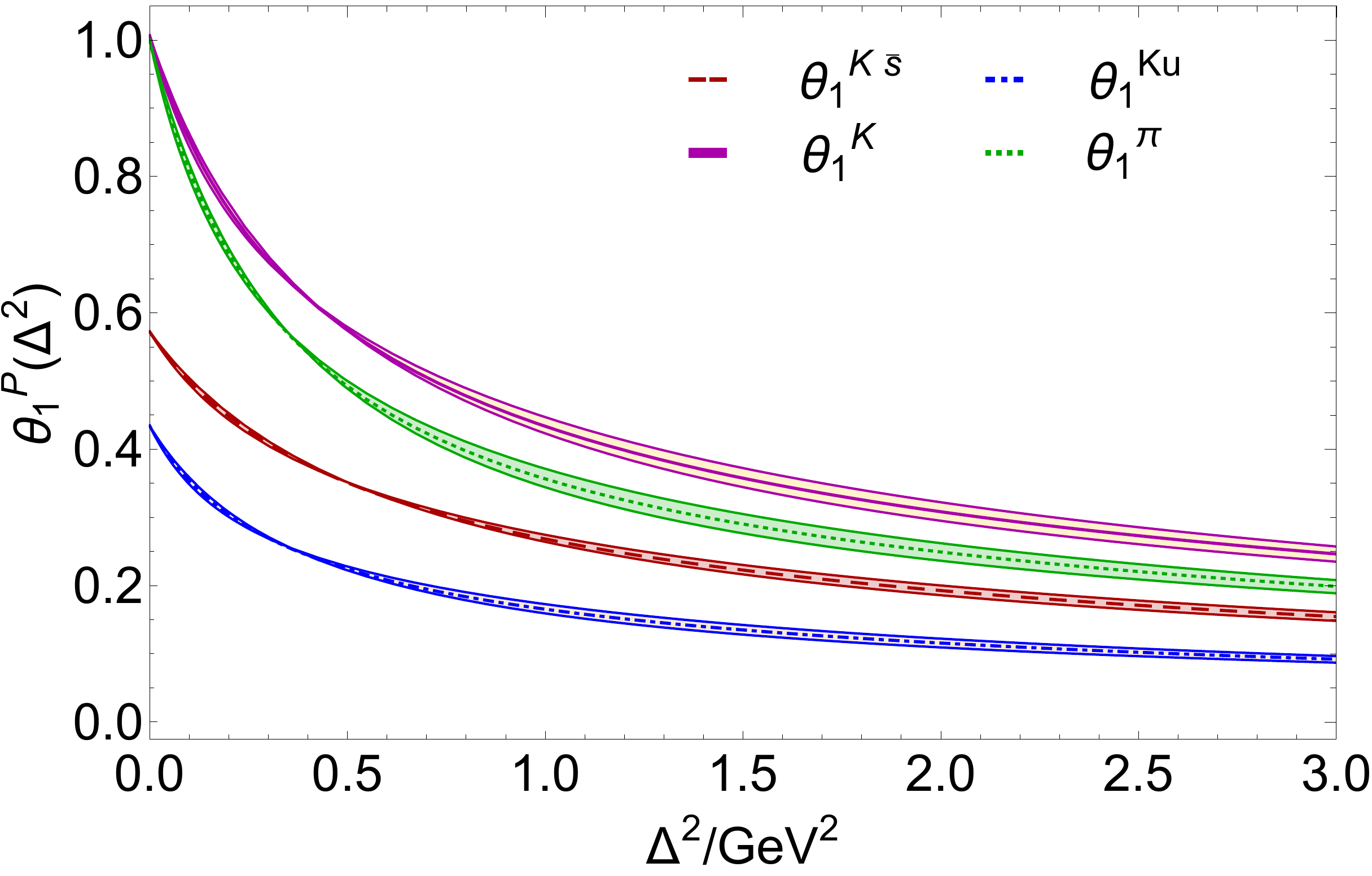}}
\caption{\label{fig:theta1P}
Pressure distribution form factors, $\theta_1$, for the pion and kaon, along with a $\zeta=\zeta_{\cal H}$ flavour separation of the latter, all computed using Eq.\,\eqref{EqKaonT1} with $\theta_2$ from Fig.\,\ref{fig:theta2P}.  The band associated with each curve describes the model uncertainty explained in connection with Eq.\,\eqref{EqKaonT1}.}
\end{figure}

Focusing on the $\Delta^2$-dependence of the $u$-in-$\pi$ extension, the algebraic formula reveals that it is weighted by the
dressed-mass of the active valence degree-of-freedom, \emph{i.e}., $4 M_u^2$ for the $u$-in-$\pi$ GPD.  This is readily generalised for a kaon extension by assuming that $u$-in-$K$ involves $4 M_u^2$ and $\bar s$-in-$K$, $4 M_s^2$.
(These masses are listed in Table~\ref{tab:params}.)
Given the simplicity of the approach we vary $M \to M (1\pm 0.1)$ so as to estimate a model uncertainty.  As the last step, we exploit the insights developed in Ref.\,\cite{Zhang:2020ecj} to arrive at:
\begin{align}
%
%
 \theta_1^{{\mathsf P}_{\mathpzc q}}&(\Delta^2)  = c_1^{{\mathsf P}_\mathpzc q} \theta_2^{{\mathsf P}_{\mathpzc q}}(\Delta^2)+ \int_{-1}^1 \! dx\, \nonumber \\
& \times  x\, \left[ H_{\mathsf P}^{\mathpzc q}(x,1,0) P_{M_{\mathpzc q}}(\Delta^2) - H_{\mathsf P}^{\mathpzc q}(x,1,-\Delta^2)\right]\,, \label{EqKaonT1}
\end{align}
where
$c_1^{\pi_u} = 1$,
$c_1^{K_{u,\bar s}}=(1 \pm_{u}^{\bar s} 0.08)$ reflects slight violation of the soft-pion constraint \cite{Polyakov:1999gs} in the kaon channel, as calculated following Ref.\,\cite{Zhang:2020ecj},
and
\begin{equation}
P_{M}(\Delta^2)= 1/(1+y \ln(1+y))\,,
\end{equation}
$y=\Delta^2/[4 M^2]$, where the $\ln(1+y)$ piece is included to express the scaling violation that characterises quantum field theories in four dimensions.

The pressure distributions produced by Eq.\,\eqref{EqKaonT1} are depicted in Fig.\,\ref{fig:theta1P}, including a $\zeta=\zeta_{\cal H}$ flavour separation for the $K$: at this scale, $\theta_1^{K_u}(\Delta^2) + \theta_1^{K_{\bar s}}(\Delta^2) = \theta_1^K(\Delta^2)$.  Following the pattern set by the electromagnetic form factors and $\theta_2$, $\theta_1^K$ is stiffer than $\theta_1^\pi$.

Adapting the usual expression for form factor radii:
\begin{equation}
[r_{\mathsf P}^{\theta_2}]^2  = \theta_2^{{\mathsf P}_u}(0)[r_{{\mathsf P}_u}^{\theta_2}]^2+\theta_2^{{\mathsf P}_{\bar s}}(0)[r_{{\mathsf P}_{\bar h}}^{\theta_2}]^2 ,
\label{RadiusDefinition}
\end{equation}
augmented by Eq.\,\eqref{EqRadius}; then using Eq.\,\eqref{eq:gaussH}, obtained with factorised LFWF \emph{Ans\"atze}, one arrives at algebraic expressions for the mass-squared radii:
{\allowdisplaybreaks
\begin{subequations}
\begin{align}
[r_{{\mathsf P}_u}^{\theta_2}]^2 & = \frac{3 r_{\mathsf P}^2}{2{\mathpzc x}_{\mathsf P}^2} \langle x^2 (1-x) \rangle_{{\mathsf P}_{\bar h}}\,, \\
[r_{{\mathsf P}_{\bar h}}^{\theta_2}]^2 & = \frac{3 r_{\mathsf P}^2}{2{\mathpzc x}_{\mathsf P}^2}
(1-{\mathpzc d}_{\mathsf P})\langle x^2 (1-x) \rangle_{{\mathsf P}_u}\,,
\end{align}
\end{subequations}
%
where ${\mathpzc x}_{\mathsf P}^2$ is given in Eq.\,\eqref{definex}.
%
}

Consider the $\pi$, for which ${\mathpzc x}_{\,\pi}^2=(3/2)\langle x^2\rangle_{\pi_u}^{\zeta_{\cal H}}$.  Then
\begin{equation}
\label{radiusratio}
[r_\pi^{\theta_2}]^2 = 2
[r_{\pi_u}^{\theta_2}]^2 = r_\pi^2 \frac{2 \langle x^2(1-x)\rangle_{\pi_u}^{\zeta_{\cal H}}}{\langle x^2\rangle_{\pi_u}^{\zeta_{\cal H}}}.
\end{equation}
The ratio of moments in Eq.\,\eqref{radiusratio} is always a positive number less than unity because for any valence-quark DF that is positive-definite and even on $x\in(0,1)$:
\begin{align}
\frac{2 \langle x^2(1-x)\rangle_{\pi_u}^{\zeta_{\cal H}}}{\langle x^2\rangle_{\pi_u}^{\zeta_{\cal H}}}
& = \frac{1/2 - \langle x^2\rangle_{\pi_u}^{\zeta_{\cal H}}}{\langle x^2\rangle_{\pi_u}^{\zeta_{\cal H}}}
= \frac{\langle x(1-x)\rangle_{\pi_u}^{\zeta_{\cal H}}}{\langle x^2\rangle_{\pi_u}^{\zeta_{\cal H}}}\,;
\end{align}
and $0<\langle x(1-x)\rangle_{\pi_u}^{\zeta_{\cal H}}<\langle x^2\rangle_{\pi_u}^{\zeta_{\cal H}}$, where the first bound is plain for even functions and the second follows from the Cauchy-Schwarz inequality, which entails $\langle x^2\rangle_{\pi_u}^{\zeta_{\cal H}}> [\langle x\rangle_{\pi_u}^{\zeta_{\cal H}}]^2=1/4$.  Generalisation of the argument to the kaon is straightforward.  Hence, based on factorisable LFWFs, the meson's mass-squared radius is always smaller than its charge radius.  The reliability of such \emph{Ans\"atze} for integrated quantities suggests that the result is also true for PTIR LFWFs.

\begin{figure}[t!]
\vspace*{3.5ex}

\leftline{\hspace*{0.5em}{\large{\textsf{A}}}}
\vspace*{-5ex}
\centerline{\includegraphics[clip, width=0.415\textwidth]{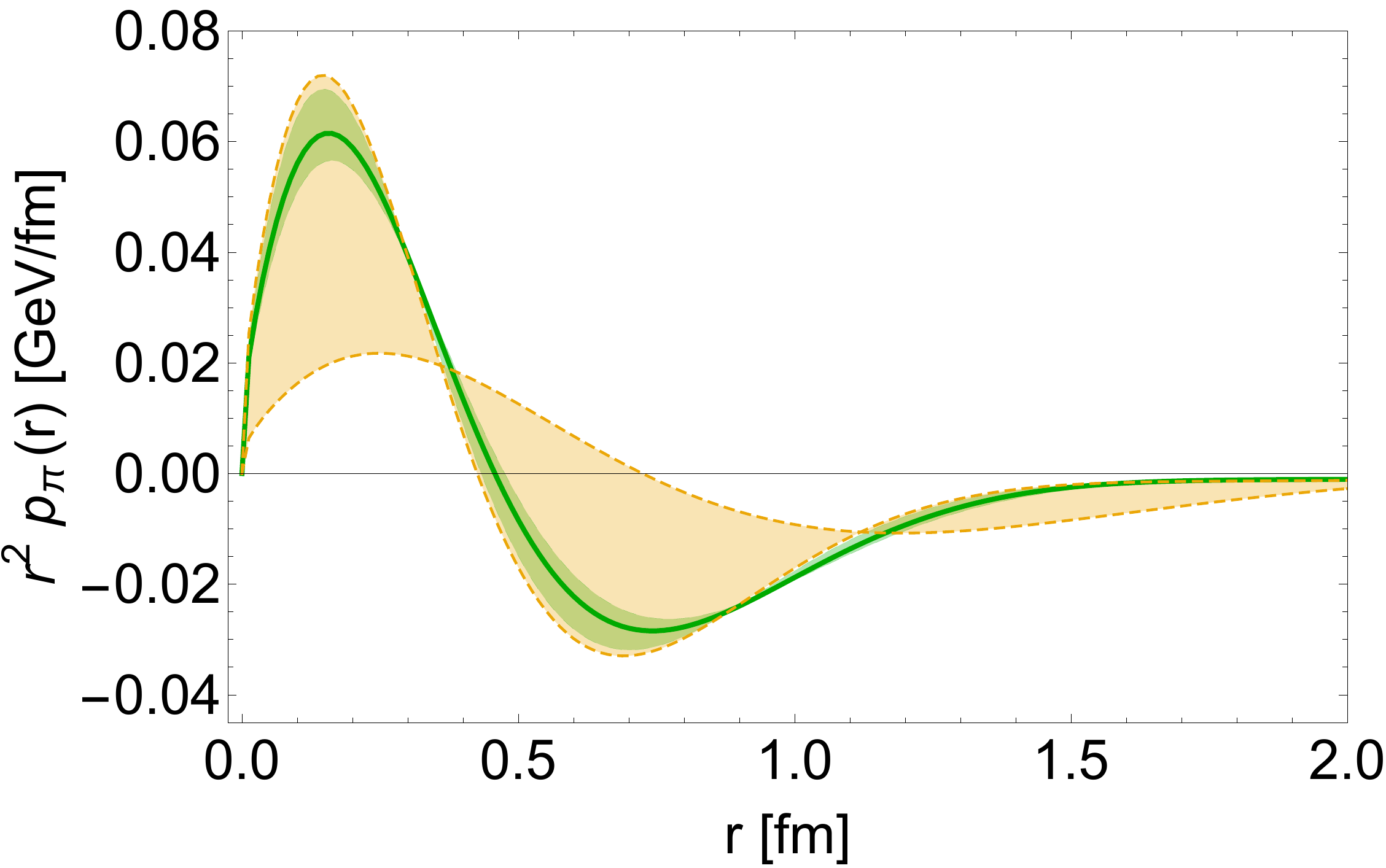}\hspace*{2.2ex}}
\vspace*{6ex}

\leftline{\hspace*{0.5em}{\large{\textsf{B}}}}
\vspace*{-5ex}
\centerline{\includegraphics[clip, width=0.4\textwidth]{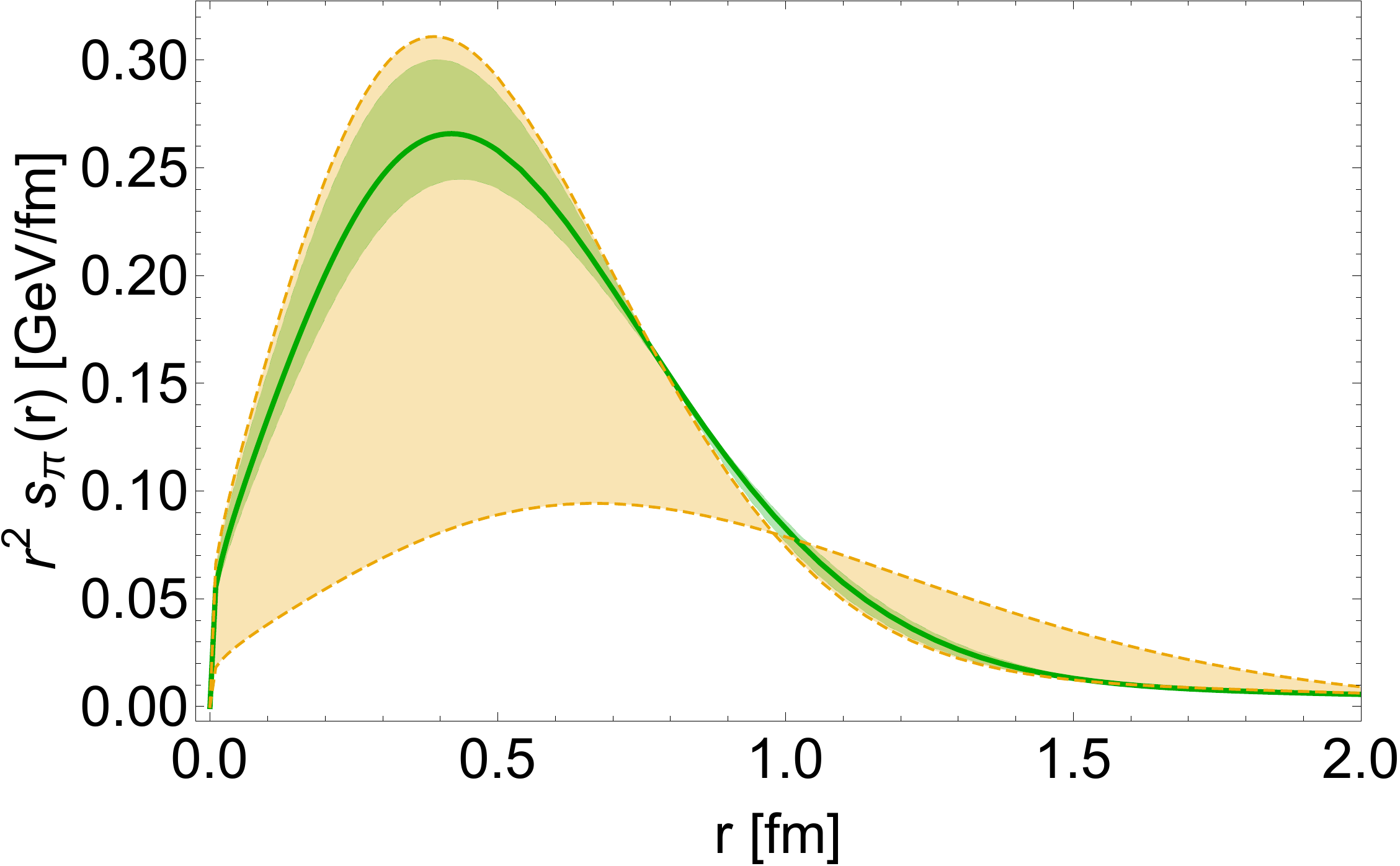}}
\vspace*{6ex}

\leftline{\hspace*{0.5em}{\large{\textsf{C}}}}
\vspace*{-5ex}
\centerline{\includegraphics[clip, width=0.4\textwidth]{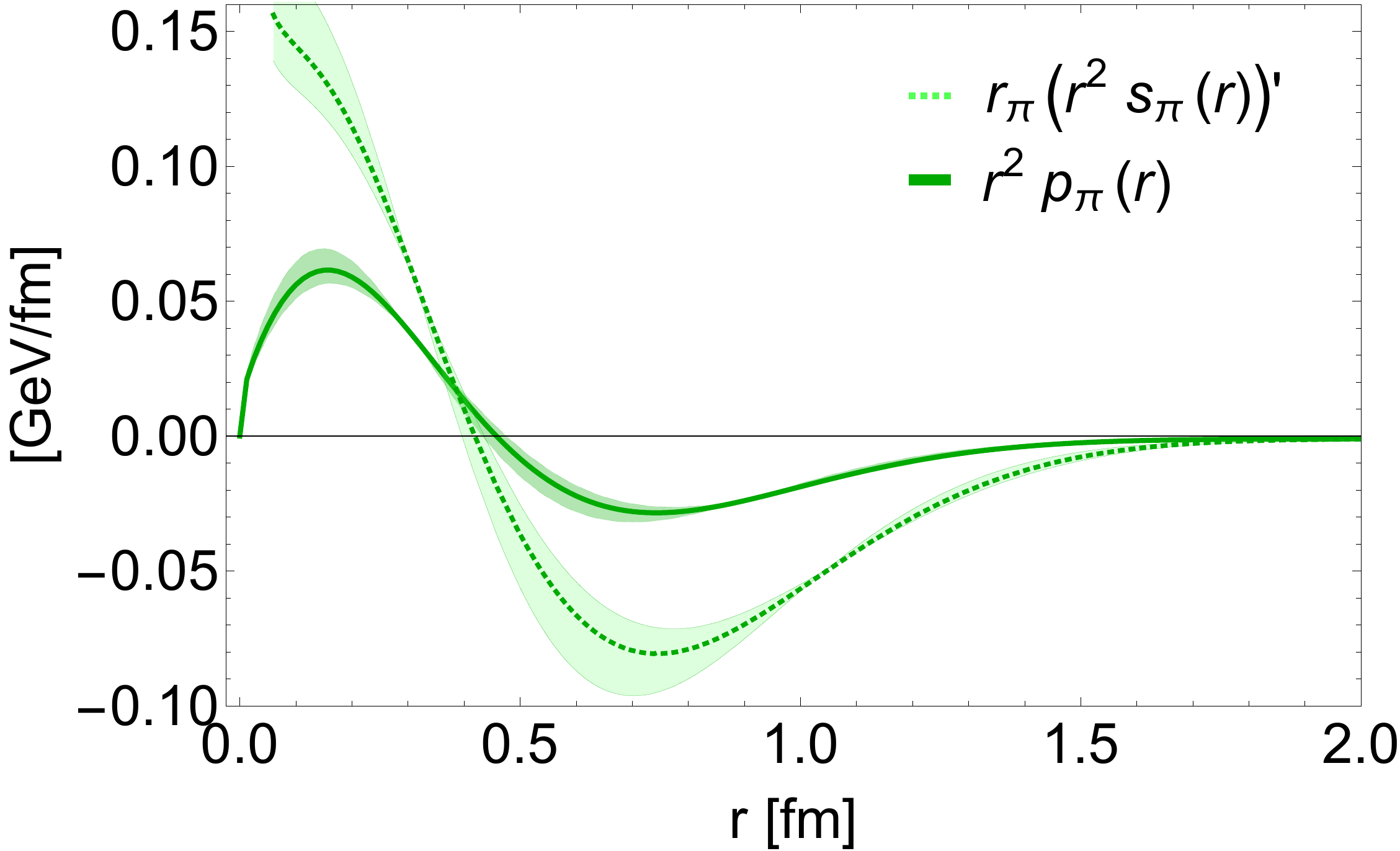}}
%
\caption{\label{fig:r2p} 
Pion profiles.  \emph{Upper panel}\,--\,{\sf A}: pressure.  \emph{Middle panel}\,--\,{\sf B}: shear force.
Green curve and band computed from obvious analogues of Eqs.\,\eqref{EqKaonT1}, \eqref{EqPressure}, with the terms in the second line of Eq.\,\eqref{EqKaonT1} evaluated using $M= M_u (1 \pm 0.1)$.
Gold band: SCI results \cite{Zhang:2020ecj}.
\emph{Lower panel}\,--\,{\sf C}:
Pion pressure distribution in Panel~{\sf A} (solid green curve) compared with the derivative of the related shear force distribution in Panel~{\sf B}.  (The $r_\pi$ factor ensures matching units.)  Both curves have near coincident zeros, revealing that shear forces are maximal in the neighbourhood of the radius whereat the pressure changes sign and confinement effects become dominant, Eq.\,\eqref{confinementradius}.
}
\end{figure}

These general statements are borne out by the LFWFs described herein -- using Eq.\,\eqref{eq:param} and Table~\ref{tab:DFs}:
\begin{equation}
\label{kaontheta2radii}
\begin{array}{c|c||c|c}
r_\pi^{\theta_2}/r_\pi & r_{K}^{\theta_2}/r_{K} 
& r_{K_u}^{\theta_2}/{\bar r}_{K} & r_{K_{\bar s}}^{\theta_2}/{\bar r}_{K}\\\hline
0.81 & 0.78 
& 0.84 & 0.72
\end{array}\,,
\end{equation}
where $\bar{r}_K^2 = r_K^2/2$.  A symmetry-preserving treatment of a vector$\,\otimes\,$vector contact interaction (SCI) yields \cite{Zhang:2020ecj}: $r_\pi^{\theta_2}/r_\pi=0.89$.

Working instead with the $\theta_1$ (pressure) form factors:
\begin{equation}
\label{kaontheta1radii}
\begin{array}{c|c||c|c}
r_\pi^{\theta_1}/r_\pi & r_K^{\theta_1}/r_K 
& r_{K_u}^{\theta_1}/{\bar r}_{K}
    & r_{K_{\bar s}}^{\theta_1}/{\bar r}_{K}\\\hline
1.18(6) & 1.19(6) 
& 1.25(7) & 1.13(5)
\end{array}\,,
\end{equation}
where the uncertainties propagate those explained in connection with Eq.\,\eqref{EqKaonT1}.  The SCI returns \cite{Zhang:2020ecj}: $r_\pi^{\theta_1}/r_\pi=1.88(13)$.

In all known cases, the SCI radii ordering matches that produced by our \emph{Ans\"atze}; and these results accord with those extracted from measurements of $\gamma^\ast \gamma \to \pi^0 \pi^0$ \cite{Kumano:2017lhr}.
In the latter connection, one could extend our approach somewhat in order to compute two-pion and -kaon generalised distribution amplitudes \cite{Diehl:1998dk, Diehl:2000uv}.
Notably, the separation of baryon number exposed by Eq.\,\eqref{SepBaryon} and Fig.\,\ref{fig:IPDGPDs} is also manifest in the gravitational radii.


\subsection{Breit-frame pressure distributions}
Following Refs.\,\cite{Polyakov:2002yz, Polyakov:2018zvc}, Breit-frame pressure profiles can be calculated for each meson; and with $u$-in-$K$ as an example, one has the following pressure, $p(r)$, and shear force, $s(r)$, distributions \cite{Zhang:2020ecj, Zhang:2021mtn}:
{\allowdisplaybreaks
\begin{subequations}
\label{EqPressure}
\begin{align}
p_K^u(r)  & =
 \frac{1}{6\pi^2 r} \int_0^\infty d\Delta \,\frac{\Delta}{2 E(\Delta)} \, \sin(\Delta r) [\Delta^2\theta_1^{K_u}(\Delta^2)] \,, \label{EqPressureA}\\
 s_K^u (r)  & =
%
\frac{3}{8 \pi^2} \int_0^\infty d\Delta \,\frac{\Delta^2}{2 E(\Delta)} \, {\mathpzc j}_2(\Delta r) \, [\Delta^2\theta_1^{K_u}(\Delta^2)] \,, \label{EqPressureB}
\end{align}
\end{subequations}
where
$2E(\Delta)=\sqrt{4 m_K^2+\Delta^2}$
and ${\mathpzc j}_2(z)$ is a spherical Bessel function.
The total pressure is a sum of the individual contributions from the valence-parton degrees-of-freedom: $p_K = p_K^u + p_K^{\bar s}$.  The total shear force is obtained similarly.
(As in Eq.\,\eqref{eq:IPDHgen}, two-dimensional Fourier transforms are sometimes preferred \cite{Miller:2010nz}; nevertheless, this results in similar profiles and magnitudes.)
}

Pressure profiles calculated using the pion form factor depicted in Fig.\,\ref{fig:theta1P} are drawn in Fig.\,\ref{fig:r2p}; and as noted elsewhere \cite{Zhang:2020ecj, Zhang:2021mtn}, they admit an intuitive interpretation.  Focusing first on Fig.\,\ref{fig:r2p}A, the pressure is large and positive on $r\simeq 0$, showing that the pion's dressed-valence constituents repel each other when their separation is small.  However, with increasing $r$, the pressure decreases, changing sign at
\begin{equation}
\label{confinementradius}
r_c^{\pi} = 0.45(3)\,{\rm fm}.
\end{equation}
The SCI predicts a similar value \cite{Roberts:2021nhw}.  This point marks the transition to a domain upon which confinement forces become the dominant influence on the pair.

Fig.\,\ref{fig:r2p}B depicts the in-pion shear pressure, which is an expression of deformation forces inside the meson.  As observed in Refs.\,\cite{Zhang:2020ecj, Zhang:2021mtn} and highlighted by Fig.\,\ref{fig:r2p}C, these forces are largest on $r \simeq r_c^\pi$; namely, in the neighbourhood of that point where attractive confinement pressure begins to overwhelm the forces driving the quark and antiquark apart.  From this perspective, $r_c^{\pi}$ defines a pressure-based pion confinement radius.

It is notable that profiles like the curve in Fig.\,\ref{fig:r2p}A can also be drawn for neutron stars.  In that case, the $r\simeq 0$ pressure is roughly $0.1\,$GeV/fm \cite{Ozel:2016oaf}.  Evidently, therefore, pions and neutron stars have near-core pressures of similar magnitudes.

\begin{figure}[t]
\vspace*{3.5ex}

\leftline{\hspace*{0.5em}{\large{\textsf{A}}}}
\vspace*{-5ex}
\centerline{\includegraphics[clip, width=0.415\textwidth]{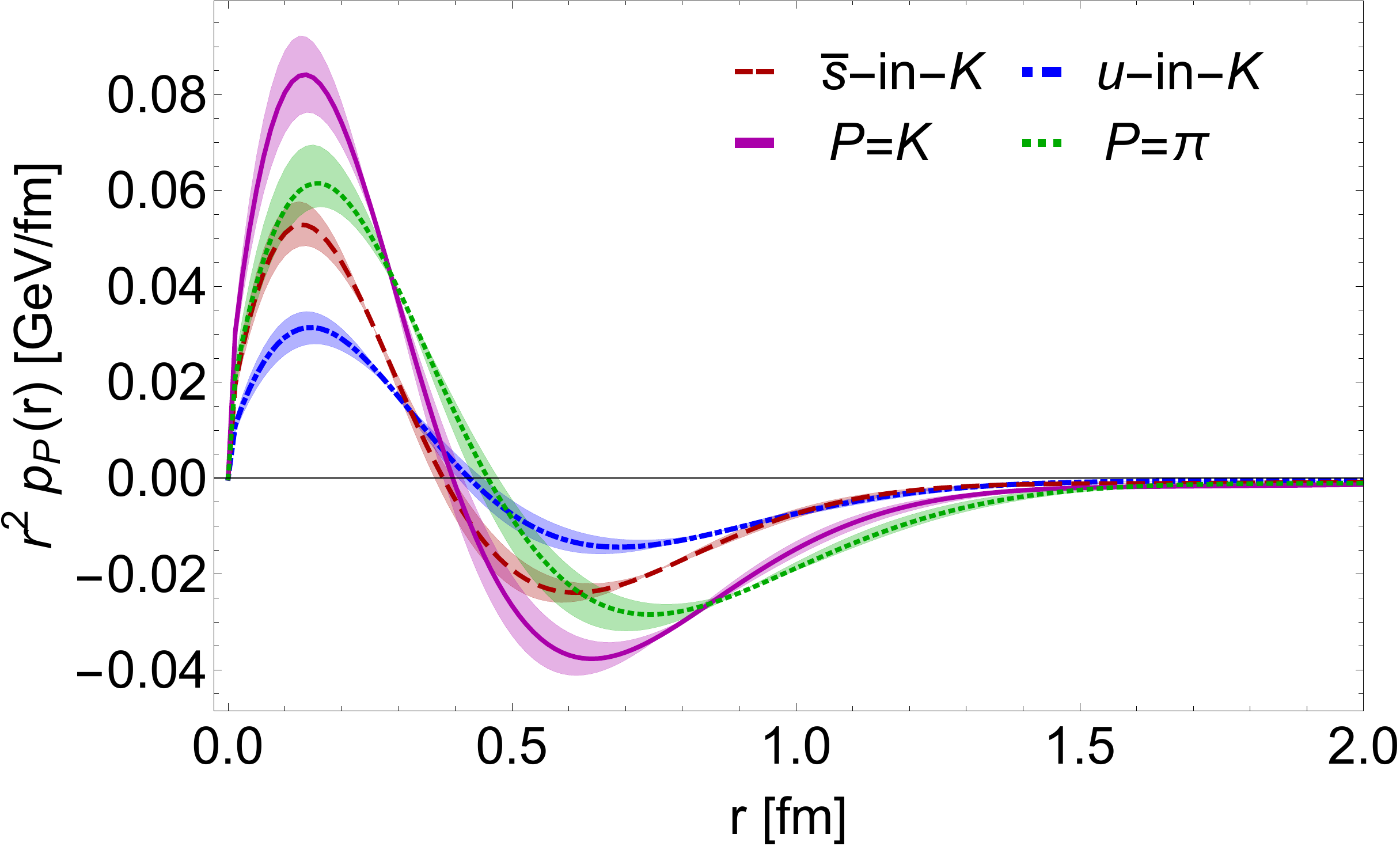}\hspace*{2.2ex}}
\vspace*{6ex}

\leftline{\hspace*{0.5em}{\large{\textsf{B}}}}
\vspace*{-5ex}
\centerline{\includegraphics[clip, width=0.4\textwidth]{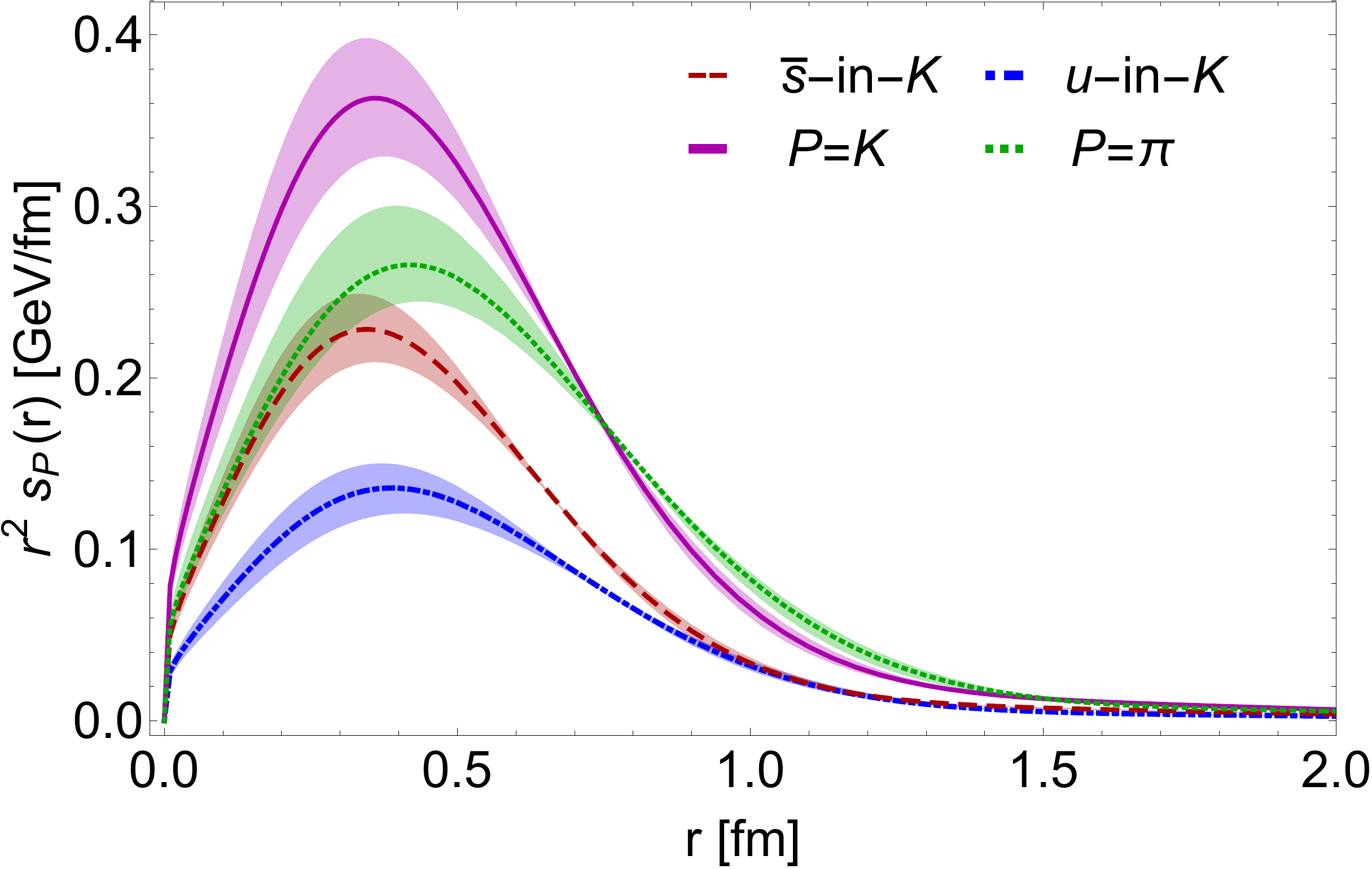}}
%
\caption{\label{fig:r2pK} 
Kaon profiles.
\emph{Upper panel}\,--\,{\sf A}: pressure, Eq.\,\eqref{EqPressureA}.
\emph{Lower panel}\,--\,{\sf B}: shear force, Eq.\,\eqref{EqPressureB}.
%
%
$K$ results obtained using Eq.\,\eqref{EqKaonT1}, with $M_u = M_u(1\pm0.1)$, $M_{\bar s} = M_{\bar s}(1\pm 0.1)$: solid magenta curve and band.
Flavour separation: $u$-in-$K$ -- dot-dashed blue curve; and $\bar s$-in-$K$ -- dashed red.
Dotted green curve and like-coloured band reproduce the pion profiles in Fig.\,\ref{fig:r2p}.
}
\end{figure}

Kaon pressure profiles are depicted in Fig.\,\ref{fig:r2pK}.  Although qualitatively similar to the pion profiles, the quantitative differences are meaningful because they reveal Higgs boson modulation of EHM.
Regarding the pressure radius, the kaon is approximately 15\% smaller than the pion.  The same level of contraction is seen when comparing the respective charge radii (Fig.\,\ref{fig:piGPD} cf.\ Fig.\,\ref{fig:KGPD}) and mass-squared radii, Fig.\,\ref{fig:theta2P}.  Moreover, the kaon core pressure is 20\% greater than that in the pion and
\begin{equation}
\int_0^\infty dr\,r^2 s_{K}(r) = 1.19(1) \int_0^\infty dr\,r^2 s_{\pi}(r)\,.
\end{equation}
Naturally, $\int_0^\infty dr\,r^2 p_{\pi,K}(r) = 0$ because the mesons are bound states.

The $\zeta=\zeta_{\cal H}$ flavour separations of the $K$ pressure profiles are also drawn in Fig.\,\ref{fig:r2pK}.  As one might have anticipated from inspection of
Figs.\,\ref{fig:FFpi}, \ref{fig:FFK}, \ref{fig:IPDGPDs}
and
Eqs.\,\eqref{bperppion}, \eqref{SepBaryon}, \eqref{kaontheta2radii}, \eqref{kaontheta1radii},
the $\bar s$-quark contributes a greater fraction of the total $K$ pressure than its partner $u$-quark,
its peak/trough magnitudes are greater,
and the associated distributions are concentrated nearer to $r=0$.
As remarked elsewhere \cite{Zhang:2021mtn}, one should expect similar effects in a flavour separation of proton pressure profiles, identifying $u_p$ with $\bar s_K$ and $d_p$ with $u_K$.  There are intimations of this in Ref.\,\cite[Fig.\,8]{Cui:2020rmu}.

Returning to Eq.\,\eqref{Mom1GPD}, we reiterate that the mass and pressure form factors are empirical observables; hence, cannot depend on the resolving scale, $\zeta$.  This is obvious with the electromagnetic form factor: defined as the zeroth moment of the hadron's GPD, it is manifestly $\zeta$-independent.  However, the gravitational form factors stem from the first moment; so, the picture is more complex in this case.  Whilst the right-hand-side of Eq.\,\eqref{Mom1GPD} is fixed, the integrand on the left-hand-side evolves with $\zeta$: on $\zeta>\zeta_{\cal H}$, the valence-quark contribution becomes a sum of valence, sea and glue contributions, the combination of which is $\zeta$-independent.  More on this in Sec.\,\ref{SecPartition}.

\section{All-orders evolution}
\label{SecAll}
\subsection{Process-independent charge}
\label{SecPI}
The $\zeta=\zeta_H$ GPD is not accessible in experiment because special kinematic conditions are required before the data can be interpreted in such terms \cite{Ellis:1991qj}.  Experiments with momentum transfers squared $Q^2 \sim \zeta_E^2 > m_N^2$ are typically needed.  Consequently, any comparison with measurements requires that the hadron scale GPDs discussed above be evolved to some appropriate $\zeta_E$.  We achieve this by adopting the all-orders evolution scheme detailed in Ref.\,\cite{Cui:2019dwv, Cui:2020dlm, Cui:2020tdf}, which uses QCD's process-independent (PI) effective charge \cite{Cui:2019dwv} to integrate the one-loop DGLAP equations.

Owing to the emergence of a nonzero gluon mass-scale \cite{Cornwall:1981zr, Aguilar:2015bud, Brodsky:2008be, Gao:2017uox, Fischer:2018sdj, Huber:2018ned, Roberts:2021nhw}, QCD's PI charge saturates in the infrared. 
The following expression provides an interpolation of the numerical result \cite{Cui:2019dwv}:
\begin{align}
\label{Eqhatalpha}
\hat{\alpha}(k^2) & = \frac{\gamma_m \pi}{\ln\left[\frac{{\mathpzc K}^2(k^2)}{\Lambda_{\rm QCD}^2}\right]}
\,,\; {\mathpzc K}^2(y) = \frac{a_0^2 + a_1 y + y^2}{b_0 + y}\,,
\end{align}
%
$\gamma_m=4/\beta_0$, $\beta_0=11 - (2/3)n_f$, $n_f=4$, $\Lambda_{\rm QCD}=0.234\,$GeV, with (in GeV$^2$)
\begin{equation}
\label{Eqhatalpha2}
\begin{array}{c|c|c}
a_0 & a_1 & b_0 \\ \hline
0.104(1) & 0.0975 & 0.121(1)
\end{array}\,.
\end{equation}
Notably, on $k^2 \gtrsim (9\Lambda_{\rm QCD})^2$, the relative difference between $\hat\alpha(k^2)$ and QCD's one-loop perturbative running coupling is less than 0.1\%.

Whilst QCD's perturbative running coupling exhibits a Landau pole at $k^2=\Lambda_{\rm QCD}^2$, this singularity is removed from the PI charge by nonperturbative gauge sector interactions.  The effect of such dynamics is apparent in Eq.\,\eqref{Eqhatalpha}, with $k^2/\Lambda_{\rm QCD}^2$ being replaced by ${\mathpzc K}^2(k^2)/\Lambda_{\rm QCD}^2$ as the argument of the logarithm.  Thus, the quantity
\begin{equation}
m_G := {\mathpzc K}(k^2=\Lambda_{\rm QCD}^2) = 0.331(2)\,{\rm GeV}
\end{equation}
defines a screening mass.  It marks a boundary \cite{Brodsky:2008be, Gao:2017uox}: modes with $k^2 \lesssim m_G^2$ are screened from interactions, the effective charge ceases to run, and the theory enters a conformal domain.  Consequently, the line $k=m_G$ is a border between soft and hard physics, leading naturally to the identification
\begin{equation}
\label{setzetaH}
\zeta_H=m_G\,.
\end{equation}

Herein we describe evolution of the $\xi=0$ valence-quark GPDs and associated IPS functions; so, we can employ the evolution scheme explained in Refs.\,\cite{Cui:2019dwv, Cui:2020dlm, Cui:2020tdf}.  Consequently, consider the $n$-th Mellin moment of the valence-quark GPD:
\begin{equation}
\langle x^n H_{\mathsf{P}}^f \rangle_{\zeta}^{\Delta^2} = \int_{-1}^1 dx x^n H_{\mathsf{P}}^f(x,0,-\Delta^2;\zeta)\;,
\end{equation}
for $f$=$u, \bar d, \bar s$, the all-orders evolution of which is defined via
\begin{equation}\label{eq:diffDGLAP}
\zeta^2 \frac{d}{d\zeta^2} \langle x^n H_{\mathsf{P}}^f \rangle_{\zeta}^{\Delta^2} = - \frac{\hat{\alpha}(\zeta^2)}{4\pi} \gamma_{ff}^n(\zeta)  \langle x^n H_{\mathsf{P}}^f \rangle_{\zeta}^{\Delta^2} \;,
\end{equation}
with
\begin{subequations}
\label{eq:gammaff}
\begin{align}
\label{eq:Pff}
&\gamma_{ff}^n(\zeta) = - \int_0^1 dz z^n P_{f \leftarrow f}(z;\zeta_H) \\
\label{eq:Puu}
&= - \int_0^1 dz z^n P_{u \leftarrow u}(z) +  \int_0^1 dz z^n {\cal D}_{f \leftarrow f}(z;\zeta_H) \;, \\
\label{eq:Dff}
&{\cal D}_{f \leftarrow f}(z;\zeta_H) = \sqrt{3} (1-2z) {\cal D}_f(\zeta)  \\
&{\cal D}_f(\zeta) = \frac{\delta_f^2}{\delta_f^2+(\zeta-\zeta_H)^2}\,.
\end{align}
\end{subequations}
Here, $P_{f \leftarrow f}$ is the $f$-quark splitting function, with $P_{u \leftarrow u}(z)$ a textbook result \cite{Ellis:1991qj}, modified to incorporate the mass-dependent correction introduced in Refs.\,\cite{Cui:2020dlm, Cui:2020tdf}, with $\delta_f=M_f-M_u$, referring to Table~\ref{tab:params} above.

Given that the hadron scale is fixed, Eq.\,\eqref{setzetaH}, then the solution of Eq.\,\eqref{eq:diffDGLAP} can be written in closed form:
\begin{equation}\label{eq:xnzeta}
\frac{\langle x^n H_{\mathsf{P}}^f \rangle_{\zeta}^{\Delta^2}}{\langle x^n H_{\mathsf{P}}^f \rangle_{\zeta_H}^{\Delta^2}} = \exp{\left[
-\frac{\gamma_0^n}{2\pi} \int_{\zeta_H}^{\zeta} \frac{dy}{y}
\hat{\alpha}(y^2) \left( 1 - a_n {\cal D}_f(y) \right)
\right]} \;,
\end{equation}
with
{\allowdisplaybreaks
\begin{subequations}
\begin{align} \label{eq:Dfn}
\gamma_0^n &= - \int_0^1 dz z^n P_{u \leftarrow u}(z) \nonumber \\
&= -\frac 4 3 \left( 3 + \frac{2}{(n+1)(n+2)} - 4 \sum_{j=1}^{n+1} \frac 1 j  \right) \;, \\
a_n &= \frac{n \sqrt{3}}{\gamma_0^n (n+1)(n+2)}  \;.
\label{eq:g0n}
\end{align}
\end{subequations}}

It is worth remarking here that $\gamma_{ff}^0(\zeta) \equiv 0$ because the splitting function conserves baryon number; hence, as stated above, the electromagnetic form factors defined by Eq.\,\eqref{eq:FF} are scale invariant.  Moreover, since ${\cal D}_{u}(y)\equiv 0$, then Eq.\,\eqref{eq:xnzeta} reproduces Ref.\,\cite[Eq.\,(18)]{Cui:2020tdf}.    

Regarding the evolution of singlet  ($S=\,$valence+sea) and glue ($g$) GPDs, it is important to recall that GPDs have nonzero support on $x\in (-1,1)$, with the $S$ GPD being an odd function on this domain and $g$ being even.  However, when evaluated at $\xi=0$, the domains on either side of $x=0$ decouple; so, for both $S$ and $g$, one can focus independently on $x>0$, whereupon one has the following all-orders evolution of the Mellin moments:
\begin{subequations}
\label{eq:DGLAPmom}
\begin{align}
&\zeta^2 \frac{d}{d\zeta^2} \langle x^n H_{\mathsf{P}}^S \rangle_{\zeta}^{\Delta^2} = - \frac{\hat \alpha(\zeta^2)}{4\pi}  \nonumber \\
&\times \left[
\gamma_{ff}^n(\zeta)  \langle x^n H_\mathsf{P}^S \rangle_{\zeta}^{\Delta^2}
+ 2 n_f \gamma_{fg}^n(\zeta)
\langle x^{n-1} H_{\mathsf{P}}^g \rangle_{\zeta}^{\Delta^2}
\right] \,, \\
&\zeta^2 \frac{d}{d\zeta^2} \langle x^{n-1} H_\mathsf{P}^g \rangle_{\zeta}^{\Delta^2} = - \frac{\hat\alpha(\zeta^2)}{4\pi}  \nonumber \\
&\times \left[
\gamma_{gf}^n(\zeta)  \langle x^n H_\mathsf{P}^S \rangle_{\zeta}^{\Delta^2}
+ \gamma_{gg}^n \langle x^{n-1} H_\mathsf{P}^g \rangle_{\zeta}^{\Delta^2}
\right] \, ,
\end{align}
\end{subequations}
where $H_\mathsf{P}^S$ and $H_\mathsf{P}^g$ are, respectively, the $\mathsf{P}$-meson quark singlet and glue GPDs, evaluated at $\xi=0$.
Notably, in the forward limit:
\begin{subequations}
\begin{align}
H_\mathsf{P}^S(&x,0,-\Delta^2)  \nonumber \\
& \stackrel{\Delta^2\to 0}{=} [\theta(x) - \theta(-x)]
\sum_{\mathpzc f} {\mathpzc f}^\mathsf{P}(|x|) + \bar{\mathpzc f}^\mathsf{P}(|x|) \,,\\
H_\mathrm{P}^g(&x,0,-\Delta^2)\,, \nonumber \\
& \stackrel{\Delta^2\to 0}{=} x g^\mathrm{P}(|x|)  \theta(x) - x g^\mathrm{P}(|x|)  \theta(-x) \,,
\end{align}
\end{subequations}
where $g^\mathsf{P}$ is the glue-in-$\mathsf{P}$ DF.

The anomalous dimensions in Eqs.\,\eqref{eq:DGLAPmom} are:
{\allowdisplaybreaks
\begin{subequations}
\label{eq:massADDGLAP}
\begin{align}
\gamma_{ff}^n(\zeta) &=  \gamma_{uu}^n \left[ 1 - a_n {\cal D}_f(y)\right]\,, \\
\gamma_{gf}^n(\zeta) &= \gamma_{gu}^n \left[ 1 + a_n \frac{\gamma_0^n}{\gamma_{gu}^n} {\cal D}_f(\zeta) \right]\,, \\
\gamma_{fg}^n(\zeta) &= \gamma_{ug}^n \left[1 + b_n {\cal D}_f(\zeta) \right]\,, \\
b_n &= \frac{\sqrt{5}}{\gamma_{ug}^n} \left[\frac{1}{1+n} - \frac{6}{2+n} + \frac{6}{3+n} \right]\,,
\end{align}
\end{subequations}
where $b_n$ is obtained via integration of ${\cal D}_{g \leftarrow f}$ in Ref.\,\cite[Eq.\,(60b)]{Cui:2020tdf}, like $a_n$ from Eq.\,\eqref{eq:Dff} above; and
\begin{subequations}
\label{eq:masslessADDGLAP}
\begin{align}
\gamma_{uu}^n &=\gamma_0^n\,, \\
\gamma_{gu}^n &= - \frac{8}{3} \left[ \frac{2}{n} - \frac{2}{n+1} + \frac{1}{n+2} \right]\,, \\
\gamma_{ug}^n &= - \left[ \frac 1 {n+1} - \frac 2 {n+2} + \frac 2 {n+3} \right]\,, \\
\gamma_{gg}^n &= -12 \left[ \frac 1 n - \frac 1 {n+1} + \frac 1 {n+2} - \frac 1 {n+3}  \right.  \nonumber \\
& \left. \qquad \qquad - \sum_{k=1}^{n+1} \frac 1 k \right] - \left[ 11 - \frac2 3 n_f \right] \;.
\end{align}
\end{subequations}}

Owing to the momentum conservation constraints discussed in Ref.\,\cite[Sec.\,7.3]{Cui:2020tdf}, the following identities follow immediately from Eqs.\,\eqref{eq:masslessADDGLAP}:
\begin{equation}
\gamma_{gu}^1=-\gamma_{uu}^1\,, \;
2 n_f \gamma_{ug}^1 + \gamma_{gg}^1=0\,.
\end{equation}
Furthermore, $b_1=0$ and
\begin{subequations}
\begin{align}
\gamma_{gf}^1(\zeta)+\gamma_{ff}^1(\zeta) &=\gamma_{gu}^1+\gamma_{uu}^1=0 \,, \\
2 n_f \gamma_{fg}^1(\zeta) + \gamma_{gg}^1 &= 2 n_f \gamma_{ug}^1 + \gamma_{gg}^1=0 \,.
\end{align}
\end{subequations}

\subsection{Benchmarking all-orders evolution}
\label{app:all-orders}
The fact that massless splitting functions are recovered in the limit of $n_f$ light quarks enables us to illustrate some important consequences of all-orders evolution.  For instance, in this case Eq.\,\eqref{eq:xnzeta} delivers:
\begin{equation}\label{eq:Mellinratios}
\frac{\langle x^n H_{\mathsf{P}}^u \rangle_{\zeta}^{\Delta^2}}{\langle x^n H_{\mathsf{P}}^u \rangle_{\zeta_H}^{\Delta^2}} =
\left( \frac{\langle x H_{\mathsf{P}}^u \rangle_{\zeta}^{\Delta^2}}{\langle x H_{\mathsf{P}}^u \rangle_{\zeta_H}^{\Delta^2}}
\right)^{\gamma_0^n/\gamma_0^1} ;
\end{equation}
and the solution of Eqs.\,\eqref{eq:DGLAPmom} is
\begin{subequations}
\label{eq:singletM}
\begin{align}
&\left(
\begin{array}{c}
\langle x^n H_{\mathsf{P}}^{\mathrm{S}} \rangle_{\zeta}^{\Delta^2} \\
\langle x^n H_{\mathsf{P}}^g \rangle_{\zeta}^{\Delta^2}
\end{array}
\right)
 = [W_n {\mathpzc E}_n W_n^{-1}]
\left(
\begin{array}{c}
\langle x^n H_{\mathsf{P}}^{\mathrm{S}} \rangle_{\zeta_H}^{\Delta^2} \\
\langle x^n H_{\mathsf{P}}^g \rangle_{\zeta_H}^{\Delta^2}
\end{array}
\right) , \\
& {\mathpzc E}_n  = \left(
\begin{array}{cc}
\left[ \frac{\langle x H_{\mathsf{P}}^u \rangle_{\zeta}^{\Delta^2}}{\langle x H_{\mathsf{P}}^u \rangle_{\zeta_H}^{\Delta^2}}
\right]^{\lambda_+^n/\gamma_0^1} & 0
\\
0 & \left[ \frac{\langle x H_{\mathsf{P}}^u \rangle_{\zeta}^{\Delta^2}}{\langle x H_{\mathsf{P}}^u \rangle_{\zeta_H}^{\Delta^2}}
\right]^{\lambda_-^n/\gamma_0^1}
\end{array}
\right),
\end{align}
\end{subequations}
%
where: owing to Eq.\,\eqref{eq:xnzeta}, the evolution matrix, ${\mathpzc E}_n$, is independent of $\Delta^2$; and $W_n$ is the modal matrix for the array of anomalous dimensions, \emph{viz}.\
\begin{equation}
\left(
\begin{array}{cc}
\gamma_{uu}^n & 2 n_f \gamma_{ug}^n \\
\gamma_{gu}^n & \gamma_{gg}
\end{array}
\right)
=
W_n
\left(
\begin{array}{cc}
\lambda_{+}^n & 0 \\
0 & \lambda_{-}^n
\end{array}
\right)
W_n^{-1} \;.
\end{equation}
For $n=1$: $\lambda_+^1=\tfrac{56}{9}$; $\lambda_-^1=0$, guaranteeing momentum conservation; and
\begin{equation}
W_1 = \left(
\begin{array}{cc}
1 & \tfrac{3}{4} \\
-1 & 1
\end{array}
\right)\,.
\end{equation}

By definition, Eqs.\,\eqref{eq:Mellinratios}, \eqref{eq:singletM} are all-orders exact.  Moreover, they do not explicitly depend on the particular choice of effective charge, $\hat\alpha$.  Our favoured form is the process-independent charge discussed in Ref.\,\cite{Cui:2019dwv}; notwithstanding that, Eqs.\,\eqref{eq:Mellinratios}, \eqref{eq:singletM} mean that once the valence-quark DF is known at $\zeta_H$ and its first moment is known at any $\zeta>\zeta_H$, then the complete valence, glue and sea DFs can readily be obtained at the new scale without explicit reference to the form of $\hat\alpha$.  (A choice for the effective charge is implicit in the $x$-dependence of the valence-quark DF at $\zeta_H$, but it need not be made specific.)

These remarks are readily elucidated by considering $\mathsf P = \pi$ and $\Delta^2=0$ in Eqs.\,\eqref{eq:Mellinratios}, \eqref{eq:singletM}; then:
%
{\allowdisplaybreaks
\begin{subequations}
\label{eq:Mellinpi}
\begin{align}
\langle x^n \rangle_u^\zeta & = \langle x^n \rangle_u^{\zeta_H} \left( \langle 2 x \rangle_u^\zeta\right)^{9\gamma_0^n/32} ,\\
\left(
\begin{array}{c}
\langle x^n \rangle_ {\mathrm{S}}^{\zeta} \\
\langle x^n \rangle_ g^{\zeta}
\end{array}
\right)
& =W_n\left(
\begin{array}{cc}
[\langle 2 x \rangle_u^{\zeta}]^{\lambda_+^n/\gamma_0^1} & 0
\\
0 & [\langle 2 x \rangle_u^{\zeta}]^{\lambda_-^n/\gamma_0^1}
\end{array}
\right) \nonumber \\
&
\quad
\times W_n^{-1} \left(
\begin{array}{c}
\langle 2 x^n \rangle_u^{\zeta_H} \\
0
\end{array}
\right) \;.
\label{eq:singletM2}
\end{align}
\end{subequations}}

\begin{figure}[t!]
\vspace*{3.5ex}

\leftline{\hspace*{0.5em}{\large{\textsf{A}}}}
\vspace*{-5ex}
\centerline{\includegraphics[clip, width=0.4\textwidth]{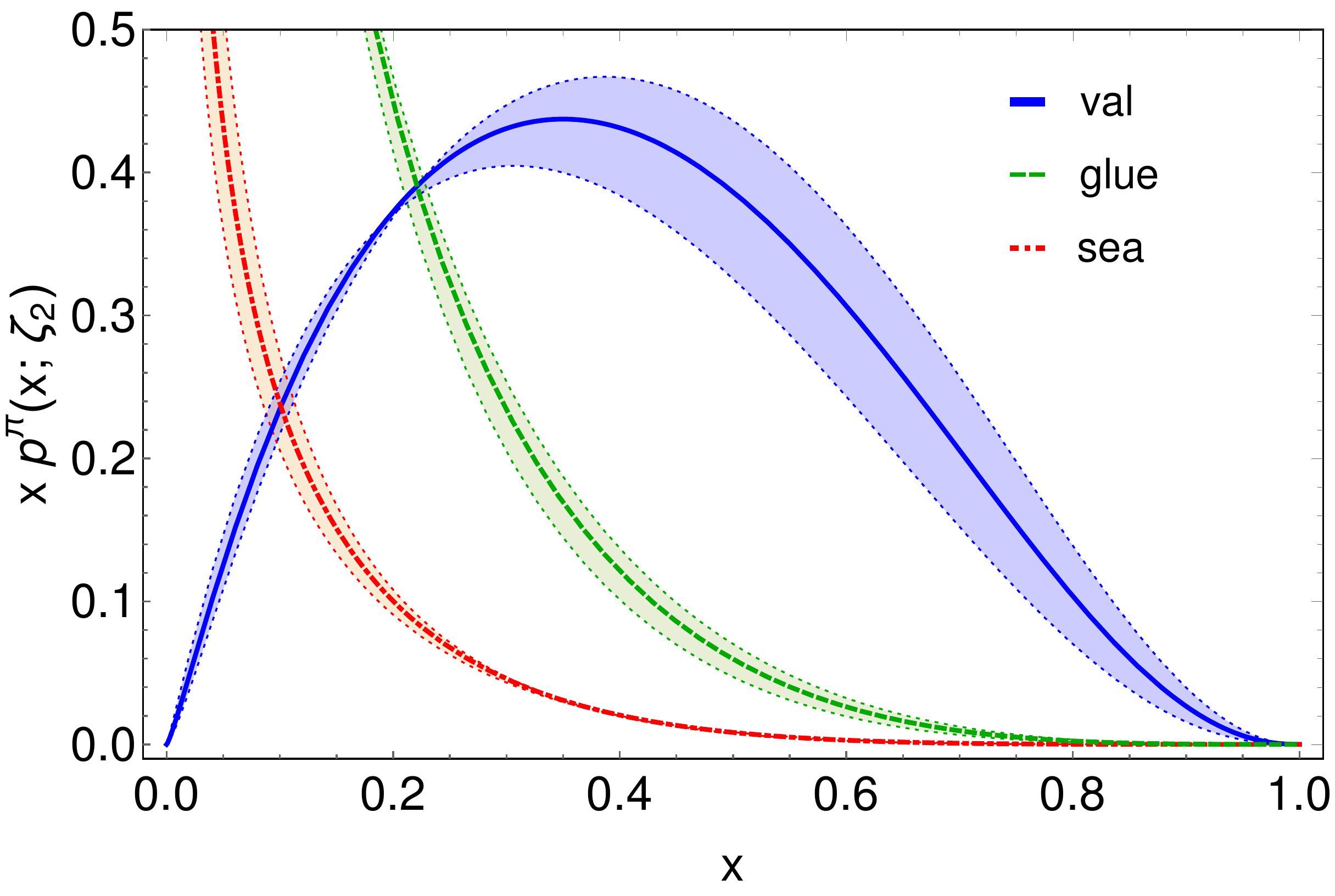}}
\vspace*{6ex}

\leftline{\hspace*{0.5em}{\large{\textsf{B}}}}
\vspace*{-5ex}
\centerline{\includegraphics[clip, width=0.40\textwidth]{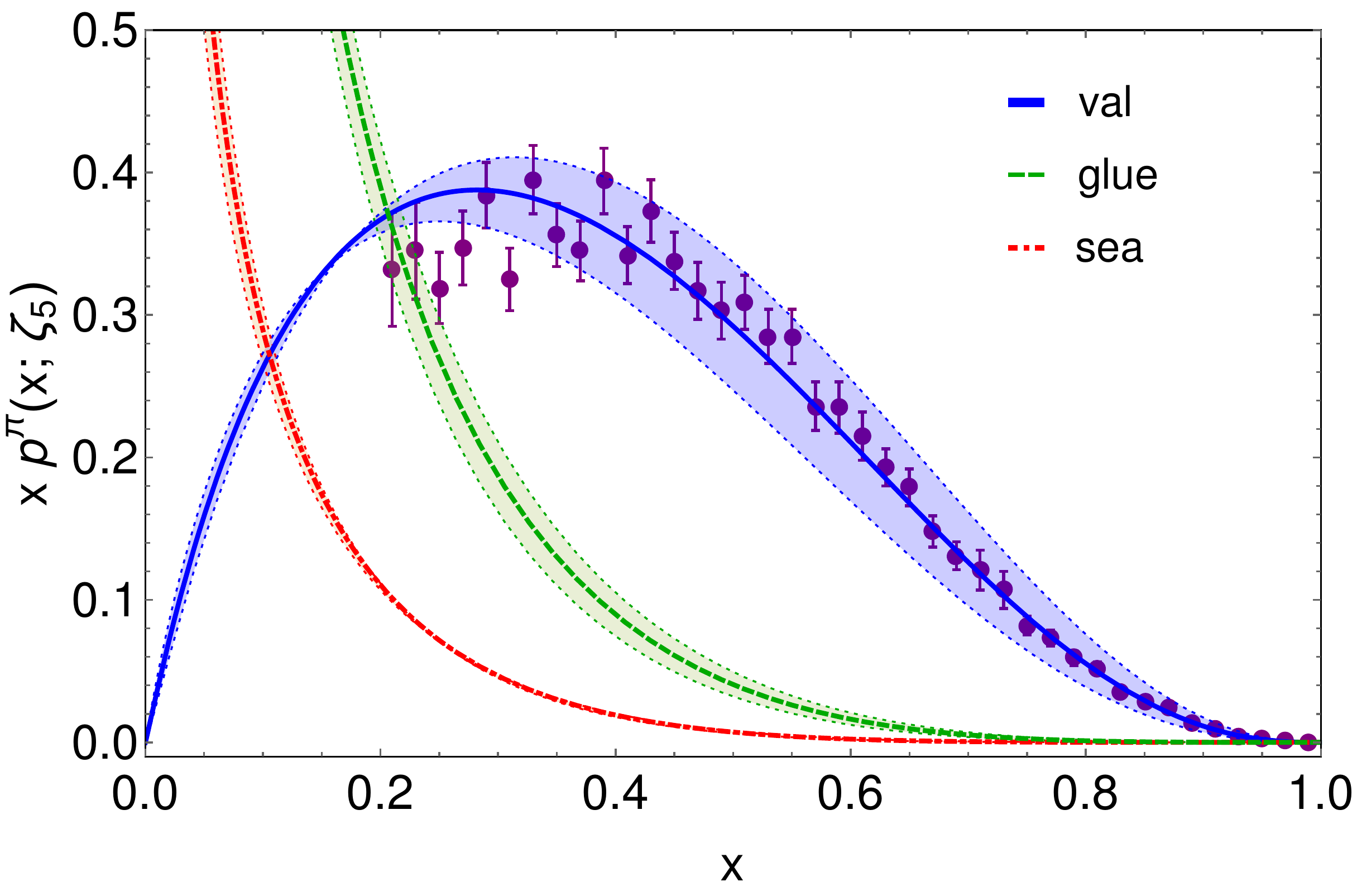}}
%
\caption{\label{FigAppB} 
Pion parton DFs obtained using Eqs.\,\eqref{eq:Mellinpi} to evolve the $\zeta=\zeta_{\cal H}$ valence-quark DF specified by Eq.\,\eqref{eq:param} and the coefficients and powers listed in Table~\ref{tab:DFs}.
\emph{Upper panel}\,--\,{\sf A}: $\zeta_{\cal H}\to \zeta_2$, seeded by $\langle 2 x \rangle_u^{\zeta_2} = 0.50(5)$, producing $\langle x \rangle_g^{\zeta_2} = 0.40(3)$, $\langle x \rangle_S^{\zeta_2} = 0.10(2)$.
 \emph{Lower panel}\,--\,{\sf B}: $\zeta_{\cal H}\to \zeta_5$, seeded by $\langle 2 x \rangle_u^{\zeta_5} = 0.42(4)$, producing $\langle x \rangle_g^{\zeta_2} = 0.45(2)$, $\langle x \rangle_S^{\zeta_2} = 0.13(2)$.
No explicit choice for the effective charge was necessary to obtain these DFs.
Data in Panel\,{\sf B}: Ref.\,\cite[E615]{Conway:1989fs}, rescaled according to the analysis in Ref.\,\cite{Aicher:2010cb}.
}
\end{figure}

Suppose the pion hadron-scale valence-quark DF is that determined by Eq.\,\eqref{eq:param}, with the coefficients and powers listed in Table~\ref{tab:DFs}.  Further, suppose that $\langle 2 x \rangle_u^{\zeta_2} = 0.50(5)$ and $\langle 2 x \rangle_u^{\zeta_5} = 0.42(4)$, $\zeta_5:=5.2\,$GeV, as inferred in a leading-logarithm, next-to-leading order (NLO) perturbative QCD fit to extant $\pi$-related Drell-Yan and prompt photon data \cite{Novikov:2020snp}.  Then, Eqs.\,\eqref{eq:Mellinpi} produce the pion DFs drawn in Fig.\,\ref{FigAppB}.

The results in Fig.\,\ref{FigAppB} are interesting for many reasons, some of which we list here.
\begin{enumerate}[label=(\emph{\roman*})]
\item The evolution outcomes are independent of the effective charge, supposing there is at least one for which all-orders evolution is a viable scheme and that the glue and sea DFs vanish at $\zeta_H$.  All dependence on the effective charge is encoded in the $\zeta_{\cal H}$ valence-quark DF.
\item The pointwise behaviours of all DFs in Fig.\,\ref{FigAppB}, valence, glue and sea, are completely determined by that of Eq.\,\eqref{eq:param}, with the coefficients and powers listed in Table~\ref{tab:DFs}, which is the valence-quark $\zeta=\zeta_{\cal H}$ DF predicted in Refs.\,\cite{Cui:2019dwv, Cui:2020dlm} solely from knowledge of the pion's leading-twist DA.
\item The $\chi^2/{\rm datum}=0.94$ agreement, evident in Fig.\,\ref{FigAppB}B, between the $\zeta=\zeta_5$ valence-quark DF predicted by the all-orders scheme and the result inferred from the data in Ref.\,\cite[E615]{Conway:1989fs} using next-to-leading-logarithm (NLL) resummation at NLO in perturbative-QCD \cite{Aicher:2010cb} is a parameter-free outcome.  It derives from a modern prediction for the pion's leading-twist DA \cite{Cui:2019dwv, Cui:2020dlm} and an estimate of $\langle 2 x \rangle_u^{\zeta_5}$ \cite{Novikov:2020snp}, both obtained without reference to the NLL reassessment of the E615 data.
\end{enumerate}

\subsection{GPD evolution}
\begin{figure}[t]
\vspace*{3.5ex}

\leftline{\hspace*{0.5em}{\large{\textsf{A}}}}
\vspace*{-5ex}
\centerline{\includegraphics[clip, width=0.4\textwidth]{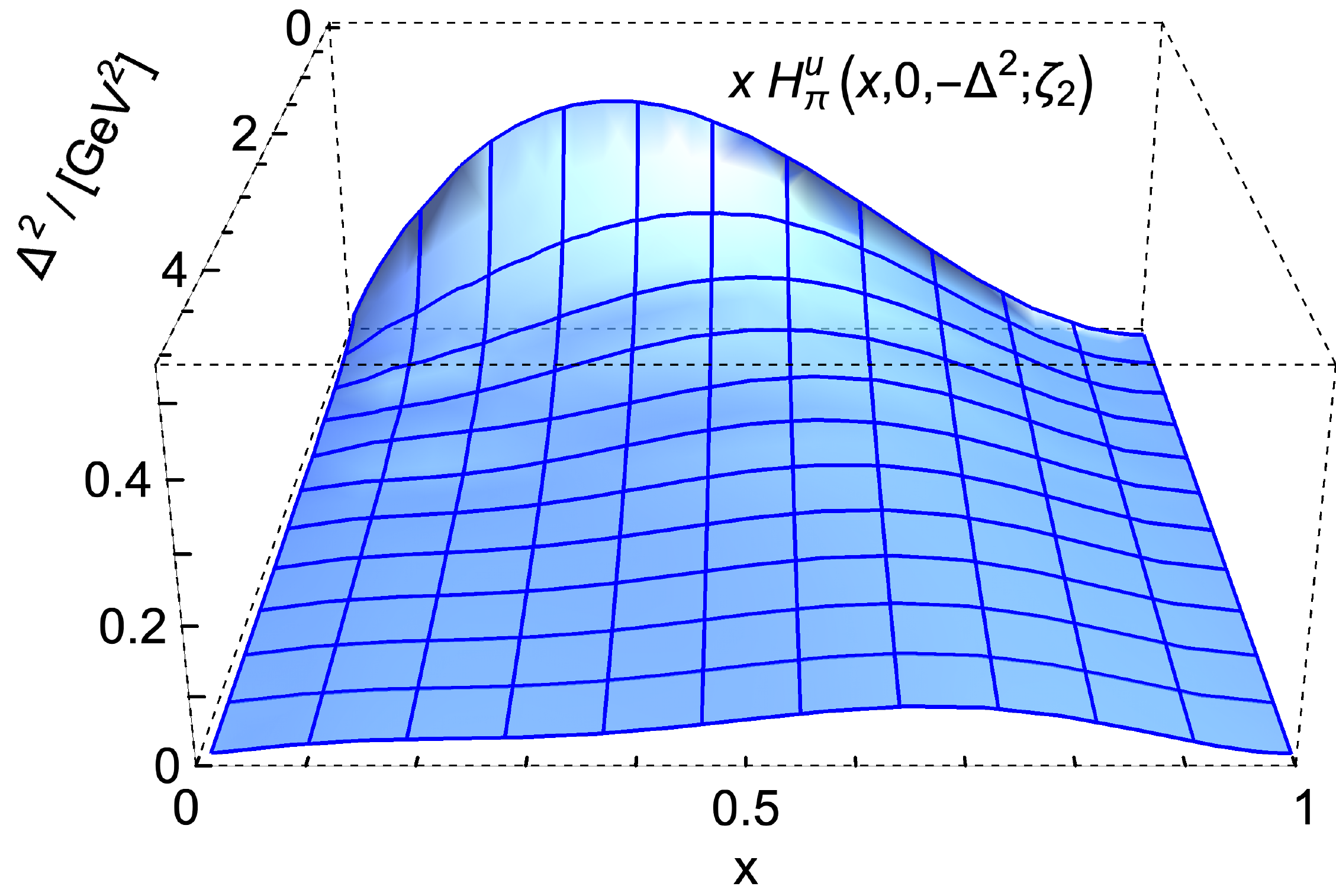}\hspace*{2.2ex}}
\vspace*{6ex}

\leftline{\hspace*{0.5em}{\large{\textsf{B}}}}
\vspace*{-5ex}
\centerline{\includegraphics[clip, width=0.4\textwidth]{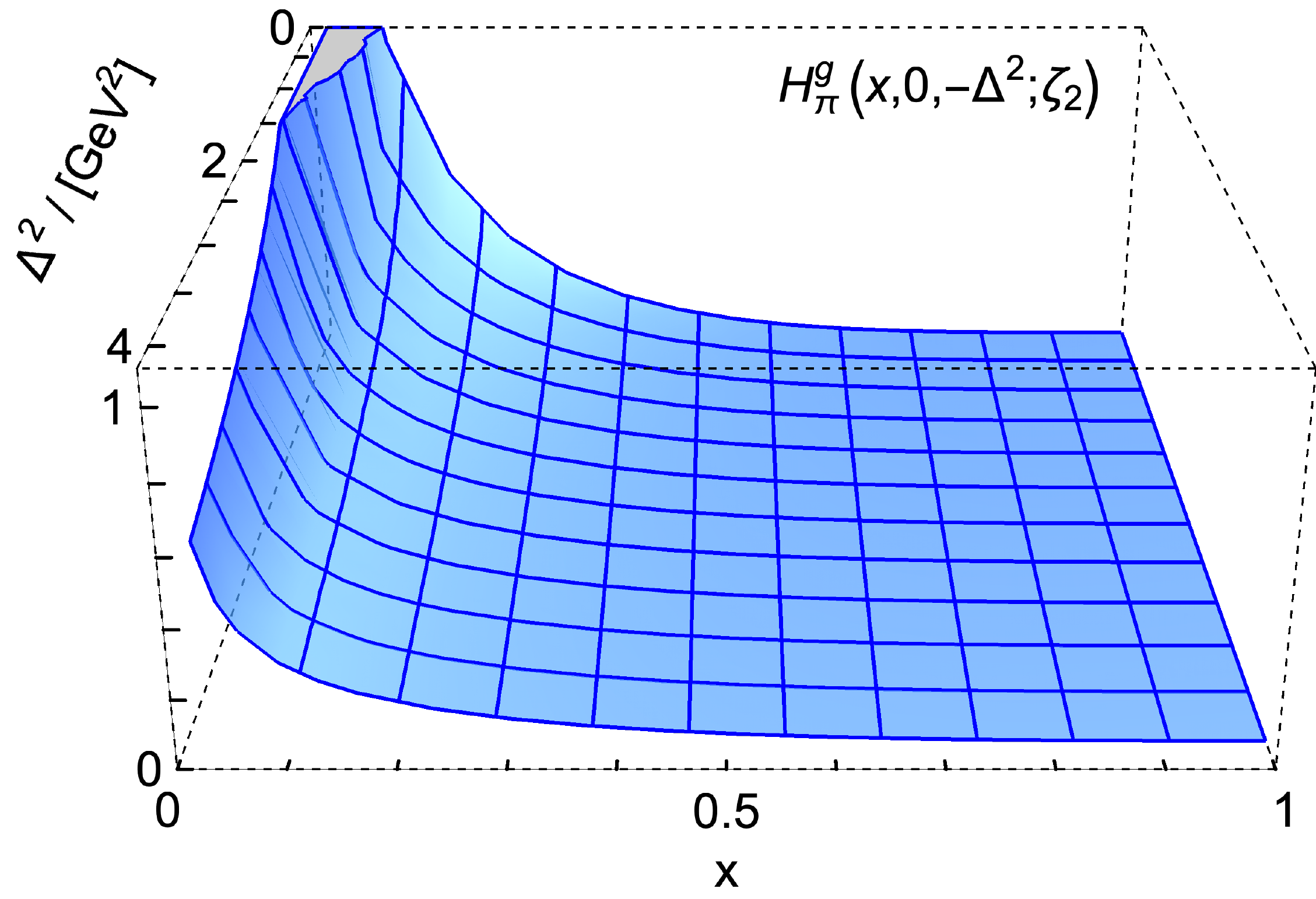}}
\vspace*{6ex}

\leftline{\hspace*{0.5em}{\large{\textsf{C}}}}
\vspace*{-5ex}
\centerline{\includegraphics[clip, width=0.4\textwidth]{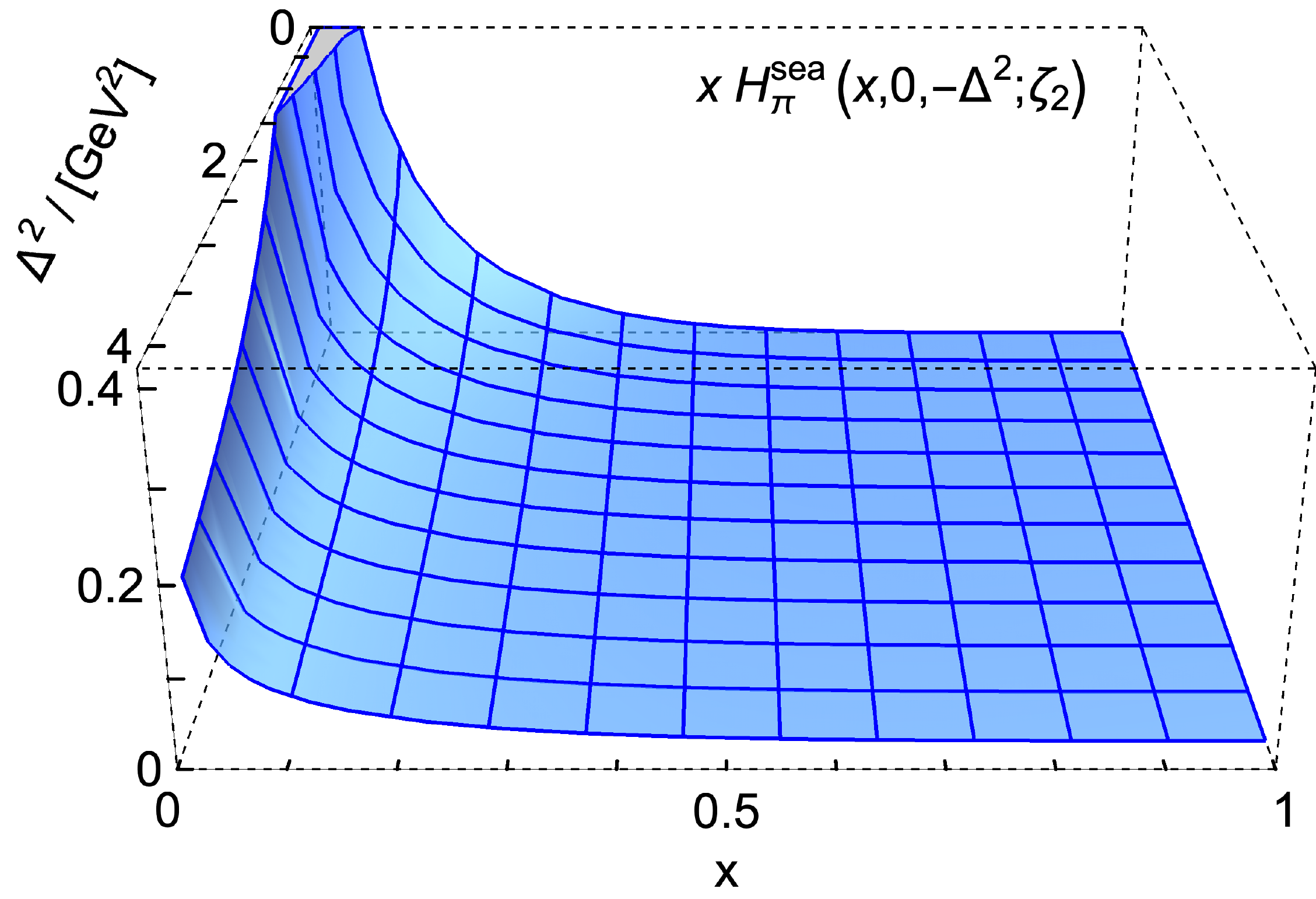}}
%
\caption{\label{fig:evolvedGPDpi} 
Pion GPDs ($\xi=0$) computed in Sec.\,\ref{SecFac}, beginning with the factorised LFWF and drawn in Fig.\,\ref{fig:piGPDFac}, evolved to $\zeta_{\cal H}\to\zeta_2$:
\emph{Upper panel}\,--\,{\sf A}: valence;
\emph{middle}\,--\,{\sf B}: glue;
and \emph{lower}\,--\,{\sf C}: sea.
}
\end{figure}

With such results obtained using the all-orders scheme, we consider it worthwhile to report evolution of all $\xi=0$ GPDs as obtained via the numerical solution of Eqs.\,\eqref{eq:diffDGLAP}\,--\,\eqref{eq:masslessADDGLAP}, running $\zeta_H$ up to a desired $\zeta$-value, and reconstructing the distributions from the evolved Mellin moments.  We use the procedure in
Ref.\,\cite{Cui:2020tdf}, determining the $x$-dependence on each $\Delta^2$ slice and therefrom building the two-dimensional evolved GPD.
In this way, beginning with the $(\xi=0,\zeta=\zeta_{\cal H})$ GPDs in Sec.\,\ref{SecFac}, one obtains the $\zeta=\zeta_2$ GPDs drawn in Fig.\,\ref{fig:evolvedGPDpi}.
Comparing the valence-quark profiles, Figs.\,\ref{fig:piGPDFac}, \ref{fig:GPDs}A with Fig.\,\ref{fig:evolvedGPDpi}A, it is apparent that:
the peak location at each $\Delta^2$-value is shifted toward $x=0$; and the evolved GPD is broader and flatter than the $\zeta_H$ profile.
The strength lost from the valence-quark GPD is transferred into the glue and sea distributions, drawn in Fig.\,\ref{fig:evolvedGPDpi}B, \ref{fig:evolvedGPDpi}C: recall, both these GPDs are zero at $\zeta_{\cal H}$.
The bulk of the support of both glue and sea GPDs is concentrated in the neighbourhood $(x\simeq 0,\Delta^2\simeq 0)$; and their fall-off with increasing $\Delta^2$ is significantly slower than with increasing $x$.

\begin{figure}[t]
\vspace*{3.5ex}

\leftline{\hspace*{0.5em}{\large{\textsf{A}}}}
\vspace*{-5ex}
\centerline{\includegraphics[clip, width=0.4\textwidth]{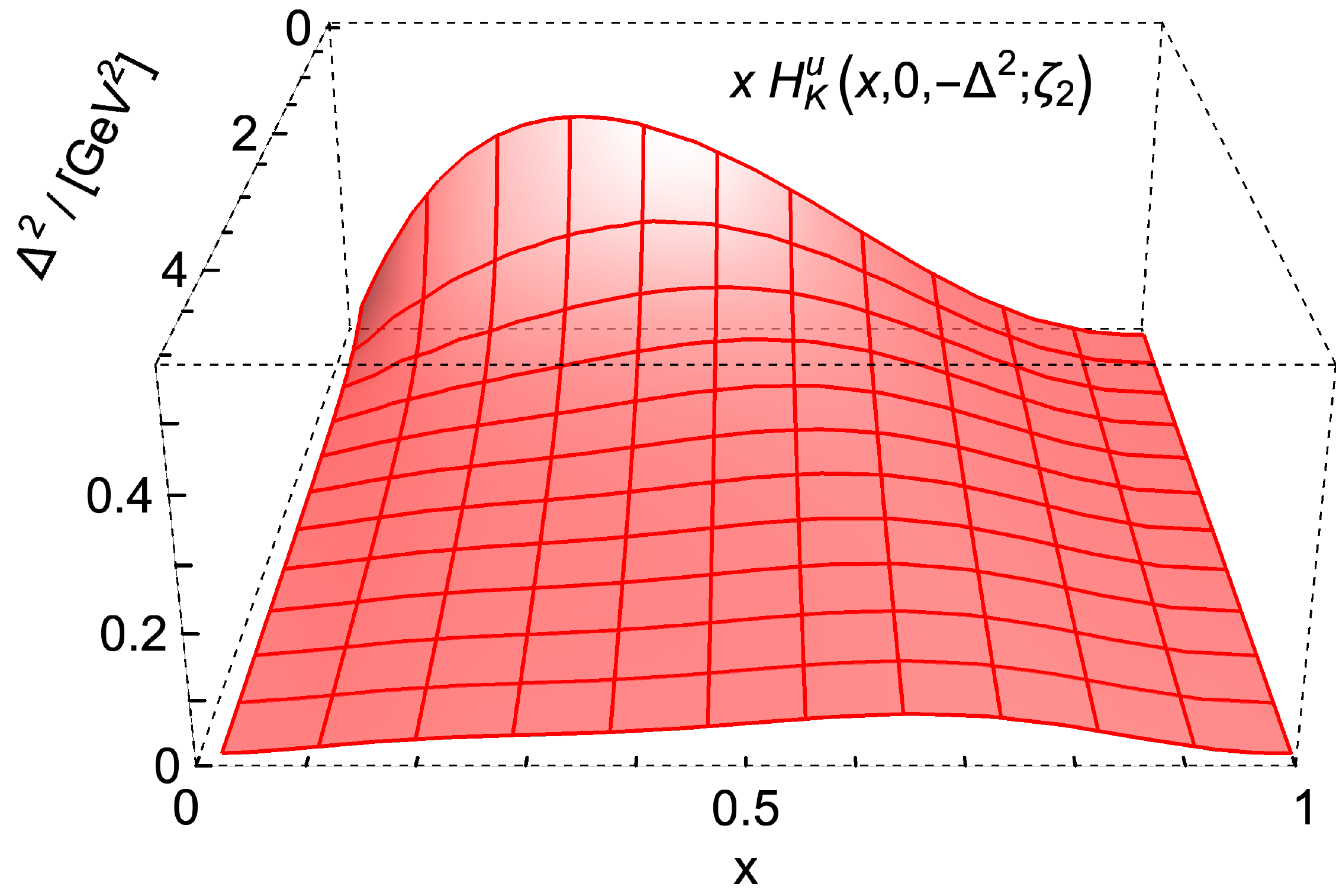}\hspace*{2.2ex}}
\vspace*{6ex}

\leftline{\hspace*{0.5em}{\large{\textsf{B}}}}
\vspace*{-5ex}
\centerline{\includegraphics[clip, width=0.4\textwidth]{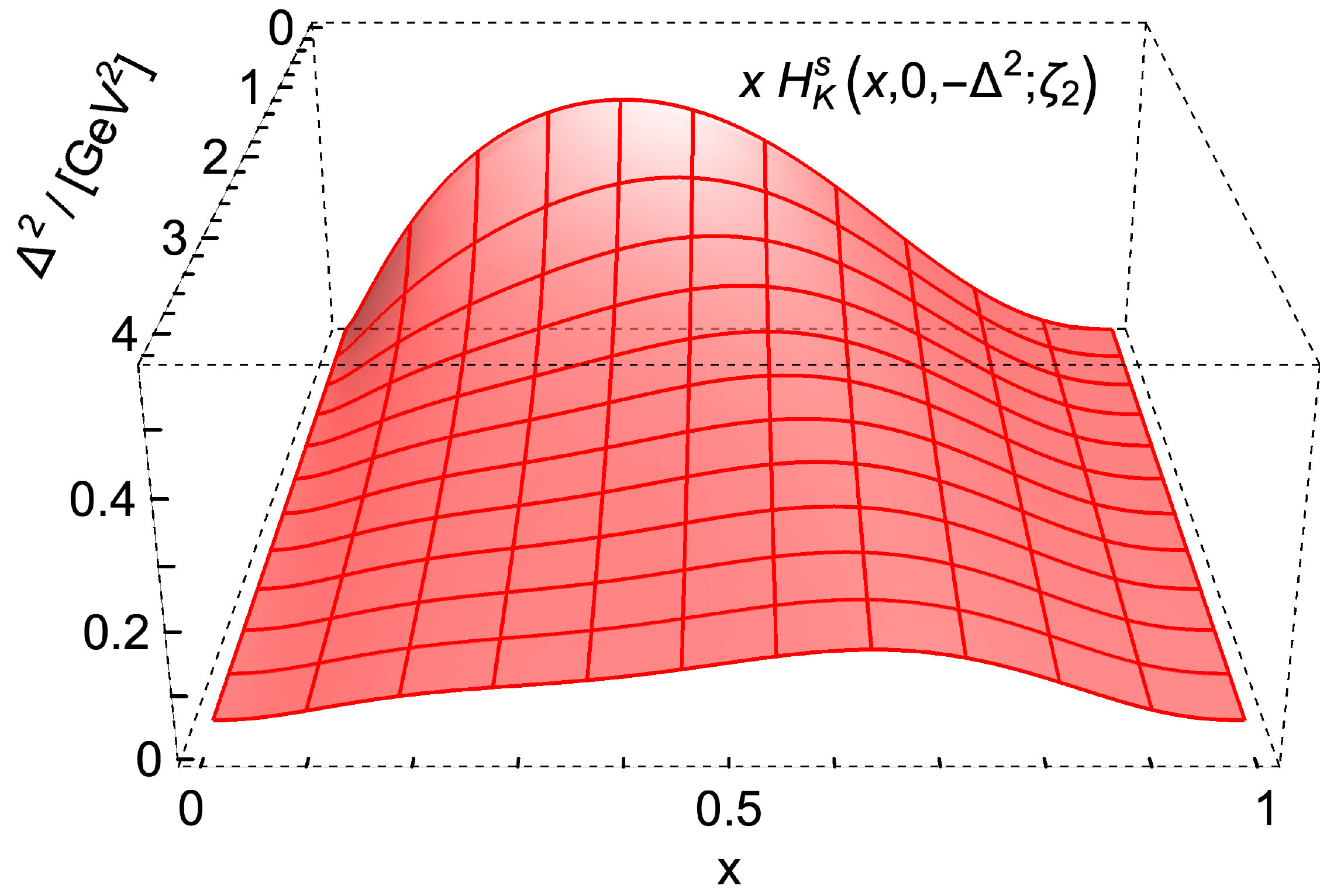}}
\vspace*{6ex}

%
\caption{\label{fig:evolvedGPDK} 
Kaon GPDs ($\xi=0$) computed in Sec.\,\ref{SecFac}, beginning with the factorised LFWF, evolved to $\zeta_{\cal H}\to\zeta_2$:
\emph{Upper panel}\,--\,{\sf A}: $u$-valence;
and \emph{lower}\,--\,{\sf B}: $\bar s$-valence.
}
\end{figure}

The evolved kaon GPDs are a qualitatively and semi-quantitatively similar; so we only draw the valence-quark profiles in Fig.\,\ref{fig:evolvedGPDK}.  The separation of baryon number is preserved under evolution to all empirically accessible scales.

\section{Partition of Meson Mass-Squared and Associated Radii}
\label{SecPartition}
\subsection{Mass-squared}
Return now to Eq.\,\eqref{Mom1GPD} and consider $\xi=0$ so as to isolate $\theta_2^{\mathsf P}(\Delta^2)$.  Owing to its connection with the expectation value of the energy-momentum tensor, $T_{\mu\nu}$, evaluated in the ${\mathsf P}$-meson \cite{Polyakov:2002yz, Polyakov:2018zvc}, the forward-limit of this expectation value produces the meson mass-squared:
\begin{equation}
\label{EPTpion}
\langle \mathsf{P}(P_{\mathsf P}) | T_{\mu\mu} | \mathsf{P}(P_{\mathsf P}) \rangle = m_{\mathsf P}^2\, \theta_2^{\mathsf P}(\Delta^2=0)  = m_{\mathsf P}^2\,,
\end{equation}
where the last equality expresses the mass-squared sum-rule: $\theta_2^{\mathsf P}(\Delta^2=0)=1$.  Connections between the expectation value on the left-hand-side in Eq.\,\eqref{EPTpion} and the QCD scale anomaly are discussed, \emph{e.g}., in Ref.\,\cite{Roberts:2016vyn}, and broader connections with EHM are canvassed in Refs.\,\cite{Aguilar:2019teb, Roberts:2020udq, Chen:2020ijn, Roberts:2020hiw, Krein:2020yor, Roberts:2021nhw, Arrington:2021biu}.

Important in proceeding here is the fact, explained above, that dressed-valence degrees-of-freedom carry all meson properties at the hadron scale, $\zeta_H$; hence,
{\allowdisplaybreaks
\begin{align}
\theta_2^{\mathsf P}&(\Delta^2) = \int_{-1}^1 \! dx\, x\,  \nonumber \\
& \times
\left[ H_{\mathsf P}^{\mathpzc u}(x,0,-\Delta^2;\zeta_{\cal H}) +
H_{\mathsf P}^{\bar {\mathpzc h}}(x,0,-\Delta^2;\zeta_{\cal H}) \right]\,. \label{AllValence}
\end{align}
Focusing on $\Delta^2=0$:
\begin{equation}
\theta_2^{\mathsf P}(0)  = \theta_2^{{\mathsf P}_u}(0) + \theta_2^{{\mathsf P}_{\bar h}}(0)
= \langle x \rangle_{\zeta_{\cal H}}^{{\mathsf{P}_u}}
 + \langle x \rangle_{\zeta_{\cal H}}^{{\mathsf{P}_{\bar h}}}\,.
\end{equation}
One may therefore say that all $m_{\mathsf P}^2$ is carried by dressed-valence degrees-of-freedom at $\zeta_H$ and the relative apportionment is given by their respective light-front momentum fractions.

Now consider shifting the resolving scale to a value $\zeta>\zeta_H$ and, for simplicity, focus on the pion:
\begin{subequations}
\begin{align}
\langle 2 x H_\pi^{u} \rangle_{\zeta_{\cal H}}^0 & \to \langle x H_\pi^{S} \rangle_{\zeta}^0 \\
& = \langle 2 x H_\pi^{{\rm valence}} \rangle_{\zeta}^0 +  \langle 2 x H_\pi^{{\rm sea}} \rangle_\zeta^0 \\
& = 2 \theta_2^{{\pi}_{\rm val}}(0;\zeta) + \theta_2^{{\pi}_{\rm sea}}(0;\zeta)\,,
\end{align}
\end{subequations}
where a direct identification of terms is intended; and
\begin{equation}
\langle H_\pi^{g} \rangle_{\zeta_{\cal H}}^0 \equiv 0 \to \langle H_\pi^{g} \rangle_{\zeta}^0 =
\theta_2^{{\pi}_{\rm g}}(0;\zeta) \neq 0\,.
\end{equation}}

Using the evolution equations in Sec.\,\ref{SecAll}, one finds ($n_f$=4 massless flavours):
{\allowdisplaybreaks
\begin{subequations}
\label{radii3}
\begin{align}
& 2 \theta_2^{{\pi}_{\rm val}}(\Delta^2;\zeta) + \theta_2^{{\pi}_{\rm sea}}(\Delta^2;\zeta) \nonumber \\
& =
2 \theta_2^{{\pi}_{\rm val}}(\Delta^2;\zeta_{\cal H}) \left[
\tfrac{3}{7}
+
\tfrac{4}{7} (\langle 2 x \rangle_u^{\zeta})^{\tfrac{7}{4}} \right],\\
%
& \theta_2^{{\pi}_{\rm g}}(\Delta^2;\zeta) \nonumber \\
& = \tfrac{4}{7} 2 \theta_2^{{\pi}_{\rm val}}(\Delta^2;\zeta_{\cal H})
 \left[1 - (\langle 2 x \rangle_u^{\zeta})^{\tfrac{7}{4}}
\right]\,.
\end{align}
\end{subequations}
It is now plain that at any scale $\zeta \geq \zeta_{\cal H}$:
\begin{align}
&2 \theta_2^{{\pi}_{\rm val}}(\Delta^2;\zeta) + \theta_2^{{\pi}_{\rm sea}}(\Delta^2;\zeta)
+ \theta_2^{{\pi}_{\rm g}}(\Delta^2;\zeta)  \nonumber \\
& =  2 \theta_2^{{\pi}_{\rm val}}(\Delta^2;\zeta_{\cal H}) = \theta_2^{\pi}(\Delta^2) \,.
\label{MassSquaredPartition}
\end{align}
}

These results can readily be generalised to the kaon and mass-dependent splitting functions.

Setting $\Delta^2=0$, there is an obvious corollary; to wit, at any scale $\zeta>\zeta_{\cal H}$ the portion of $m_{\mathsf P}^2$ carried by a given parton species is given by its light-front momentum fraction.  Thus, at $\zeta = \zeta_2$, working with the DFs in Ref.\,\cite{Cui:2020tdf}, which are reproduced by the LFWF \emph{Ans\"atze} herein:
\begin{equation}
\label{Mass2Fraction}
\begin{array}{l|cccc}
 \zeta=\zeta_2   & \multicolumn{4}{c}{\mbox{mass-squared fraction}}\\
 & u & {\bar h} & g  & {\rm sea} \\\hline
m_\pi^2 & 0.24(2) & 0.24(2) & 0.41(2) & 0.11(2) \\
m_K^2 & 0.23(2) & 0.27(2) & 0.40(2) & 0.10(2)
\end{array}\,.
\end{equation}
As the scale $\zeta$ increases beyond $\zeta_2$, the mass-squared fraction stored with the valence-quarks runs logarithmically to zero, as may be read from Eq.\,\eqref{eq:xnzeta}; and consequently, using Eqs.\,\eqref{radii3}, the glue and sea fractions, respectively, approach $4/7$ and $3/7$ at the same rate.  (\emph{N.B}.\ The values of these limiting glue and sea fractions depend on the splitting functions.)

\begin{figure}[t!]
\vspace*{3.5ex}

\leftline{\hspace*{0.5em}{\large{\textsf{A}}}}
\vspace*{-5ex}
\centerline{\includegraphics[clip, width=0.42\textwidth]{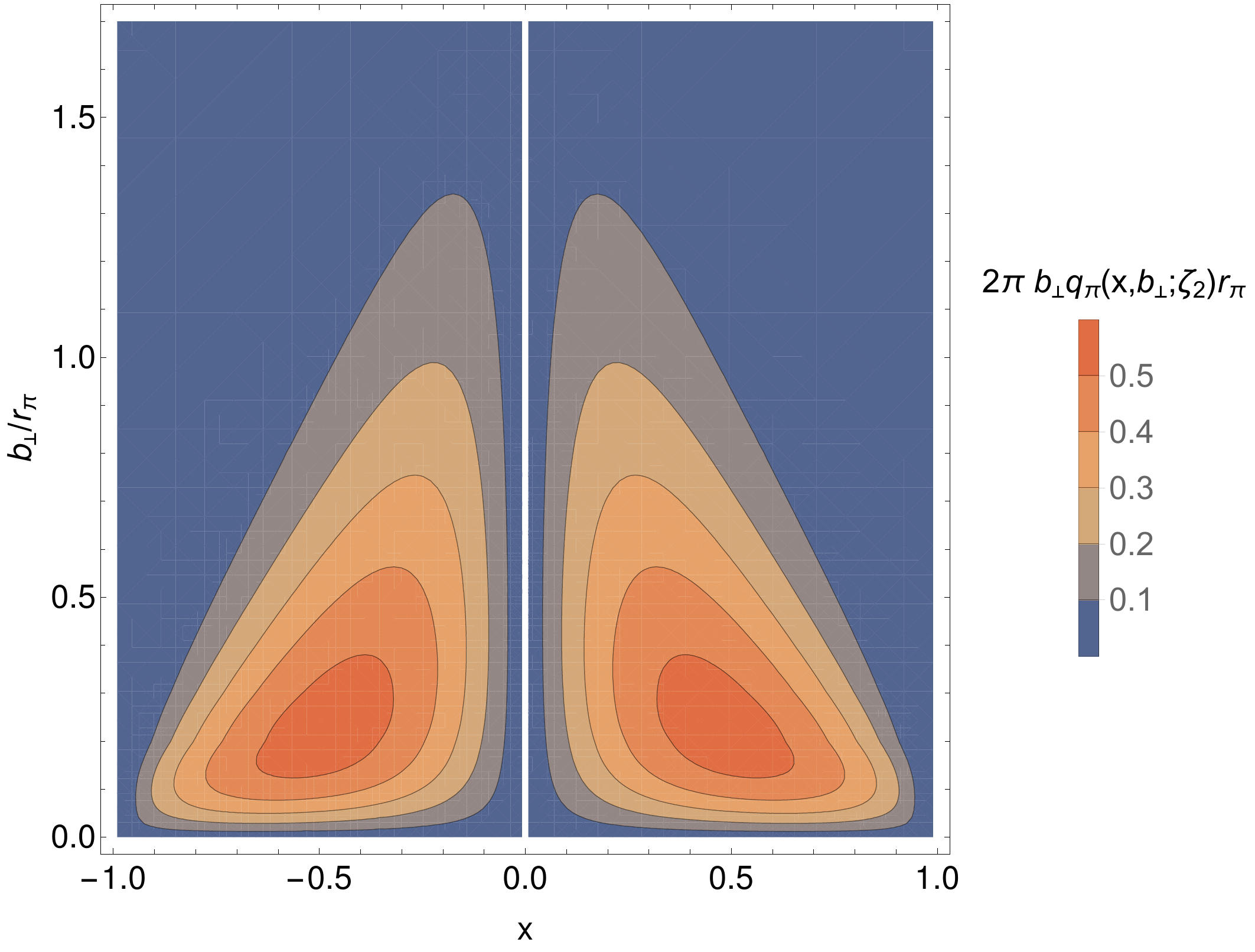}}
\vspace*{6ex}

\leftline{\hspace*{0.5em}{\large{\textsf{B}}}}
\vspace*{-5ex}
\centerline{\includegraphics[clip, width=0.42\textwidth]{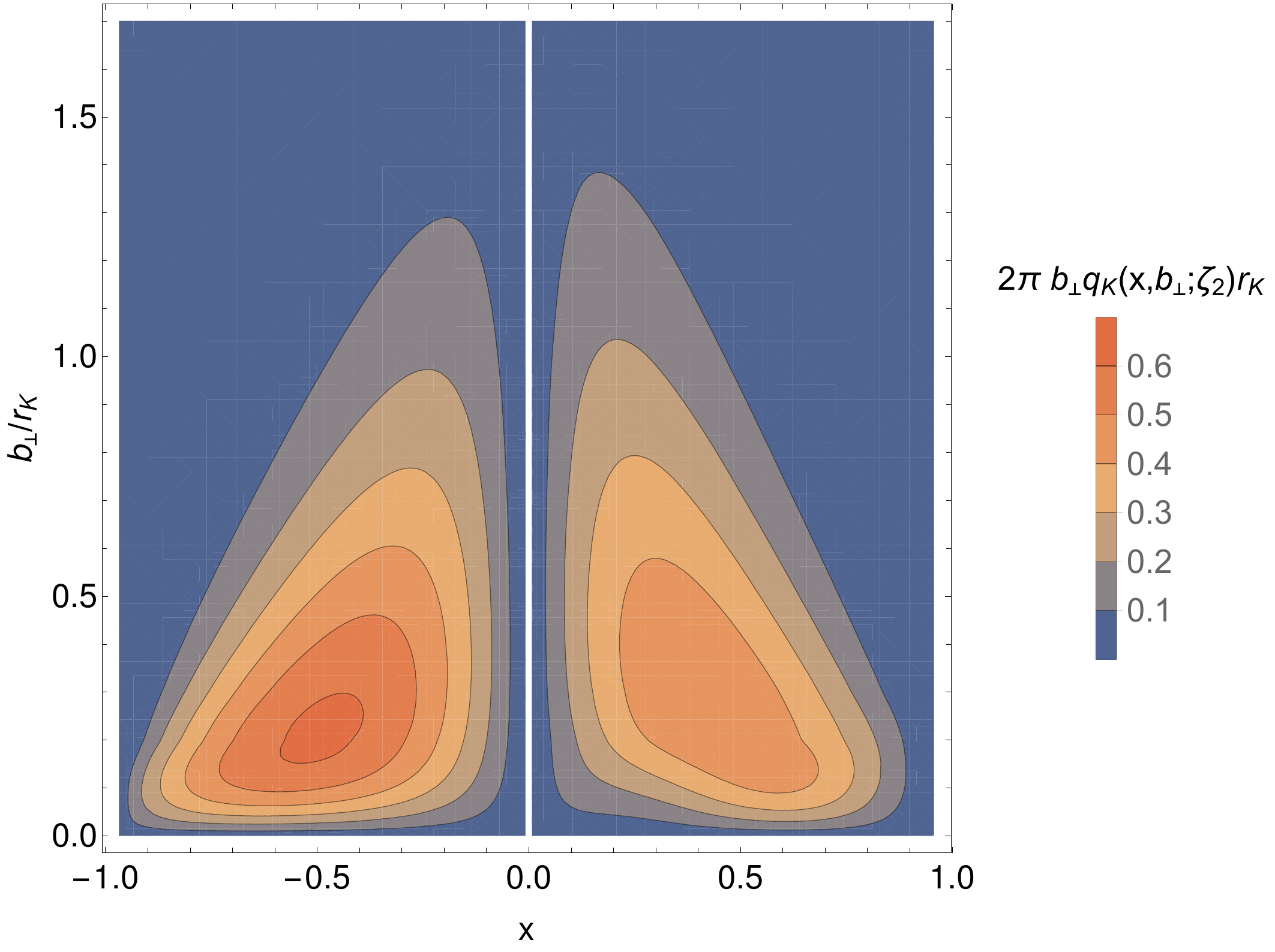}}
%
\caption{\label{fig:IPDGPDszeta2} 
\emph{Upper panel}\,--\,{\sf A}.
Pion. $u$-quark IPS distribution slices, plotted as a function of $|b_\perp|$ at four $x$ values, as computed with: PTIR LFWF (solid curves), Sec.\,\ref{SecPTIR}; and factorised LFWF (short-dashed curves), Sec.\,\ref{SecFac}.  The peak height decreases with increasing $x$.
\emph{Lower panel}\,--\,{\sf B}.  Kaon.  Analogous profiles for $u$-quark in the $K^+$.
}
\end{figure}

\subsection{Mass-squared radii}
\label{SecRadii}
Eq.\,\eqref{MassSquaredPartition} also imposes a sum rule on the contributions from the three parton classes to the total mass-squared radius.  To see this, consider the $\pi$ and recall that the following radius is observable; hence, $\zeta$-independent:
\begin{align}
[r_\pi^{\theta_2}]^2 := \left. \frac{-6}{\theta_2^\pi(0)}\frac{d \theta_2^\pi(\Delta^2)}{d\Delta^2} \right|_{\Delta^2=0}.
\end{align}

Considering the hadron scale, Eq.\,\eqref{AllValence} highlights that the dressed-valence degrees-of-freedom carry the entirety of the mass-squared distribution; so,
\begin{align}
[r_\pi^{\theta_2}]^2 = \left. \frac{-6}{\theta_2^{\pi_{\rm val}}(0;\zeta_{\cal H})}\frac{d \theta_2^{\pi_{\rm val}}(\Delta^2;\zeta_{\cal H})}{d\Delta^2} \right|_{\Delta^2=0}.
\end{align}

At any scale $\zeta>\zeta_{\cal H}$, using Eq.\,\eqref{MassSquaredPartition}:
\begin{subequations}
\label{radii2}
\begin{align}
[r_\pi^{\theta_2}]^2  & =
2 \theta_2^{\pi_{\rm val}}(0;\zeta) [r^{\theta_2}_{\pi_{\rm val}}(\zeta)]^2 \nonumber\\
&
 + \theta_2^{\pi_{\rm sea}}(0;\zeta) [r^{\theta_2}_{\pi_{\rm sea}}(\zeta)]^2
 +\theta_2^{\pi_g}(0;\zeta)  [r^{\theta_2}_{\pi_g}(\zeta)]^2 ,\\
 [r^{\theta_2}_{\pi_p}&(\zeta)]^2  = \left. \frac{-6}{\theta_2^{\pi_p}(0;\zeta)}
 \frac{d\theta_2^{\pi_p}(\Delta^2;\zeta)}{d\Delta^2}\right|_{\Delta^2=0}\,,
\end{align}
\end{subequations}
where $p = {\rm val}, {\rm sea}, g$.
The factors $\theta_2^{\pi_p}(0;\zeta)$ in Eq.\,\eqref{radii2} are determined by Eqs.\,\eqref{eq:Mellinpi} and their $\zeta=\zeta_2$ values are given in Eq.\,\eqref{Mass2Fraction}.

Reviewing Eq.\,\eqref{eq:xnzeta}, setting $n=1$, a little thought reveals that
\begin{equation}
[r^{\theta_2}_{\pi_{\rm val}}(\zeta)]^2 = [r^{\theta_2}_{\pi_{\rm val}}(\zeta_{\cal H})]^2 = [r_\pi^{\theta_2}]^2;
\end{equation}
and similarly, from Eqs.\,\eqref{radii3}, $[r^{\theta_2}_{\pi_{\rm sea}}(\zeta)]^2 =[r^{\theta_2}_{\pi_g}(\zeta)]^2 =[r_\pi^{\theta_2}]^2$.
It is now apparent that when all hadron properties are carried by the dressed-valence degrees-of-freedom at $\zeta_{\cal H}$, with all-orders evolution generating glue and sea DFs at $\zeta>\zeta_{\cal H}$, then all in-pion mass-squared distributions have the same radii.

Considering the $K$, the following quantities are $\zeta$-independent: $r_{K_u}^{\theta_2}$, $r_{K_{\bar s}}^{\theta_2}$, $r_K^{\theta_2}$, with the last quantity computed from the first two according to Eq.\,\eqref{RadiusDefinition}.  With mass-dependent evolution as described above, the kaon's valence-quark/antiquark mass-squared distributions continue to be characterised by the flavour-specific radii indicated here and the glue and sea distributions are characterised by the $r_K^{\theta_2}$.  The relative contributions to the total observable mass-squared radius are specified by the light-front momentum fractions carried by the particular species.

Exploiting these observations, one can decompose the $\pi$ and $K$ mass-squared radii thus:
\begin{equation}
\label{radiiFraction}
\begin{array}{l|cccc}
 \zeta=\zeta_2   & \multicolumn{4}{c}{\mbox{mass-squared radii partition}}\\
 & u & {\bar h} & g  & {\rm sea} \\\hline
r_\pi^{\theta_2}/r_\pi & 0.49(2) & 0.49(2) & 0.64(2) & 0.33(3) \\
r_K^{\theta_2}/r_K  & 0.47(2) & 0.52(2) & 0.63(2) & 0.32(3)
\end{array}\,.
\end{equation}
With $r_\pi^{\theta_2}/r_\pi=0.81$, Eq.\,\eqref{kaontheta2radii}, $r_\pi = 0.640(7)\,$fm \cite{Cui:2021aee}, then $\zeta=\zeta_2$ valence, glue and sea fractions are, respectively, in fm: $2\times 0.25(1)$, $0.33(1)$, $0.17(1)$.
Turning to the $K$, $r_K^{\theta_2}/r_K=0.78$, Eq.\,\eqref{kaontheta2radii}, $r_K \approx 0.53\,$fm \cite{Cui:2021aee}, then the analogous fractions are roughly, in fm: $0.21$, $0.20$, $0.26$, $0.13$.  (The quality of extant charged-kaon elastic form factor data prevent any listing of sensible uncertainties \cite{Cui:2021aee}.)

\section{Impact Parameter Space GPD -- Evolved}
\label{SecIPSEvolved}
Inserting the $\zeta_2$-evolved valence-quark GPDs drawn in Figs.\,\ref{fig:evolvedGPDpi}A, \ref{fig:evolvedGPDK} into the right-hand-side of Eq.\,\eqref{eq:IPDHgen}, one obtains the associated $\zeta_2$-evolved IPS GPDs depicted in Fig.\,\ref{fig:IPDGPDszeta2}.
To check our methods, we verified that the same results are produced by computing the Mellin moments of the $\zeta=\zeta_{\cal H}$ IPS GPD directly, evolving those moments according to Eq.\,\eqref{eq:diffDGLAP}, reconstructing the $x$-profile on each $|b_\perp|$ slice, and therefrom building the two-dimensional evolved IPS GPD.

Comparing Fig.\,\ref{fig:IPDGPDszeta2} with Fig.\,\ref{fig:IPDGPDs}, one sees that, as anticipated in Ref.\,\cite{Zhang:2021mtn} and sketched for the pion in Ref.\,\cite{Mezrag:2014jka}, evolution causes both the profiles to broaden and the maxima to drift toward $x=0$.
Each density, $2\pi |b_\perp| q^{\mathsf{P}}(x,b_\perp^2;\zeta_2)$, still has a global maximum, with locations:
\begin{subequations}
\begin{align}
\mbox{$\pi$:} & \quad (|x|,b_\perp/r_\pi)=(0.47,0.23)\,, \; {\mathpzc i}_\pi=0.55\,, \\
\mbox{$K_u$:} & \quad
(x,b_\perp/r_K)_u=(0.41,0.28)\,,\; {\mathpzc i}_K^u=0.49\,, \\
\mbox{$K_{\bar s}$:} & \quad
(x,b_\perp/r_K)_{\bar s}=(-0.48,0.22)\,, \; {\mathpzc i}_K^{\bar s}=0.61\,,
\end{align}
\end{subequations}
which may be contrasted with the results in Eqs.\,\eqref{IPSheights}.
In this case, the pion peak height equals the average of the $u$-in-$K$ and $\bar s$-in-$K$ heights, so that evolution has eliminated the relative pion excess.

All aspects of the comparisons just described are straightforwardly understood.
Under evolution, the dressed-valence degrees-of-freedom shed pieces of their ``partonic clothing'', thereby populating glue and sea distributions.
Momentum conservation therefore requires that the peak location move away from $x=1$.
This entails that the light-front momentum fraction carried by the valence degrees-of-freedom is reduced, so they play a smaller part in defining the centre of transverse momentum, in consequence of which the associated $|b_\perp|$ values are larger, leading to a broader distribution.
These features are all apparent in the comparison between Figs.\,\ref{fig:IPDGPDszeta2} and \ref{fig:IPDGPDs}.

\section{Summary and Perspective}
\label{epilogue}
The basic inputs for this survey were hadron-scale ($\zeta_{\cal H}$) pion and kaon valence-quark distribution functions (DFs), computed and shown to explain existing data elsewhere \cite{Cui:2020dlm, Cui:2020tdf, Roberts:2021nhw}.  Therefrom, we constructed algebraic \emph{Ans\"atze} for the light-front wave functions (LFWFs) of these mesons and subsequently generalised parton distributions (GPDs), $H(x,\xi,-\Delta^2;\zeta_{\cal H})$, defined throughout the DGLAP domain: $|x|\geq\xi$.  The GPDs may be judged realistic because they provide good descriptions of pion and kaon observables not used in their construction [Figs.\,\ref{fig:FFpi}, \ref{fig:FFK}].

Transforming these GPDs into impact parameter space (IPS), numerous insights were developed by exploiting their algebraic character [Sec.\,\ref{SecIPSGPD}].  For example, broadening of the valence-quark DFs, owing to the dynamics responsible for the emergence of hadron mass (EHM), is similarly expressed in $\langle b_\perp^2(x;\zeta_{\mathpzc H}) \rangle_{q}^{\mathsf P}$, the longitudinal light-front distribution of
the mean-square transverse light-front extent of the valence-$q$ parton in the $\mathsf P$-meson [Fig.\,\ref{fig:b2x}].  Moreover, there is a separation of baryon number in the kaon, with the $\bar s$ quark, on average, lying closer to the kaon's centre of transverse momentum than the $u$ quark [Fig.\,\ref{fig:IPDGPDs}].  The size of this displacement is determined by the scale of Higgs boson modulation of EHM.

Extension of our GPDs onto the ERBL domain, $|x|<\xi$, enabled calculation of the two gravitational form factors of both the pion and kaon and, subsequently, associated Breit-frame pressure distributions [Sec.\,\ref{SecPressure}].  Again capitalising on algebraic simplicity, we demonstrated that a meson's mass-squared form factor, $\theta_2$, is necessarily stiffer than its electromagnetic form factor.  On the other hand, available examples indicate that the pressure form factor, $\theta_1$, is typically softer.  In all cases, heavier objects have stiffer form factors [Figs.\,\ref{fig:theta2P}, \ref{fig:theta1P}].

Considering the pressure profiles obtained from $\theta_{1,2}$, the kaon is more compact than the pion and the near-core pressures in both these pseudo-Nambu-Goldstone bosons are commensurate with that thought to exist within neutron stars [Figs.\,\ref{fig:r2p}, \ref{fig:r2pK}].  Here, too, the magnitudes of differences between $\bar s$- and $u$-quark profiles are fixed by Higgs boson modulation of EHM.

For any comparison with empirical results relating to $\pi$ and $K$ GPDs, which may become available in future, the hadron scale GPDs must be evolved to scales $\zeta > m_p$, where $m_p$ is the proton mass.  We accomplished this using an all-orders scheme that delivers parameter-free predictions [Sec.\,\ref{SecAll}].  The approach was benchmarked against available data on the pion's valence-quark DF, delivering a prediction in agreement with an existing analysis [Fig.\,\ref{FigAppB}].  Thus validated, we employed the scheme to deliver parameter-free predictions for the valence, glue, and sea GPDs of the pion and kaon on the DGLAP domain [Figs.\,\ref{fig:evolvedGPDpi}, \ref{fig:evolvedGPDK}]: both glue and sea GPDs have maximal support in the neighbourhood $(x\simeq 0,\Delta^2\simeq 0)$.

The all-orders evolution scheme also enables one to arrive at predictions for the fraction of $m_{\pi,K}^2$ carried by different parton species at any resolving scale.  Eq.\,\eqref{Mass2Fraction} lists $\zeta=2\,$GeV results, whereat the mass-squared fraction carried by glue and sea combined matches that stored in the valence degrees-of-freedom.  This contrasts markedly with the apportionment at $\zeta_{\cal H}$, where all hadron properties are lodged entirely with the dressed valence degrees-of-freedom [Sec.\,\ref{SecGFF}].  Analogous results for the mass-squared radii are discussed in connection with Eq.\,\eqref{radiiFraction}: at $\zeta_2$, roughly one-half of each meson radius is contributed by glue and sea degrees-of-freedom.

The likewise $\zeta_2$-evolved $\pi$ and $K$ IPS GPDs contain no surprises [Sec.\,\ref{SecIPSEvolved}].  With respect to the $\zeta_{\cal H}$ results, they are dilated and flattened, and their maxima float closer to $x=0$, owing to evolution-induced unclothing of the dressed valence degrees-of-freedom.

In future, it is worth looking harder at GPD extension onto the ERBL domain.  Whereas we have chosen to employ algebraic approximations when inverting Radon transforms, it may be profitable to use our realistic DGLAP-domain GPD \emph{Ans\"atze} as testbeds in the development of reliable numerical methods.
Of at least equal importance is construction of realistic GPDs for the nucleon.  The nucleon distribution amplitudes built elsewhere \cite{Mezrag:2017znp}, based upon a dynamical quark+diquark picture of nucleon structure, might provide a useful starting point.

\begin{acknowledgments}
We are grateful for constructive comments from
F.~De~Soto, C.~Mezrag, 
J.~Segovia and J.-L.~Zhang.
Work supported by:
National Natural Science Foundation of China (grant nos.\ 12135007 and 11805097);
Jiangsu Provincial Natural Science Foundation of China (grant no.\ BK20180323);
Spanish Ministry of Science and Innovation (MICINN) (grant no.\ PID2019-107844GB-C22);
Junta de Andaluc\'ia (contract nos.\ P18-FR-5057, UHU-1264517);
and
University of Huelva (grant no.\ EPIT-2019).
\end{acknowledgments}

\appendix
\setcounter{equation}{0}
\setcounter{figure}{0}
\setcounter{table}{0}
\renewcommand{\theequation}{\Alph{section}.\arabic{equation}}
\renewcommand{\thetable}{\Alph{section}.\arabic{table}}
\renewcommand{\thefigure}{\Alph{section}.\arabic{figure}}

\section{PTIR LFWF in the chiral limit}
\label{app:PTIRfac}
Consider the chiral limit, so that $m_{\mathsf P}^2=0$, $M_u = M_{\bar h}$, $v_{\mathsf P}=0$.  In this case, Eqs.\,\eqref{Inputs} simplify greatly and, consequently, Eq.\,\eqref{psiLFWF} becomes:
\begin{subequations}
\begin{align}
\psi_{{\mathsf P}}^{u}(x,k_\perp^2;\zeta_{\mathpzc H}) & =
[{\mathpzc u}^{\mathsf P}(x;\zeta_{\mathpzc H})]^{\tfrac{1}{2}}
\tilde\psi_{\mathsf P}^u(k_\perp^2;\zeta_{\mathpzc H})\,, \\
\tilde\psi_{\mathsf P}^u(k_\perp^2;\zeta_{\mathpzc H}) & =
4\sqrt{3} \pi \frac{M_u^3}{\left(k_\perp^2+M_u^2\right)^2} \,, \label{tildepsi}
\end{align}
\end{subequations}
where we have used Eqs.\,\eqref{eq:DA}, \eqref{eq:uMzetaH}.  Evidently, the chiral limit LFWF assumes a factorised (separable) form; so, the corresponding $u$-quark GPD is given by Eq.\,\eqref{eq:Hfac}:
\begin{align} \label{eq:Hfac-app}
&H^u_{\mathsf{P}}(x,\xi,t;\zeta_H) = \nonumber \\
&\theta(x_-) \sqrt{u^{\mathsf{P}}\left(x_-;\zeta_H \right) u^{\mathsf{P}}\left(x_+;\zeta_H \right)} \; \Phi_{\mathsf{P}}^u\left( z; \zeta_H \right)\,.
\end{align}
Now inserting Eq.\,\eqref{tildepsi} into Eq.\,\eqref{eq:Phi}, one arrives at the following compact result:
\begin{align}
&\Phi_{\rm P}^u(z;\zeta_H)= \frac{6 M_u^6}{\displaystyle \left(z + 4M_u^2\right)^3}
\left( \rule[0cm]{0cm}{0.6cm} 10 + \frac{z}{M_u^2} + \frac{8(z+M_u^2)}{z} \right.  \nonumber \\
&\times
\left. \left[
\sqrt{\frac{z+4M_u^2}{z}} \arctanh{\left(\sqrt{\frac{z}{z+4M_u^2}}\right)} -1\right]
\right) \,.
\label{eq:PhiA}
\end{align}

One can expand $\Phi_{\rm P}(z;\zeta_H)$ around $z=0$ to obtain
\begin{equation}
\Phi_{\rm P}(z;\zeta_H) = 1 - \frac 3 5 \frac{z}{M_u^2} + {\cal O}(z^2)\,;
\end{equation}
then, according to Eq.\,\eqref{eq:dphiMdz}, specialised to the pion,
\begin{equation}\label{eq:dphidz}
\left. - \,\frac d {dz} \Phi_\pi(z;\zeta_H) \right|_{z=0} = \frac 3 {5 M_u^2} =
\frac{r_\pi^2}{6 \langle x^2 \rangle_u^{\zeta_H}} \,.
\end{equation}
Thus the framework yields a very simple result for the pion charge radius in the chiral limit:
\begin{equation}\label{eq:rpi}
r_\pi = \sqrt{\frac{18}{5} \langle x^2 \rangle_u^{\zeta_H}} \, \frac 1 {M_u} = 0.66 \; \mbox{\rm fm} \,,
\end{equation}
using Eq.\,\eqref{PDFeqPDA2} and Tables~\ref{tab:params}, \ref{tab:DFs}.
The result in Eq.\,\eqref{eq:rpi} presents a fair match with contemporary values inferred from experiment: \cite{Zyla:2020zbs, Cui:2021aee}: $0.659(4)\,$fm, $0.640(7)\,$fm, respectively.

The mathematical structure of Eq.\,\eqref{eq:rpi} is also interesting.  It indicates that the electromagnetic radius of the pion is determined by the second light-front momentum-fraction Mellin moment of the valence-quark DF, which measures its dilation, in units determined by the scale of the chiral-limit light-quark dressed-mass.

In closing this appendix, it is worth highlighting that the GPD constructed using Eqs.\,\eqref{eq:Hfac-app}, \eqref{eq:PhiA} is almost pointwise identical to the pion GPD depicted in Fig.\,\ref{fig:piGPD}, which was obtained from the PTIR LFWF \emph{Ansatz}, as a comparison between Figs.\,\ref{fig:piGPD} and \ref{fig:piGPDFac} would suggest.

\bibliographystyle{elsarticle-num-names}
\bibliography{../../../../CollectedBiB}

\end{document}